\documentclass[aps,prc,twocolumn,showpacs,floatfix,nofootinbib,preprintnumbers,groupedaddress,amsmath,amssymb,longbibliography,amsproc]{revtex4-1}
\usepackage{epsfig,dsfont,amssymb,amsmath,amsthm,amsfonts,amsbsy,mathrsfs}
\usepackage{graphicx}
\usepackage{amsmath}
\usepackage{amssymb}%
\usepackage{dcolumn}
\usepackage{multirow}
\usepackage[colorlinks=true,linkcolor=blue,citecolor=blue,urlcolor=blue]{hyperref}
\usepackage{bbold}

\graphicspath{{.}{./figs/}}

\usepackage{stackengine}

\usepackage[disable]{todonotes}

\newcommand{\be}{\begin{equation}}
\newcommand{\ee}{\end{equation}}
\newcommand{\ba}{\begin{eqnarray}}
\newcommand{\ea}{\end{eqnarray}}

\newcommand{\MeV}{\text{MeV}}

\newcommand{\ai}{\textit{ab initio}}

\newcommand{\ph}[1]{$#1p$-$#1h$}

\stackMath
\newcommand\tsup[2][2]{%
 \def\useanchorwidth{T}%
  \ifnum#1>1%
    \stackon[-.5pt]{\tsup[\numexpr#1-1\relax]{#2}}{\scriptscriptstyle\sim}%
  \else%
    \stackon[.5pt]{#2}{\scriptscriptstyle\sim}%
  \fi%
}

\DeclareMathAlphabet{\mathpzc}{OT1}{pzc}{m}{it}

\begin{document}

\title{Angular-momentum projection in coupled-cluster theory: structure of $^{34}$Mg}

\author{G.~Hagen}
\affiliation{Physics Division, Oak Ridge National Laboratory, Oak Ridge, TN 37831, USA}
\affiliation{Department of Physics and Astronomy, University of Tennessee, Knoxville, TN 37996, USA}

\author{G.~R. Jansen}
\affiliation{National Center for Computational Sciences, Oak Ridge National Laboratory, Oak Ridge, TN 37831, USA}
\affiliation{Physics Division, Oak Ridge National Laboratory, Oak Ridge, TN 37831, USA}

\author{J.~G. Lietz}
\affiliation{National Center for Computational Sciences, Oak Ridge National Laboratory, Oak Ridge, TN 37831, USA}

\author{S.~J.~Novario}
\altaffiliation[Present address: ]{Theoretical Division, Los Alamos National Laboratory, Los Alamos, New Mexico 87545, USA}
\affiliation{Physics Division, Oak Ridge National Laboratory, Oak Ridge, TN 37831, USA}
\affiliation{Department of Physics and Astronomy, University of Tennessee, Knoxville, TN 37996, USA}

\author{Z. H. Sun}
\affiliation{Physics Division, Oak Ridge National Laboratory, Oak Ridge, TN 37831, USA}
\affiliation{Department of Physics and Astronomy, University of Tennessee, Knoxville, TN 37996, USA}

\author{T.~Papenbrock}
\affiliation{Physics Division, Oak Ridge National Laboratory, Oak Ridge, TN 37831, USA}
\affiliation{Department of Physics and Astronomy, University of Tennessee, Knoxville, TN 37996, USA}

\author{T. Duguet}
\affiliation{IRFU, CEA, Universit{\'e} Paris-Saclay, F-91191 Gif-sur-Yvette, France}
\affiliation{KU Leuven, Instituut voor Kern- en Stralingsfysica, 3001 Leuven, Belgium}

\author{A. Tichai}
\affiliation{Institut f{\"u}r Kernphysik, Technische Universit{\"a}t Darmstadt, Darmstadt, Germany}
\affiliation{Max-Planck-Institut f{\"u}r Kernphysik, Heidelberg, Germany}
\affiliation{ExtreMe Matter Institute EMMI, GSI Helmholtzzentrum f{\"u}r Schwerionenforschung GmbH, Darmstadt, Germany}

\begin{abstract}
Single-reference coupled-cluster theory is an accurate and affordable
computational method for the nuclear many-body problem. For open-shell nuclei, the reference state typically breaks rotational invariance and angular momentum must be restored as a
good quantum number.  
We perform angular-momentum projection after variation and employ the disentangled coupled-cluster formalism and a Hermitian 
approach. We compare our results with benchmarks for $^{8}$Be and $^{20}$Ne
using a two-nucleon interaction from chiral effective field theory and for
$pf$-shell nuclei within the traditional shell model. We 
compute the rotational band in the exotic nucleus $^{34}$Mg and find agreement with data. 
\end{abstract}


\maketitle

\section{Introduction}

While angular momentum, parity, and the numbers of neutrons and protons are good quantum numbers of atomic nuclei, mean-field states often break  symmetries of the nuclear Hamiltonian~\cite{schmidt1989,bender2003,sheikh2021}. This is a blessing and a burden: On the one hand, the actual breaking of symmetries by the mean field corresponds to the emergent symmetry breaking in atomic nuclei; it informs us about deformation and superfluidity (in the case of breaking of angular momentum and particle numbers, respectively) and identifies the corresponding Nambu-Goldstone modes as low-energy degrees of freedom. 
Symmetry-breaking product states also are the starting point of 
single-reference methods (such as coupled-cluster theory~\cite{kuemmel1978,hagen2014}, in-medium similarity renormalization group~\cite{tsukiyama2011,hergert2016,stroberg2017,stroberg2019,stroberg2021,heinz2021}, Green's function / Gorkov approaches~\cite{dickhoff2004,soma2013}, and perturbation theory~\cite{holt2014,tichai2016,hu2016,tichai2018,tichai2018ncsmpt,tichai2020}) that capture dynamical correlations beyond the mean field~\cite{tichai2018,novario2020,soma2021}. 
On the other hand, the restoration of broken symmetries is necessary when one wants to obtain precise ground-state energies, excited states, or transition matrix elements between states with definite quantum numbers. 

One can, of course, address phenomena related to emergent symmetry
breaking without actually breaking any
symmetry~\cite{caurier2005,caurier2007,caprio2013,jansen2014,bogner2014,caprio2015,maris2015}.
However, such exact computations typically scale exponentially with increasing  number of active nucleons because the emergence of a 
new low-energy scale associated with, e.g., collective rotational excitations in intrinsically
deformed nuclei requires the superposition of $A$-particle--$A$-hole excitations in a nucleus with mass number $A$. The successful computation of
such states is then limited to light nuclei or small shell-model spaces. Monte Carlo methods with angular momentum projection extended such computations to somewhat larger model spaces~\cite{shimizu2012}. 
In contrast, symmetry-adapted approaches~\cite{dytrych2013,dytrych2020}
and effective theories~\cite{papenbrock2011,coelloperez2015,chen2017,papenbrock2020,alnamlah2021} are
simpler because they employ the degrees of freedom that are relevant
at such low energies. The effective theories have advantages and
disadvantages: they have less predictive power because low-energy constants
such as the moment of inertia, for instance, must be taken from data or microscopic
computations. However, they allow us to estimate uncertainties and
reveal the simple patterns that complex systems exhibit based on their
symmetries and symmetry breaking.

While nuclear density-functional theory has been the main workhorse to compute nuclei across the nuclear
chart~\cite{bender2003,paar2007,erler2012,goriely2009,niksic2011,erler2012,shen2019},
polynomially scaling computational
methods~\cite{dickhoff2004,hagen2014,hergert2016,tichai2020} based on
nucleon-nucleon and three-nucleon interactions from effective field
theories of quantum
chromodynamics~\cite{vankolck1999,epelbaum2008,machleidt2011} are advancing
steadily towards heavier nuclei~\cite{binder2013b,hagen2017,morris2018,holt2021,novario2020,hu2021}. In
this paper we employ single-reference coupled-cluster
theory~\cite{kuemmel1978,bartlett2007,hagen2014}, start from a
deformed but axially symmetric reference state~\cite{novario2020}, and perform angular momentum projection.
Alternative methods for angular momentum projection are the  multi-reference in-medium similarity renormalization group method~\cite{Hergert:2014iaa} and many-body perturbation theory based on a reference state obtained via the projected generator coordinate method~\cite{Frosini:2021fjf,Frosini:2021sxj,Frosini:2021ddm}.

Several works discuss symmetry projection within coupled-cluster theory~\cite{duguet2015,qiu2017,Duguet:2015yle,qiu2017,qiu2018,qiu2019,tsuchimochi2018,mizusaki2021}. In particular the disentangled cluster formalism of Refs.~\cite{qiu2017,qiu2018,qiu2019} performs angular-momentum and particle-number projections. Here, the simplest approach is to insert the projected coupled-cluster state into the Schr\"odinger equation and project from the left onto the symmetry-broken reference state. In standard coupled-cluster theory, however, one would start from a bi-variational energy functional of the projected Hamiltonian, where the bra state consists of a linear superposition of particle-hole excitations~\cite{bartlett2007}. Such an approach is more accurate in general~\cite{qiu2017}. In both approaches the energy expression 
is non-Hermitian and the ket state is the usual exponential wavefunction ansatz. 

There are coupled-cluster methods where the bra state is treated more on equal footing with the ket state, such as the extended coupled cluster~\cite{arponen1983,arponen1987}, the quadratic coupled cluster~\cite{vanvoorhis2000,byrd2002},
the expectation-value coupled-cluster~\cite{bartlett1988}, and the variational coupled-cluster~\cite{szalay1995} methods. 
While these methods are more accurate than the standard coupled-cluster method they come at a significantly higher computational cost. In this work we also follow a middle way using a Hermitian energy functional in the projection, inspired by the variational coupled-cluster method~\cite{szalay1995}. In this approach the rotation operator can be treated exactly but the exponential coupled-cluster state must be truncated. 

This paper is organized as follows. In Sect.~\ref{sec:corr} we discuss the role of static and dynamical correlations in angular momentum projection. Section~\ref{refstates} introduces the coupled-cluster method and discusses reference states. In Sect.~\ref{sec:proj} we
present the theoretical derivations of angular-momentum projections after variation. Here, both the non-Hermitian and Hermitian projection methods will be described. In Sect.~\ref{sec:eft} we construct collective Hamiltonians within an effective theory and discuss the resolution-scale dependence and size consistency in projections. In Sect.~\ref{sec:results} we show results for angular-momentum projection of the nuclei $^8$Be, $^{20}$Ne, and $^{34}$Mg. Section~\ref{sec:pf} deals with angular momentum projection of coupled-cluster computations 
in the traditional shell model. We present a discussion and summary in
Sect.~\ref{sec:sum}.

\section{Static and dynamical correlations}
\label{sec:corr}

The restoration of rotational symmetry of
a nucleus with mass number $A$ involves $A$-particle--$A$-hole
(\ph{A}) excitations, because it requires the rotation of a deformed
nucleus. This makes it challenging to keep size extensivity~\cite{duch1994} together with computational affordability. Fortunately,
the problem is somewhat less daunting, because the computation of a
nuclear ground-state energy involves dynamic and non-dynamic (or
static) correlations, see, e.g., Ref.~\cite{ramoscordoba2016} for a
recent discussion of this topic.

Dynamic correlations mix a dominant configuration and a large number 
of configurations carrying small individual weights but yielding a 
significant energy contribution. An example is provided by the
dominant (symmetry breaking) Hartree-Fock reference state and its \ph{2} and \ph{3} excitations. 
While the number of relevant configurations is large, it only grows polynomially with mass number and 
model-space size for any targeted precision. As dynamical correlations bring in the lion's share of 
nuclear binding one needs size-extensive methods to capture them accurately. 

In contrast, static correlations are caused by a number of equally important configurations. 
Any rotation of a deformed reference state, for example, yields a configuration that is 
degenerate in energy. Mixing these states, e.g. via angular momentum projection, lowers the 
rotational zero-point energy. As we will see below, this energy gain decreases with increasing 
mass number. For heavy deformed nuclei the energy gain from projection is of the order of the 
nuclear level spacing near the ground state.

Based on this discussion, we can decompose the ground-state energy as
\be
\label{Edecomp}
E = E_{\rm ref} + \Delta E + \delta E \ .
\ee
Here $E_{\rm ref}$, $\Delta E$, and $\delta E$ denote the energy of the
symmetry-breaking reference state, the energy associated with dynamical
correlations and the static energy from angular momentum
restoration, respectively. These energies scale as $E_{\rm ref}\propto A$,
$\Delta E\propto A$ and as $\delta E \propto A^{-\eta}$ where
$\eta>0$. We estimate that $\eta\approx 1/6$, based on the scaling $\delta
E\sim \langle J^2\rangle/\Theta$ where $\Theta \sim A^{5/3}$ expresses
the scaling of the moment of inertia and $\langle J^2\rangle\sim
A^{3/2}$ is the scaling of the angular momentum for the unprojected
reference state found empirically in projected mean-field
calculations~\cite{bertsch2019}. Because of its smallness and scaling with $A$, one does not need size-extensive methods 
to compute $\delta E$. In contrast, we compute $E_{\rm ref}$ and $\Delta E$ via the symmetry-breaking and 
size-extensive Hartree-Fock and coupled-cluster methods, respectively.

\section{Coupled-cluster method and reference states}

\label{refstates}

We use single-particle states $|p\rangle$ where $p=(n,\pi,j_z,t_z)$ denotes a quantum number $n$, parity $\pi$, angular momentum projection $j_z$, and isospin projection $t_z$. We have $|p\rangle
\equiv c_p^\dagger|0\rangle$ where $|0\rangle$ is the vacuum and $c_p^\dagger$ creates a nucleon. The creation operators and 
corresponding annihilation operators $c_p$ obey the usual anti-commutation relations for fermions.

The Hamiltonian is
\be
\label{ham}
H = \sum_{pq} \varepsilon_{pq} c^\dagger_p c_q +{1\over 4}\sum_{pqrs} v_{pqrs} c^\dagger_p c^\dagger_q c_s c_r \ .
\ee
While three-nucleon forces are unavoidable in nuclear physics~\cite{fujita1957,bedaque1999,epelbaum2009,hammer2013}, 
we omitted them for simplicity and to benchmark with other methods that employ the same interaction. 

Our calculations start from the symmetry-unrestricted product state
\begin{equation}
| \Phi \rangle \equiv \prod_{i=1}^A c_i^\dagger |0\rangle \, .
\end{equation}
This state breaks angular momentum but we assume that its projection $J_z$ is conserved. We also assume that parity, isospin projection, and mass number $A$ are conserved, and that we deal with even-even nuclei. Then, the reference $|\Phi\rangle$ is also invariant under time reversal (and occupied single-particle states come in degenerate Kramer pairs), invariant under rotations by an angle $\pi$ around any axis that is perpendicular to the symmetry axis (denoted as ${\cal R}$-parity~\cite{bohr1975}), and fulfills $J_z| \Phi \rangle=0$.

We discuss three choices for the reference state. The Hartree-Fock state, which minimizes the energy, is the first. The second reference is obtained as follows. We minimize the energy under the constraint of a fixed expectation value $q_{20}$ of the mass quadrupole operator
\be
Q_{20} = \sqrt{16\pi\over 5} \sum_{i=1}^{A} r_i^2 Y_{20}(\theta_i,\phi_i) \ .
\label{Q20}
\ee
Such a constrained Hartree-Fock calculation employs the Routhian
\be
R \equiv H - \lambda Q_{20} \, ,
\ee
where the Lagrange multiplier $\lambda$ is adjusted to obtain the desired value  $q_{20}=\langle Q_{20}\rangle$. We use the augmented Lagrangian method~\cite{staszscak2010} for this purpose and generate
an energy curve as a function  $q_{20}$. Performing an angular-momentum projection after variation at fixed $q_{20}$ then yields a new curve whose minimum we seek. The corresponding product state is the Hartree-Fock restricted-variation-after-projection (HF-RVAP) reference~\cite{Rod05,Ripoche:2017ydv}.
The third reference state is obtained from a variation-after-projection Hartree-Fock (HF-VAP) calculation~\cite{ringschuck,hoyos2012}. The HF-VAP reference state constitutes the optimal (i.e. lowest energy) product state that can be obtained while including the projection onto zero angular momentum.

Once the single-particle basis is determined we normal-order the Hamiltonian~(\ref{ham}) with respect to the vacuum state and write
\begin{equation}
\label{HNO}
    H= E_{\rm ref} + H_N \ .
\end{equation}
Here, $E_{\rm ref}=\langle \Phi|H|\Phi\rangle=\sum_{i=1}^A\varepsilon_{ii}+{1\over 2}\sum_{i,j=1}^Av_{ijij}$ is the vacuum energy and the normal-ordered Hamiltonian is
\be
\label{ham}
H_N = \sum_{pq} f_{pq} \left\{c^\dagger_p c_q\right\} +{1\over 4}\sum_{pqrs} v_{pqrs} \left\{c^\dagger_p c^\dagger_q c_s c_r\right\} \ .
\ee
The curly brackets denote normal ordering and the Fock-matrix elements are  $f_{pq}=\varepsilon_{pq}+\sum_{i=1}^A v_{piqi}$. The papers~\cite{ripoche2020,Frosini:2021tuj} proposed how to deal with three-body forces in the normal-ordered two-body approximation~\cite{hagen2007a,roth2012} in symmetry-breaking situations.  

Coupled-cluster theory parameterizes the ground state as 
\begin{align}
|\Psi \rangle \equiv e^T | \Phi \rangle \ . 
\label{eq:CC}
\end{align}
The cluster operator 
\begin{align}
T \equiv T_1 + T_2 + ...+ T_A 
\label{eq:CCansatz}
\end{align}
consists of \ph{n} excitation operators
\begin{align}
T_n = \frac{1}{(n!)^2} \sum_{\substack{a_1 ... a_n \\ i_1 ... i_n} }
   t^{a_1...a_n}_{i_1...i_n} c^\dagger_{a_1} \cdots
  c^\dagger_{a_n} c_{i_n}  \cdots c_{i_1} \, .  \label{clusteroperators}
\end{align}
Here and in what follows, $i,j,k,\ldots$ label occupied single-particle states while $a,b,c,\ldots$ refer to unoccupied states. Generic states are labeled as $p,q,r,\ldots$. Coupled-cluster theory is a powerful method because the numerically inexpensive singles and doubles (CCSD) approximation $T\approx T_1 + T_2$
yields about 90\% of the correlation energy for closed-shell nuclei; the inclusion of triples $T=T_1+T_2+T_3$, still numerically affordable via perturbative methods, yields about 98\%. In CCSD, the cluster amplitudes are computed for a given Hamiltonian and reference state by solving the well-known coupled-cluster equations~\cite{shavittbartlett2009}
\begin{align}
  \langle \Phi_i^a |e^{-T}H_Ne^T|\Phi\rangle &= 0 \ , \nonumber\\
  \langle \Phi_{ij}^{ab} |e^{-T}H_Ne^T|\Phi\rangle &= 0 \ .
\end{align}
Here, $|\Phi_{i_1\cdots i_n}^{a_1\cdots a_n}\rangle \equiv c^\dagger_{a_1}\cdots c^\dagger_{a_n}c_{i_n}\cdots c_{i_1}|\Phi\rangle$ is a \ph{n} excitation of the reference state. The energy associated with the (dynamical) CCSD correlations is
\begin{equation}
    \Delta E_{\rm CCSD}= \langle \Phi|e^{-T}H_Ne^T|\Phi\rangle  \ .
\end{equation}
As $\langle \Phi|e^{-T} = \langle\Phi|$ and $\langle\Phi|\Psi\rangle=1$ we can rewrite the total energy as
\begin{equation}
\label{Ecorr}
E=E_{\rm ref}+\Delta E_{\rm CCSD}= \frac{\langle \Phi|H|\Psi\rangle}{\langle \Phi|\Psi\rangle} \ .
\end{equation}
This energy expression will be modified below for the evaluation of the static correlation energy ($\delta E$) associated with angular-momentum projection.

For any truncation of the cluster operator the state $|\Psi\rangle$ breaks the same symmetries as the reference state $|\Phi\rangle$. 
Only the full expansion~(\ref{eq:CCansatz}) restores the symmetries broken by the reference. In what follows we restore the rotational invariance 
via angular-momentum projection. In addition to CCSD , we will also use the CCD approximation $T=T_2$. In the Hartree-Fock basis there is little difference between CCD and CCSD (because singles excitations are small). This is different in other single-particle bases. 

The angular momentum projection methods we employ come at a considerable higher computational cost compared to the unprojected coupled-cluster calculations, and thus require us to work in smaller model spaces. The residual basis dependence can to a large extent be eliminated by including the singles-excitations in the projection, i.e. by using CCSD rather than CCD. In this work we also  employ a natural-orbital basis obtained from second-order many-body perturbation theory~\cite{tichai2019,novario2020}. 
In this approach we compute the Hartree-Fock state in a large harmonic oscillator basis, and in a final step truncate the normal-ordered Hamiltonian that enter the coupled-cluster computations to a smaller model-space~\cite{hoppe2021, kortelainen2021}. This alleviates the dependence on the oscillator frequency. We note, however, that such a truncation of the normal-ordered Hamiltionian breaks rotational invariance to a small extent.

\section{Angular momentum projection}
\label{sec:proj}
Throughout this Section we use the Hamiltonian in the form of Eq.~(\ref{HNO}) [rather than Eq.~(\ref{ham})] because this facilitates the evaluation of the matrix elements we need to compute within the angular momentum projection. 

\subsection{Projection operator}

The operator 
\begin{equation}
  \label{projector}
P_{J} = {1\over 2}\int\limits_0^\pi {\rm d}\beta \sin\beta d^J_{00}(\beta) R(\beta) \ ,
\end{equation}
projects a state with $J_z=0$ onto total angular momentum $J$. 
Here,
\begin{equation}
R(\beta) \equiv e^{-i\beta J_y} 
\end{equation}
denotes the operator that rotates by the angle $\beta$ around the $y$ axis, and 
$d^{J}_{MK}(\beta)$ is related to the Wigner $D$-function function via~\cite{varshalovich88a}
\begin{align}
D^J_{MK}(\alpha,\beta,\gamma)\equiv e^{-iM\alpha}d^J_{MK}(\beta)e^{-iK\gamma} \, .
\end{align}
The ${\cal R}$ symmetry allows us to restrict the domain of
integration in Eq.~(\ref{projector}) to the interval
$[0,\pi/2]$ because
\begin{align}
  R(\pi-\beta)|\Phi\rangle = R(\beta)|\Phi\rangle \ .
\end{align}
In a spherical single-particle basis $\{| p \rangle \equiv  \vert \alpha_p j_p m_p\rangle\}$ of the one-body Hilbert space, matrix elements of the rotation operator are given by
\be
\langle \alpha_p j_p m_p |R(\beta)|\alpha_q j_q m_q\rangle =  \delta_{\alpha_p\alpha_q} \delta_{j_p j_q} d_{m_p m_q}^{j_p}(\beta) \ . \label{reducedwigner}
\ee

\subsection{Non-Hermitian projection formalism}
\label{app:disentangled}

\subsubsection{Projected energy}

The disentangled cluster formalism~\cite{qiu2017,qiu2018,qiu2019} (in its simplest implementation) is based on the standard non-Hermitian energy expression of coupled-cluster theory. In this approach the energy projected onto angular momentum $J$ after variation is 
\begin{equation}
  \label{pavCC}
E^{(J)}  \equiv \frac{ \langle \Phi | P_J H  |\Psi \rangle }{\langle \Phi | P_J | \Psi \rangle} \ .
\end{equation}
In this expression we used that $P_J$ commutes with $H$. The projected energy is a natural modification of Eq.~(\ref{Ecorr}). 
We insert Eq.~\eqref{projector} into Eq.~\eqref{pavCC} and find
\begin{align}
E^{(J)} = \frac{\int\limits_0^\pi {\rm d}\beta \sin\beta d_{00}^J(\beta){\cal H}(\beta)}
  {\int\limits_0^\pi {\rm d}\beta \sin\beta d_{00}^J(\beta){\cal N}(\beta)} \ .
\label{eq:proj_energy}
\end{align}
Here, 
\begin{align}
  {\cal N}(\beta) &\equiv \langle \Phi|R(\beta) |\Psi\rangle \, ,
\label{eq:nkernel}
\end{align}
and 
\begin{align}
  {\cal H}(\beta) &\equiv \langle \Phi|R(\beta)H |\Psi\rangle \, ,
\label{eq:hkernel}
\end{align}
are the norm and Hamiltonian kernels, respectively.  The ${\cal R}$ symmetry implies
\begin{align}
  \label{Rkernels}
  {\cal N}(\pi-\beta)&={\cal N}(\beta) \, , \nonumber\\
  {\cal H}(\pi-\beta)&={\cal H}(\beta) \, .
\end{align}
We also have ${\cal H}(0)= E_{\rm ref}+\Delta E$ and ${\cal N}(0)=1$.
With view on Eq.~(\ref{Edecomp}) the static correlation energy associated with angular momentum projection is then $\delta E=E^{(0)}- E$. 

\subsubsection{Disentangled cluster formalism}

Thouless' theorem~\cite{thouless1960} allows us to write~\cite{qiu2017}
\begin{equation}
  \label{thouless}
\langle \Phi|R(\beta) = \langle	\Phi|R(\beta)|\Phi\rangle \langle\Phi| e^{V(\beta)} \ .
\end{equation}
Here the de-excitation operator is
\begin{equation}
V(\beta) = \sum_{ia} V_a^i(\beta) c^\dagger_i c_a \, .  \label{thoulessoperator}
\end{equation}
Expressions for the matrix elements $V_a^i$ and mean-field norm kernel $\langle \Phi|R(\beta)|\Phi\rangle$ can be found in Ref.~\cite{qiu2017}. 
In what follows we suppress the explicit dependence of various quantities, e.g. $V(\beta)$, on the angle $\beta$.

By virtue of Eq.~(\ref{thouless}) the kernels become
\begin{subequations}
\begin{align}
  \label{kernels2}
  {\cal N}(\beta) &= \langle \Phi|R(\beta)|\Phi\rangle \langle\Phi| e^Ve^T|\Phi\rangle \, , \\
  {\cal H}(\beta) &= \langle \Phi|R(\beta)|\Phi\rangle \langle\Phi| e^V H e^T|\Phi\rangle \, ,
\end{align}
\end{subequations}
and we are left with the evaluation of the reduced kernels
\begin{subequations}
\begin{align}
  \mathpzc{n}(\beta) &\equiv \langle\Phi| e^Ve^T|\Phi\rangle \, ,\\
  \mathpzc{h}(\beta) &\equiv \langle\Phi| e^V H e^T|\Phi\rangle \, .
\end{align}%
\label{smallkernels}%
\end{subequations}
Inserting the identity operator $1=e^{-V}e^V$ introduces the similarity transformation~\cite{duguet2015,qiu2017}
\begin{equation}
  \label{HVsim}
\overline{H} \equiv e^V H e^{-V} \, . 
\end{equation}
As $V$ is a one-body operator, $\overline{H}$ has the same particle rank as $H$, i.e. no many-body forces of higher rank are induced by the
similarity transformation. Contrary to $H$, the operator $\overline{H}$ does not commute with $J_z$ and this significantly increases the number of non-zero matrix elements that need to be stored. 
Using $\overline{H}$ yields
\begin{equation}
\label{kernels3}
  \mathpzc{h}(\beta) = \langle\Phi| \overline{H}e^V e^T|\Phi\rangle \, .
\end{equation}

The exact evaluation of $e^V e^T$ in Eq.~(\ref{kernels3}) is exponentially expensive. To make progress we introduce the disentangled cluster operator 
\begin{equation}
    W(\beta)= W_0 +W_1 +W_2 +\cdots +W_A
\end{equation}
via
\begin{align}
  \label{disentangled}
e^V e^{T} |\Phi\rangle \equiv e^{W_0 +W_1 +W_2 +\cdots}|\Phi\rangle \ . 
\end{align}
Here $W_k$ is a $k$-body excitation operator. In contrast to $T$, the operator $W(\beta)$ also contains a zero-body term $W_0(\beta)$. Even if $T$ is truncated within the CCSD approximation $T = T_1+T_2$, the operator $W(\beta)$ contains up to $A$-body operators. To keep the computation affordable one needs to truncate at some rank $k\ll A$.  Then neither $e^V$ nor $e^T$ are exactly contained in the truncated $e^W$ (note however that $e^{T_1}$ will be treated exactly as it only amounts to a rotation of the basis as will be shown below). As a consequence the ${\cal R}$ symmetry of the kernels [see Eq.~\eqref{Rkernels}] is lost and the projection of the Hamiltonian $H= a \hat{J}^2$ only leads to approximate values  $E^{(J)}\approx a J(J+1)$. As we will see below, however, the disentangled approach is quite accurate. The truncation is a controlled approximation and can be systematically improved. 

The matrix elements of the operator $W(\beta)$ are computed  by solving the following set of ordinary differential
equations~\footnote{Equation~\eqref{dgl} corrects typographical errors found in Ref.~\cite{qiu2017}.}
\begin{align}
  \label{dgl}
  {{\rm d}\over{\rm d}\beta} W_0 &= X_a^iW_i^a \ , \nonumber\\
  {{\rm d}\over{\rm d}\beta} W_i^a	&= X_c^k\left(W_{ik}^{ac}-W_k^aW_i^c \right) \, , \nonumber\\
  {{\rm d}\over{\rm d}\beta} W_{ij}^{ab} &= X_c^k \left[W_{ijk}^{abc}-P(ij)W_{kj}^{ab}W_i^c -P(ab)W_{ij}^{cb}W_k^a  \right] \ , \nonumber\\
      & \vdots 
\end{align}
where
\begin{equation}
  \label{Xmat}
X^i_a \equiv \left({{\rm d}V \over{\rm d}\beta}\right)^i_a \ , 
\end{equation}
and where $P(ab)$ denotes the permutation operator. Expressions of the matrix elements $X_i^a$ can be found in Ref.~\cite{qiu2017}. The initial conditions for
$\beta=0$ are $W_0(0)=0$ and $W_k(0)=T_k$ for all $k$.

\subsubsection{Approximation schemes}

In this work we truncate $W=W_0+W_1+W_2$ at the two-body level. 
The Baker-Campbell-Hausdorff expansion yields
\begin{subequations}
\label{reducedkernelsfinal}
\begin{align}
  \mathpzc{n}(\beta) &= e^{W_0} \, , \\
  \mathpzc{h}(\beta) &= e^{W_0} \langle \Phi| \overline{H} \left(1+W_1+W_2+{W_1^2\over 2}\right)|\Phi\rangle \, .
\end{align}
\end{subequations}

In Ref.~\cite{qiu2017} the disentangled formalism was developed for
the coupled-cluster doubles (CCD) approximation $T = T_2$. Those authors noted that the extension to
$T = T_1+T_2$ could be achieved by inserting the identity $e^{T_1}e^{-T_1}$ to the left and right of the rotation operator in the original expression of the approximate kernels such that
\begin{subequations}
\label{eq:kernel-mod}
\begin{align}
  {\cal N}(\beta) &\equiv \langle \Phi|\breve{R}(\beta)e^{T_2}|\Phi\rangle \, , \\
  {\cal H}(\beta) &\equiv \langle \Phi|\breve{R}(\beta)\breve{H}e^{T_2}|\Phi\rangle \, .
\end{align}
\end{subequations}
Here the similarly transformed operator
\be
\label{T1simO}
\breve{O} \equiv e^{-T_1} O e^{T_1}
\ee
amounts to a change of the single-particle basis. Given the form of the approximate kernels in Eq.~\eqref{eq:kernel-mod}, all equations at play for $T = T_2$ can be reused by substituting $R(\beta)$ and $H$ by $\breve{R}(\beta)$ and $\breve{H}$, respectively, and by using $W_1(\beta=0)=0$ as the initial condition. We checked that the two equivalent ways to include $T_1$ lead to identical results.

\subsection{Hermitian projection formalism}

\subsubsection{Projected energy}

The Hermitian ansatz for the projected-after-variation energy functional
\be
\label{EI}
E^{(J)}  = \frac{\langle\Psi|P_J H|\Psi\rangle}{\langle\Psi|P_J |\Psi\rangle} \, ,
\ee
where $\vert \Psi \rangle$ is the coupled-cluster state~(\ref{eq:CC}), can be rewritten as
\ba
E^{(J)} = \frac{\int\limits_0^{\pi/2}{\rm d}\beta \sin\beta d^J_{00}(\beta) {\cal H}_{H}(\beta)}
{\int\limits_0^{\pi/2}{\rm d}\beta \sin\beta d^J_{00}(\beta) {\cal N}_{H}(\beta)} \ , 
\ea
using the norm and Hamiltonian kernels 
\begin{equation}
\begin{aligned}
  \label{kern1}
  {\cal N}_{H}(\beta) &\equiv \langle\Psi|R(\beta) |\Psi\rangle \, , \\
  {\cal H}_{H}(\beta) &\equiv \langle\Psi|R(\beta) H |\Psi\rangle \, .
\end{aligned}
\end{equation}
As in the non-Hermitian case, the coupled-cluster energy is recovered in absence of the projector~\cite{shavittbartlett2009}.

\subsubsection{Approximation schemes}

In the following the coupled-cluster state~(\ref{eq:CC}) is truncated at the CCSD level, $T=T_1+T_2$. Because of the exponential ansatz in the bra and ket, the exact evaluation of the projected energy~(\ref{EI}) is not feasible and we need to make an approximation. We truncate the power series of $e^{T_2}$ term by its first three terms and approximate $|\Psi\rangle$ as
\ba
\vert \Psi_{\rm SQD} \rangle \equiv
e^{T_1}\left(1+T_2 +{1\over 2}T_2^2 \right)
|\Phi\rangle \, ,
\label{sqd}
\ea
which leads to the singles quadratic doubles (SQD) approximation. Further dropping the $T^2_2$ term leads to the singles linear doubles (SLD) approximation
\ba
\vert \Psi_{\rm SLD} \rangle \equiv
e^{T_1}\left(1+T_2 \right) |\Phi\rangle \, . \label{sld}
\ea
The SLD and SQD approximations are attractive because they treat the rotation $R(\beta)$ and $\exp(T_1)$ exactly via similarity transformations. This approach is similar to the spin-extended formalism of Ref.~\cite{tsuchimochi2018}. The SLD ansatz is computationally inexpensive because it neglects the  $T_2^2$ term.

\subsubsection{Amplitude equations}

The symmetry-breaking amplitudes $t^a_i$ and $t^{ab}_{ij}$ employed in the SQD approximation~(\ref{sqd}) are obtained by left-projecting the Schr\"odinger equation onto \ph{1} and \ph{2} excitations of the reference state; this yields
\begin{subequations}
\begin{align}
\label{sqd_eq}
\langle \Phi^a_i \vert (H-E) e^{T_1}\left(1+T_2 +{1\over 2}T_2^2 \right)
  |\Phi\rangle &= 0 \ , \\
\langle \Phi^{ab}_{ij} \vert (H-E) e^{T_1}\left(1+T_2 +{1\over 2}T_2^2 \right)
  |\Phi\rangle &= 0 \ ,    
\end{align}
\end{subequations}
\be
E = \langle\Phi|H e^{T_1}\left(1+T_2+{1\over 2}T_2^2 \right) |\Phi\rangle \ .
\ee
These are the standard CCSD equations. In the SLD approximation, the amplitudes are obtained from solving a configuration interaction problem truncated at doubles excitations starting from  $\breve{H}$, i.e.
\be
\breve{H}\left(1+T_2\right) |\Phi\rangle  = E \left(1+T_2\right) |\Phi\rangle\ ,
\ee
and projecting from the left with $ \langle \Phi^{ab}_{ij} \vert $ we obtain a linear problem in $T_2$ that can be solved iteratively. Contrary to SQD (or CCSD), the SLD approach is variational but not size extensive. Due to the lack of size extensivity the SLD (dynamical) correlation energy will not be accurate for $^{20}$Ne and $^{34}$Mg, but the SLD energy gain ($\delta E_{SLD}$) from angular-momentum projection can accurately be computed with methods that lack size extensivity since it is small and decreases with increasing mass.

\subsubsection{Hamiltonian and norm kernels}
\label{HCCkernels}

The norm and Hamiltonian kernels~(\ref{kern1}) are
\begin{subequations}
\label{kern}
\begin{align}
  {\cal N}_{H}(\beta) &\equiv \langle\Phi|U^\dagger e^{T_1^\dagger} R(\beta) e^{T_1} U |\Phi\rangle \ , \\
  {\cal H}_{H}(\beta) &\equiv \langle\Phi|U^\dagger e^{T_1^\dagger} R(\beta) H e^{T_1} U |\Phi\rangle \, .
\end{align}
\end{subequations}
Here we introduced the excitation operator
\ba
U \equiv 1+T_2 +{1\over 2}T_2^2  \ . \label{DiscOpU}
\ea
We note that for the SQD approach ${\cal H}_{H}(0)/{\cal N}_{H}(0)\ne E_{\rm SQD}$, and that 
${\cal H}_{H}(0)/{\cal N}_{H}(0)>  E_{\rm SQD}$ because the Hermitian approach lacks size extensivity when the exponential $e^T$ is truncated. Nevertheless, the Hermitian approach is useful to compute energy differences $E^{(J)}-E^{(0)}$ and the energy gain $\delta E = E^{(0)}-{\cal H}_{H}(0)$ from angular-momentum projection because 
the magnitudes of these quantities decrease with increasing mass number.

Focusing first on the Hamiltonian kernel and inserting $1=\exp{(T_1)}\exp{(-T_1)}$ allows us to rewrite 
\ba
   {\cal H}_{H}(\beta) &= \langle\Phi|U^\dagger {\cal R}(\beta)  \breve{H} U |\Phi\rangle \, .
\ea
Here
\be
{\cal R}(\beta) \equiv e^{T_1^\dagger} R(\beta) e^{T_1} \ ,
\ee
is a product of exponentiated one-body operators, each of which induces a transformation of the 
single-particle basis that can be handled exactly.

Inserting the identity ${\cal R}(\beta) {\cal R}^{-1}(\beta)$ and using Thouless' theorem leads to
\ba
 {\cal H}_{H}(\beta) &=& \langle\Phi|{\cal R}(\beta)|\Phi\rangle   \nonumber\\
   & \times& \langle\Phi|e^{V} {\cal R}^{-1}(\beta)U^\dagger {\cal R}(\beta) \breve{H} U |\Phi\rangle \ .
\ea
Here, the operator ${V}(\beta)$ results from Eq.~\eqref{thouless} 
by replacing $R(\beta)\to {\cal R}(\beta)$. (This has to be kept in mind for the rest of this Subsection.)   

We insert the identity $1=\exp{(-V)}\exp{(V)}$ three times and use 
$\exp{(V)}|\Phi\rangle = |\Phi\rangle$. This yields  
\ba
\label{mainH}
   {\cal H}_{H}(\beta) & = & \langle\Phi|{\cal R}(\beta)|\Phi\rangle
   \langle\Phi|\widetilde{U}^\dagger \overline{H} \, \overline{U}
   |\Phi\rangle  \ ,
\ea
where
\ba
 \widetilde{U}^\dagger & \equiv& e^V {\cal R}^{-1}(\beta)U^\dagger
{\cal R}(\beta) e^{-V} \ . \label{STROTU}
\ea
Because the similarity transformation~\eqref{HVsim} is presently applied to $\breve{H}$ and not to $H$, 
the notation $\overline{\breve{H}}$ should have been used in Eq.~(\ref{mainH}). 
For simplicity we continue to use the lighter notation $\overline{H}$. The operators in Eq.~(\ref{mainH}) 
all depend on $\beta$ and therefore do not commute with $J_z$. This increases the storage demands.  

Equation~(\ref{mainH}) requires us to perform a number of basis transformations followed by taking expectation 
values of a product consisting of 
up to three two-body operators. To do this, we insert 
a resolution of the identity in terms of the reference state and its particle-hole excitations 
\ba
\label{mainH2}
   {\cal H}_{H}(\beta) & = & \langle\Phi|{\cal R}(\beta)|\Phi\rangle
   \sum_\mu \langle\Phi|\widetilde{U}^\dagger \vert \mu\rangle \langle
   \mu \vert \overline{H} \, \overline{U}
   |\Phi\rangle \ .
\ea
The sum truncates to the finite set of states $\vert \mu
\rangle \in \left\{ \vert \Phi\rangle, \vert
    \Phi^{a}_i\rangle, \vert \Phi^{ab}_{ij}\rangle, \vert \Phi^{abc}_{ijk}\rangle, \vert
    \Phi^{abcd}_{ijkl}\rangle \right\}$ including up to \ph{4} excitations. The norm kernel is dealt with in a similar fashion, i.e. 
\ba
\label{mainN}
   {\cal N}_{H}(\beta) & = & \langle\Phi|{\cal R}(\beta)|\Phi\rangle
   \langle\Phi| \widetilde{U}^\dagger \overline{U} |\Phi\rangle \nonumber \\
   & = & \langle\Phi|{\cal R}(\beta)|\Phi\rangle
   \sum_{\mu'} \langle\Phi| \widetilde{U}^\dagger \vert \mu'\rangle \langle
   \mu' \vert \overline{U} \vert \Phi\rangle \ ,
\ea
where the sum now truncates to the smaller set $\vert \mu'
\rangle \in \left\{ \vert \Phi\rangle, \vert
    \Phi^{a}_i\rangle, \vert \Phi^{ab}_{ij}\rangle \right\}$ of up to \ph{2} excitations.

\subsubsection{Algebraic expressions}
\label{algexpr}

Let us work out the algebraic expression of the norm and Hamiltonian
kernels given in Eqs.~(\ref{mainN}) and (\ref{mainH2}), respectively. We focus on 
\begin{equation}
  \label{expand_ket}
  \overline{U} |\Phi\rangle = \left(1 + \overline{T}_2 + {1\over 2}
        \overline{T}_2^2 \right)|\Phi\rangle \ . 
\end{equation}
Here, 
\begin{equation}
\overline{T}_2 = \overline{T}_2^{(0)} + \overline{T}_2^{(1)}  + \overline{T}_2^{(2)} \, , \label{NOSTT2}
\end{equation}
is decomposed into zero, one, and two-body operators (as indicated by the subscript) and the normal-ordered components are 
\begin{subequations}
\label{MESTT2}
\begin{align}
 \overline{T}_2^{(0)} & = {1\over 2}\sum_{ij}  \overline{t}^{ij}_{ij} \, ,
                         \\
 \overline{T}_2^{(1)} &=   \sum_{ipq} \overline{t}^{ip}_{iq}
                         \lbrace c^\dagger_pc_q \rbrace \equiv   \sum_{pq} \tau^{p}_{q}
                         \lbrace c^\dagger_pc_q \rbrace \, ,
                         \\
 \overline{T}_2^{(2)} & =  {1\over 4}\sum_{pqrs} \overline{t}^{pq}_{rs}
                         \lbrace c^\dagger_pc^\dagger_qc_sc_r \rbrace \, .
\end{align}
\end{subequations}
We introduced
$\tau^{p}_{q}\equiv\sum_i\overline{t}^{ip}_{iq}$ and $\lbrace \ldots \rbrace$ denotes the normal ordering. 
By virtue of Thouless' theorem, the amplitudes of the similarity-transformed cluster operators can 
be obtained via a transformation of the single-particle basis, see App.~\ref{sub-sim} for details.

Only the excitation part of the operators~\eqref{NOSTT2} matters in Eq.~\eqref{expand_ket}.  Thus, we limit ourselves to the excitation part of $\overline{U}$ and rewrite Eq.~\eqref{expand_ket} as
\ba
\overline{U} |\Phi\rangle = \left( {\overline{U}^{(0)}}
  +{\overline{U}^{(1)}} + {\overline{U}^{(2)}} + {\overline{U}^{(3)}} + {\overline{U}^{(4)}}\right)  |\Phi\rangle \, .
\ea
Again, we used subscripts $(k)$ to denote a $k$-body operator. The amplitudes of the excitation operators
\ba
\overline{U}^{(k)} = \sum_{i_1\cdots i_k} \sum_{a_1 \cdots a_k} \overline{u}^{a_1\cdots
  a_k}_{i_1 \cdots i_k} c^\dagger_{a_1}\cdots c^\dagger_{a_k}
c_{i_k}\cdots c_{i_1} \, ,
\ea
can be expressed in terms of those introduced in Eq.~\eqref{MESTT2}. The corresponding algebraic expressions for
 \begin{subequations}
 \begin{align}
 \label{b-matele}
 \overline{U}^{(0)} &= \langle\Phi\vert \overline{U}\vert \Phi\rangle \, , \\
 \overline{u}^a_i & =  \langle\Phi^a_i\vert \overline{U}\vert \Phi\rangle \, , \\
 \overline{u}^{ab}_{ij} & =  \langle\Phi^{ab}_{ij}\vert \overline{U}\vert \Phi\rangle \, \\
 \overline{u}^{abc}_{ijk} & =  \langle\Phi^{abc}_{ijk}\vert \overline{U}\vert \Phi\rangle \, , \\
 \overline{u}^{abcd}_{ijkl} & =  \langle\Phi^{abcd}_{ijkl}\vert \overline{U}\vert \Phi\rangle \, ,
 \end{align}
 \end{subequations}
are presented in App.~\ref{MEDEXCOP}.  The matrix elements associated with the bra state
$\langle\Phi| \widetilde{U}^\dagger$ follow from the above derivation by exchanging particle and hole indices, 
and by replacing the matrix elements of
the similarity-transformed operator $\overline{T}_2$ by those of $\widetilde{T}_2$ defined 
similarly to Eq.~\eqref{STROTU}. With these matrix elements at hand, the norm kernel~\eqref{mainN} can be  evaluated.

To evaluate the Hamiltonian kernel~\eqref{mainH2} we also need the matrix elements 
 \begin{subequations}
  \label{h-matele}
 \begin{align}
 \overline{X}^{(0)} &\equiv \langle \Phi \vert \overline{H}\overline{U} \vert \Phi \rangle \, , \\
 \overline{x}^a_i               &  \equiv \langle\Phi^a_i\vert \overline{H} \overline{U}\vert \Phi\rangle \, , \\
 \overline{x}^{ab}_{ij}        & \equiv  \langle\Phi^{ab}_{ij}\vert \overline{H} \overline{U}\vert \Phi\rangle \, , \\
 \overline{x}^{abc}_{ijk}    & \equiv  \langle\Phi^{abc}_{ijk}\vert \overline{H} \overline{U}\vert \Phi\rangle \, , \\
 \overline{x}^{abcd}_{ijkl} & \equiv  \langle\Phi^{abcd}_{ijkl}\vert \overline{H} \overline{U}\vert \Phi\rangle \, ,
 \end{align}
\end{subequations}
in terms of those of the normal-order pieces of $\overline{H}$ and $\overline{U}$. These expressions are presented in App.~\ref{MESTHU}.

Using these ingredients, the kernels can finally be computed as
\begin{subequations}
\label{finalformkernels}
\begin{align}
\frac{{\cal H}_{H}(\beta)}{\langle\Phi|{\cal R}(\beta)|\Phi\rangle} &= \widetilde{U}^{\dagger(0)}  \overline{X}^{(0)} + \sum_{ai} \widetilde{u}^i_a \overline{x}^a_i  \nonumber\\
& + {1\over 4}\sum_{abij} \widetilde{u}^{ij}_{ab} \overline{x}^{ab}_{ij}   +  {1\over 36}\sum_{abcijk} \widetilde{u}^{ijk}_{abc}
    \overline{x}^{abc}_{ijk}  \nonumber \\
                & + {1\over 576}\sum_{abcdijkl}
                  \widetilde{u}^{ijkl}_{abcd} \overline{x}^{abcd}_{ijkl} \, , \label{finalformkernelsH} \\
\frac{{\cal N}_{H}(\beta)}{\langle\Phi|{\cal R}(\beta)|\Phi\rangle} &= \widetilde{U}^{\dagger(0)}  \overline{U}^{(0)} + \sum_{ai} \widetilde{u}^i_a \overline{u}^a_i  \nonumber\\
& + {1\over
    4}\sum_{abij} \widetilde{u}^{ij}_{ab} \overline{u}^{ab}_{ij}   
   +  {1\over 36}\sum_{abcijk} \widetilde{u}^{ijk}_{abc}
    \overline{u}^{abc}_{ijk}  \nonumber \\
                & + {1\over 576}\sum_{abcdijkl}
                  \widetilde{u}^{ijkl}_{abcd} \overline{u}^{abcd}_{ijkl} \, . \label{finalformkernelsN}
\end{align}
\end{subequations}
Naively estimated, the inclusion of 
$T_2^2$ in the bra and ket states of
the Hamiltonian kernel comes at a considerable cost
of $n_u^6n_o^4$ computational cycles, with $n_u$ the number of
unoccupied and $n_o$ the number of occupied states, respectively. However, $\overline{U}^{(4)}$ is a product of two disconnected terms and this reduces the cost to $n_u^5n_o^4$ because intermediates can be introduced. In Eq.~\eqref{finalformkernelsH} for example, we first compute a three-body intermediate by contracting  $\overline{T}_2^{(2)}$ with $\overline{H}$ and this is then followed by a contraction with the second $\overline{T}_2^{(2)}$. We note that the SLD and the disentangled approaches used in this work scale as $n_u^4n_o^2$ (which is the familiar CCSD scaling), but they require significant larger computational resources than the unprojected CCSD solution because $J_z$ is not a good quantum number in the projection.    

\section{Collective Hamiltonian and effective theory}
\label{sec:eft}

\subsection{Collective Hamiltonian}
The expressions derived so far allow us to compute the  collective Hamiltonian~\cite{wawong1975,ringschuck,broglia2000,yannouleas2007}. Let $\Omega\equiv(\theta,\phi)$ consist of a polar angle and azimuth, and let  
\begin{equation}
    |\Omega\rangle \equiv e^{-i\phi J_z}e^{-i\theta J_y}|\Psi\rangle
\end{equation}
be a rotation of the state $|\Psi\rangle$. Thus, in the state $|\Omega\rangle$ the symmetry axis of the nucleus 
points into the direction of the radial unit vector $\mathbf{e}_r(\theta,\phi)$. The matrix elements 
\ba
\begin{aligned}
{\cal H}(\Omega',\Omega) & \equiv \langle \Omega'|H|\Omega\rangle \ ,  \\
{\cal N}(\Omega',\Omega) & \equiv \langle \Omega'|\Omega \rangle
\end{aligned}
\ea
are related to the kernels~(\ref{kern1}) via
\ba
\begin{aligned}
\label{HcollKern}
  {\cal H}(\Omega',\Omega) &= {\cal H}_H(\beta) \ , \\
  {\cal N}(\Omega',\Omega) &= {\cal N}_H(\beta) \ ,
\end{aligned}
\ea
where
$\cos\beta=\mathbf{e}_r(\theta,\phi)\cdot\mathbf{e}_r(\theta',\phi')$.
The collective Hamiltonian then becomes
\be
\label{Hcoll}
H_{\rm coll}(\Omega',\Omega) = \left[{\cal N}^{-{1\over 2}}{\cal
  H}{\cal N}^{-{1\over 2}}\right](\Omega',\Omega) \ .
\ee
To compute the matrix elements~(\ref{HcollKern}) we distribute
a few tens of angles $\Omega$ evenly over the sphere,
following Ref.~\cite{deserno2004}. 
The diagonalization of $H_{\rm coll}$ then yields the low-lying collective spectrum, and the corresponding collective state can be used to compute other observables or transition matrix elements of interest. An alternative approach to the collective Hamiltonian is via effective field theory (EFT).  

\subsection{Effective theory}
\label{subsec:eft}

At lowest energies, the only degree of freedom for an axially symmetric even-even nucleus is the orientation $\mathbf{e}_r(\theta,\phi)$ of its symmetry axis~\cite{papenbrock2011,papenbrock2020}, where $(\theta,\phi)$ define the polar and azimuth angles. Following arguments from spontaneous~\cite{weinberg_v2_1996} and
emergent symmetry
breaking~\cite{broglia2000,yannouleas2007,papenbrock2014} the
effective Hamiltonian can only contain the angular derivative
\be
\boldsymbol{\nabla}_\Omega \equiv \mathbf{e}_\theta \frac{\partial}{\partial\theta}
+ \mathbf{e}_\phi {1\over\sin\theta}\frac{\partial}{\partial\phi} \ .
\ee
Here, $\mathbf{e}_\theta$ and $\mathbf{e}_\phi$ denote the polar and
azimuth unit vectors that, together with the radial unit vector
$\mathbf{e}_r$, form the body-fixed coordinate system. 
The leading-order effective Hamiltonian is~\cite{papenbrock2011}
\be
\label{HEFT1}
H_{\rm EFT} = E^{(0)} + a\left(-i\boldsymbol{\nabla}_\Omega\right)^2  \, .
\ee
Here, the low-energy coefficients $E^{(0)}$ and $a$ are the ground-state
energy and the rotational constant, respectively. Thus, the
constraints from emergent symmetry breaking ensure that the effective
Hamiltonian~(\ref{HEFT1}) can be matched to the collective Hamiltonian~(\ref{Hcoll}) by adjusting $E^{(0)}$ and $a$. 
The low-energy coefficient $E^{(0)}$ contains short-range physics, and one needs size-extensive methods to compute it. 
In contrast, the rotational constant $a$ is the inverse moment of inertia
and depends on the nuclear shape and -- for heavy nuclei -- also on
the degree of superfluidity. Thus, computing $a$ does not require
high-resolution methods.  As we will see below, mean field methods (which
are quite inaccurate regarding $E^{(0)}$) yield reasonable values
for $a$ in the nuclei we compute.

Introducing the angular momentum $\hat{J} = \mathbf{e}_r\times (-i
\boldsymbol{\nabla}_\Omega)$ allows us to rewrite the effective Hamiltonian~(\ref{HEFT1}) as
\be
\label{HEFT}
H_{\rm EFT} = E^{(0)} + a\hat{J}^2 \ .
\ee
This is the Hamiltonian of the rigid-rotor model, and its spectrum is 
\be
\label{spec}
E^{(J)} = E^{(0)} +a J(J+1) \ .
\ee
Taking the expectation value of the effective Hamiltonian~(\ref{HEFT}) in the unprojected wave function yields
\be
E = E^{(0)} + a\langle J^2\rangle \ .
\ee
Thus, the ground-state energy gain from projection is
\be
\label{gain}
\delta E \equiv E^{(0)} - E = -a\langle J^2\rangle \ ,
\ee
delivering the result of \textcite{peierls1957}.  

To check the consistency of these arguments we return to
Eqs.~(\ref{spec}) and (\ref{gain}) and find
\be
\label{EJ}
E^{(J)} -E^{(0)} = -{\delta E}\frac{J(J+1)}{\langle J^2\rangle}
\ee
for the excitation energy of the state with spin $J$. To estimate
the ground-state energy gain from projection we use
\be
\label{estimate}
\delta E = (E^{(0)} -E^{(2)})\frac{\langle J^2\rangle}{6} \, ,
\ee
where $E^{(0)}$ and $E^{(2)}$ result from, e.g., angular momentum projection.  Taking instead the energy
difference $E^{(0)}-E^{(2)}$ from experimental data (i.e. 3~MeV in $^8$Be, 0.6~MeV in the
island of inversion), one immediately obtains an estimate for the energy
gain from symmetry projection by employing the expectation value
$\langle J^2\rangle$ of the symmetry-unrestricted state.  The energy
gain from projection is much smaller than the energy from dynamical
correlations, which allows us to employ approximations that lack size extensivity. 
It is also clear from Eq.~\eqref{estimate} that the gain from symmetry restoration is 
proportional to the expectation value $\langle J^2\rangle$ of the unprojected
state. Thus, including more correlations in that state is expected to
reduce the value $\langle J^2\rangle$ and the energy gain from
projection.

The approach via effective field theory also enables us to estimate
uncertainties. The first step involves the identification of a
breakdown scale. For simplicity we use here a breakdown spin $J_{\rm
  b}$ where new physics enters; the corresponding breakdown energy,
relative to the ground-state energy $E^{(0)}$, is then $\Lambda_{\rm b} = a
J_{\rm b}(J_{\rm b}+1)$. The subleading correction to the
Hamiltonian~(\ref{HEFT}) is the term $b(\hat{J}^2)^2$, which adds the
contribution $b[J(J+1)]^2$ to the spectrum~(\ref{spec}), see, e.g.,
Ref.~\cite{papenbrock2011}. At the breakdown scale, the subleading
contribution becomes significant, and this allows us to estimate the
unknown coefficient $b$. A simple estimate can be obtained by assuming
that the leading and subleading contributions are approximately equal
in size at the breakdown scale; this yields $b\approx a/[J_{\rm
    b}(J_{\rm b}+1)]$. A more realistic assumption is that the
subleading term is of the same size as the leading-order level spacing
$E^{(J_{\rm b}+1)}-E^{(J_{\rm b}-1)}=2a (J_{\rm b}+1)$ at the breakdown
scale; this yields the estimate
\be
b\approx \frac{2a}{J_{\rm b}^2(J_{\rm b}+1)} \ .
\ee
With this assumption about the breakdown scale, the uncertainty at spin $J\lesssim J_{\rm b}$ is
\be
\label{uncert}
\Delta E^{(J)} \approx 2a\frac{\left[J(J+1)\right]^2}{J_{\rm b}^2(J_{\rm b}+1)} \ .
\ee
Using more information than merely the breakdown scale (or making
more assumptions), would allow us to quantify uncertainties~\cite{schindler2009,furnstahl2014c,coelloperez2015b}.  Here,
we will contend ourselves with the estimate~(\ref{uncert}). Long sequences of data would also allow us to empirically
determine the breakdown scale, e.g. via Lepage
plots~\cite{alnamlah2021}.

The rotational
constant $a$ is inverse proportional to the moment of inertia which --
for a liquid drop -- is proportional to $A^{5/3}$ where $A$ is the
mass number. Taken together with the empirical relation $\langle
J^2\rangle \sim A^{3/2}$ from mean-field
calculations~\cite{bertsch2019} then implies that the energy gain from
symmetry projection is small and decreases with increasing mass
number. Long-wavelength physics dominates angular momentum projection. We can also see this in an equation-of-motion approach
where the (non-normalized) excited $J^\pi=2^+$ state is 
\be
|{2^+} \rangle = Q^{(p)}_{20}|{0^+}\rangle \ . \label{schematic2+}
\ee
Here the charge quadrupole operator $Q^{(p)}_{20}$ is defined as in Eq.~\eqref{Q20} but the sum is over protons. This long-range operator is insensitive to short-range physics. The equation-of-motion approach to excited states then yields
\be
\left[H,Q^{(p)}_{20}\right]|{0^+}\rangle = (E^{(2)}-E^{(0)})Q^{(p)}_{20}|{0^+}\rangle \ \, . \label{EOM}
\ee
Employing  the effective Hamiltonian~(\ref{HEFT}) in Eq.~\eqref{EOM} yields $E^{(2)}-E^{(0)} = 6a$ 
because $Q^{(p)}_{20}$ is a spherical tensor of rank two. Thus, rotational excitation energies are 
low-resolution observables and can be computed with low-resolution methods. 

The quadrupole operator connects harmonic oscillator shells that differ by two units of $\hbar\omega$. 
Thus, the equation-of-motion approach also shows that microscopic computations of the states with 
angular momentum $J=2,4,6,\ldots$ via symmetry projection requires model spaces that exceed the Fermi 
surface by $J$ shells.

A future goal consists of computing quadrupole moments of projected states and transition matrix elements between 
projected states. For now, we characterize collectivity via a phenomenological approach. 
The quadrupole deformation parameter for a charged liquid drop is~\cite{loebner1970}
\be
\label{beta2}
\beta_2 \equiv {\sqrt{5\pi}  \langle Q^{(p)}_{20} \rangle  \over 3Z R_0^2} \, ,
\ee
where $R_0=1.2 A^{1/3}$~fm is the empirical charge radius. 
The NNLO$_{\rm opt}$ potential yields too small radii for all but the lightest nuclei~\cite{dytrych2015,kanungo2016,duguet2017,burrows2020}. To compute the quadrupole deformation parameter we therefor replace $R_0^2$ by the charge radius squared ($R_{ch}^2$). Computing expectation values in the unprojected ground-state, the deformation parameter then becomes 
\be
\label{beta2}
\beta_2 \equiv {\sqrt{5\pi}  \langle Q^{(p)}_{20} \rangle  \over 3Z  R_{ch}^2 } \, .
\ee

Superfluid systems that break particle number can similarly be addressed in an effective theory. One then
replaces the orientation of the symmetry axis by the gauge angle and
$a\hat{J}^2$ by the pairing rotational term $b(\hat{A}-\langle
\hat{A}\rangle)^2$ where $b$ is a low-energy coefficient and $\hat{A}$ the number operator. The
energy gain~(\ref{gain}) from particle-number projection is then related to the particle-number variation of the
unprojected state. For an estimate in medium-mass nuclei we note that
$b\approx 0.1$~MeV in tin isotopes~\cite{brink-broglia2005} and that
$\langle(A-\langle A\rangle)^2\rangle\approx 16-36$ in
nickel isotopes~\cite{tichai2020}. Thus, particle-number projections do also not seem to require size extensive methods.

\section{Results for $^8$Be, $^{20}$Ne, and $^{34}$Mg } 
\label{sec:results}

We compute the nuclei $^8$Be, $^{20}$Ne, and $^{34}$Mg using the nucleon-nucleon interaction NNLO$_{\rm opt}$~\cite{ekstrom2013}. 
For $^8$Be, we benchmark with the no-core shell model (NCSM)~\cite{caprio2015,maris2015} and for $^{20}$Ne with the symmetry-adapted NCSM~\cite{dytrych2020}. These benchmarks are close to experimental data, and the accuracy of the NNLO$_{\rm opt}$ potential makes it interesting to compute the structure of the neutron-rich nucleus $^{34}$Mg.

The Hamiltonian~(\ref{ham}) is the sum of the intrinsic kinetic energy (to decouple the center-of-mass mode~\cite{hagen2009a,hagen2010b,parzuchowski2017}) 
and the two-body potential. 
We use a single particle basis that consists of the eigenstates $\{| p \rangle \equiv  \vert n l j j_z t_z\rangle\}$ with energy $(2n+l)\hbar\omega$ of the spherical harmonic oscillator (omitting the zero-point energy). States with an energy up to and including $N_{\rm max}\hbar\omega$ are in the basis. 

For the construction of the reference state, shells are filled according to the nuclear shell model~\cite{mayer1955}. The partially occupied subshell at the Fermi surface is filled by occupying Kramer-degenerate pairs of states with
increasing values $|j_z|=1/2, 3/2, \ldots$ of the angular-momentum projection. This leads to prolate deformation. 
The various reference states we used are described in Sec.~\ref{refstates}.

\begin{figure}[htb!]
  \includegraphics[width=0.49\textwidth]{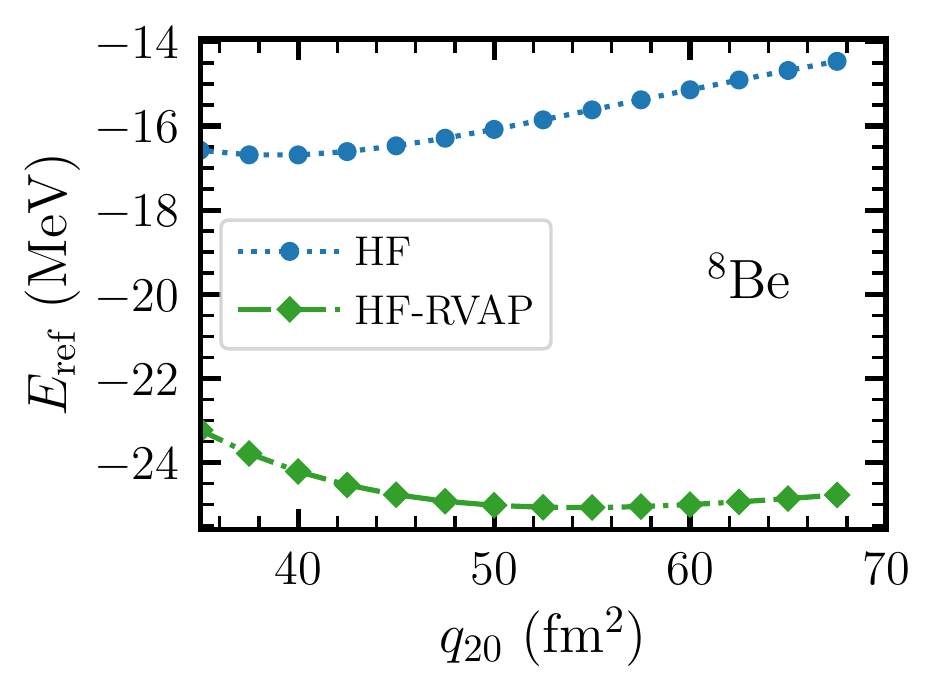}
  \caption{Energies of $^8$Be from symmetry-unrestricted Hartree-Fock  (blue circles)
    and HF-RVAP (green diamonds) as a function of the mass quadrupole moment $q_{20}$. Calculations are performed with $\hbar \omega = 24$~MeV.
    \label{fig:phfpes}}
\end{figure}

Figure~\ref{fig:phfpes} shows the energy of the reference states from Hartree-Fock and from HF-RVAP for $^8$Be as a function of quadrupole moment.  The projection shifts the minimum towards a larger value of the quadrupole moment.

We found that the Hermitian SLD and SQD approximations are insensitive to the choice
of the reference state presumably because we treat the singles excitations $e^{T_1}$ exactly. As we will see below these approaches also meet benchmarks from NCSM calculations~\cite{caprio2015,dytrych2020}.  
The disentangled approach yields somewhat too compressed spectra, particularly for the CCSD approximation. The CCSD spectra exhibit a small spread with respect to the reference state. The level spacing increases as we go from Hartree Fock to HF-RVAP to HF-VAP, and the CCSD results also depend mildly on the chosen oscillator spacing. We attribute the compressed spectra to the simple energy expression~(\ref{pavCC}), which is not a bi-variational energy functional. In contrast, results for the CCD approximation are somewhat more accurate when compared to the benchmarks. This, however, limits us to the Hartree-Fock basis where singles excitations ($\exp{(T_1)}$) are small.

\subsection{Operator kernels}

\begin{figure*}[t!]
  \includegraphics[width=0.95\columnwidth]{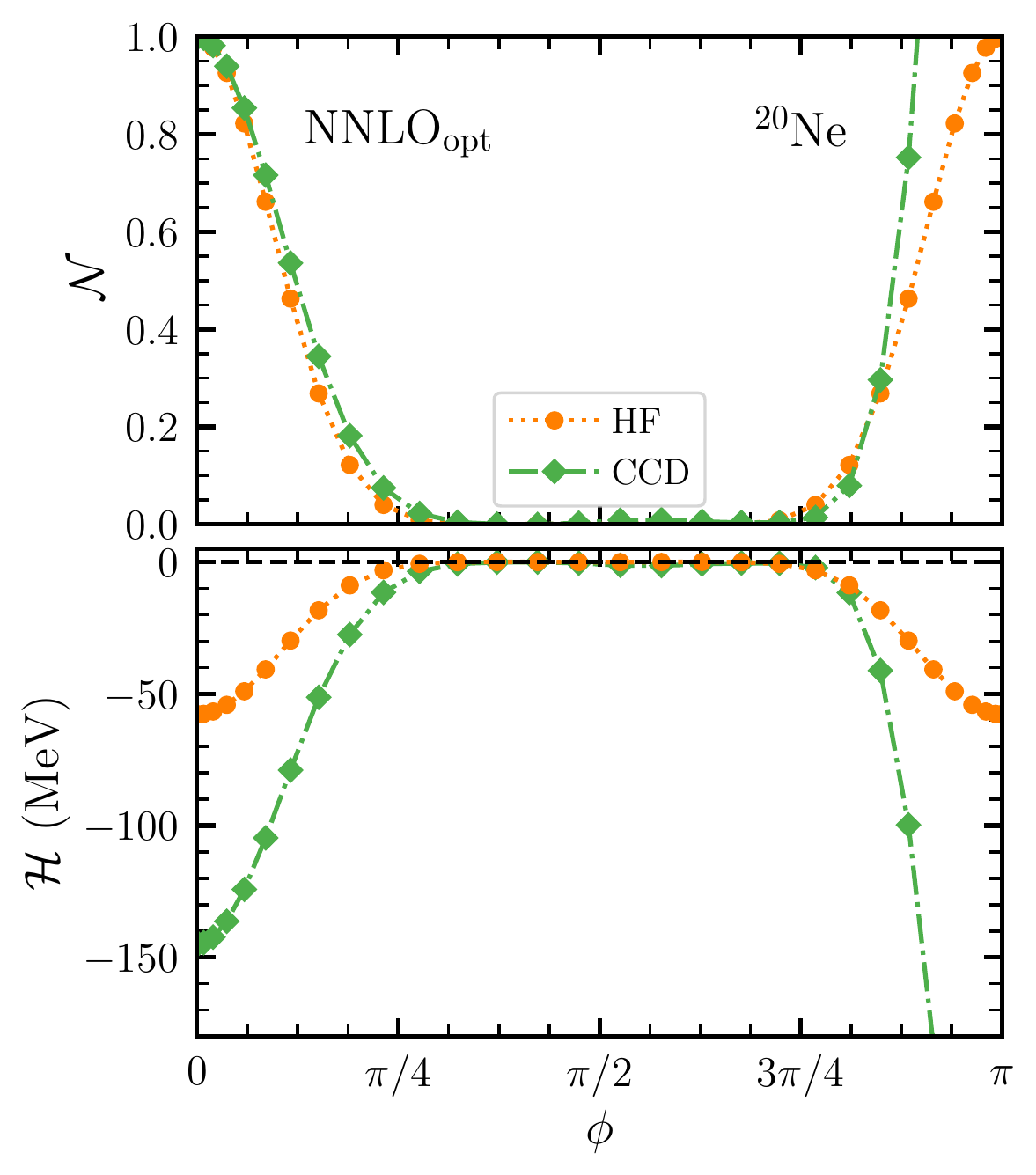}
  \includegraphics[width=0.95\columnwidth]{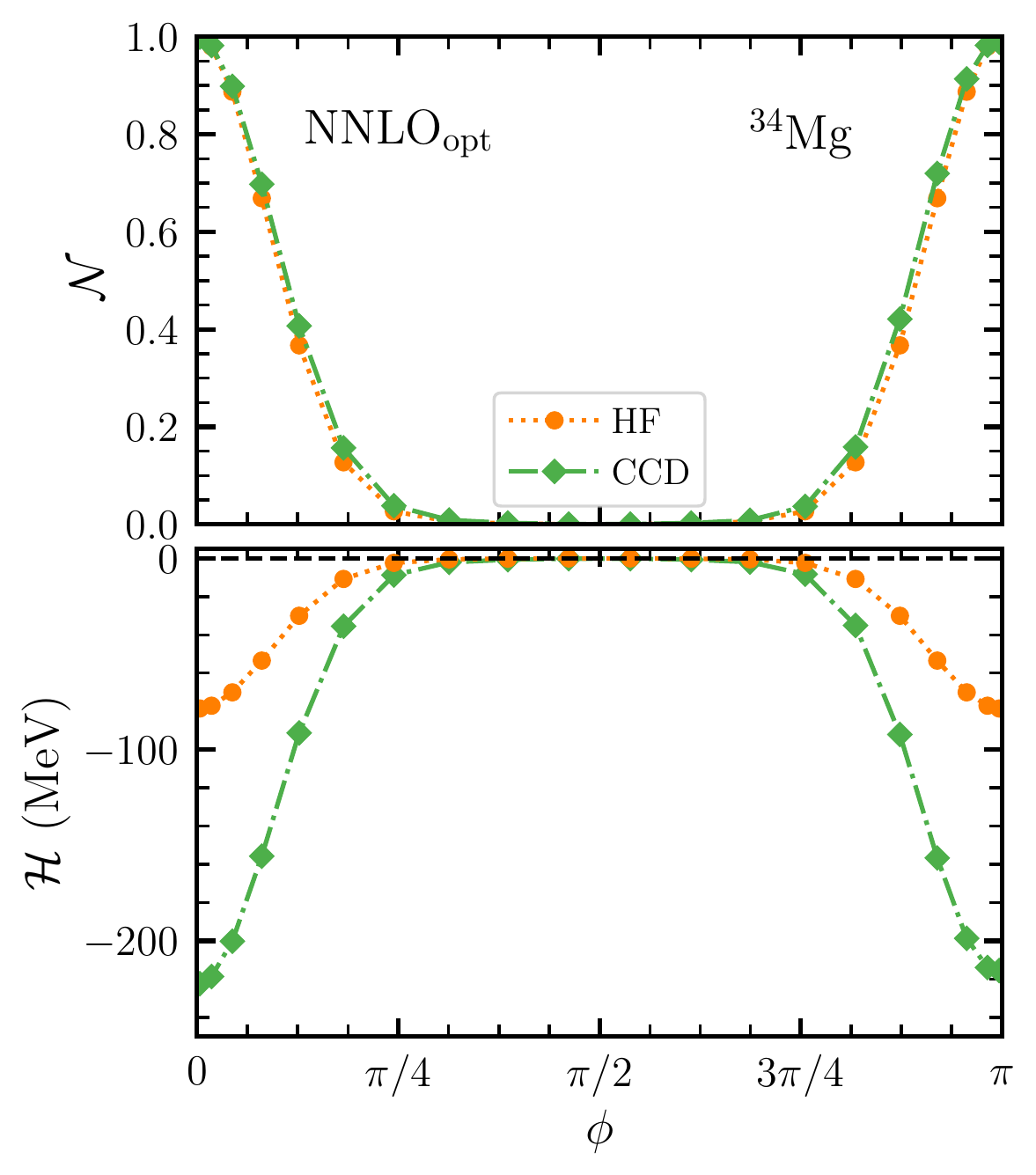}
  \caption{Kernels for $^{20}$Ne (left panels) and $^{34}$Mg (right panels) as a function of the rotation angle. The norm kernels (top panels) and Hamiltonian kernels (bottom panels) are shown for Hartree Fock (orange circles) and CCD (green diamonds) calculations. We employed the Hartree-Fock basis and a model-space with $N_{\rm max} = 7$ and $\hbar\omega = 24$~MeV.
    \label{fig:ne20kernels}}
\end{figure*}

Figure~\ref{fig:ne20kernels} show the norm and Hamiltonian kernels in $^{20}$Ne and $^{34}$Mg, respectively, as a function of rotation angle $\beta$ for the disentangled formalism and compares them to Hartree-Fock kernels. Here we employed the Hartree-Fock basis and a model-space with $N_{\rm max} = 7$ and $\hbar\omega = 24$~MeV. At $\beta=0$ the Hamiltonian kernels yield the energies of the unprojected Hartree-Fock and CCD states.
The Hartree-Fock kernels exhibit the ${\cal R}$ parity and are symmetric upon reflection at $\beta=\pi/2$.  The disentangled formalism approximates the rotation operator and does not keep the ${\cal R}$ parity. This is clearly seen for the case of $^{20}$Ne while less so for $^{34}$Mg. To avoid problems we assume ${\cal R}$ parity and limit the domain of integration in Eq.~\eqref{eq:proj_energy} to $[0,\pi/2]$. This ensures that odd angular momenta are excluded from the ground-state rotational band.

In the absence of symmetry breaking, the kernels are constant and do not depend on $\beta$. Therefore, one expects that operator kernels flatten out at higher truncation levels~\cite{qiu2019}. The upper panels of Fig.~\ref{fig:ne20kernels} show that this is indeed the case. An alternative view is as follows: The curvature of the norm kernel at $\beta=0$ is proportional to the expectation value $\langle \hat{J}^2\rangle$ of the unprojected state, and this value should decrease with increasing sophistication of the employed many-body wave function. Figure~\ref{fig:ne20kernels} confirm this.

How do the truncations in the disentangled formalism impact the projection? To address this question we computed the expectation value of $\hat{J}^2$ in the projected states. We found that the difference to $J(J+1)$ is less than 0.4 for $J=0$, less than 4\% for $J=2$, and less than 2\% for $J=4$. Deviations where largest for $^{20}$Ne and smallest for $^{34}$Mg. We speculate that is because the $^{34}$Mg spectrum is closest to a rigid rotor.

\subsection{Spectra of $^8$Be and $^{20}$Ne}
\label{sub:8Be}

\begin{figure}[b!]
  \includegraphics[width=1.\columnwidth]{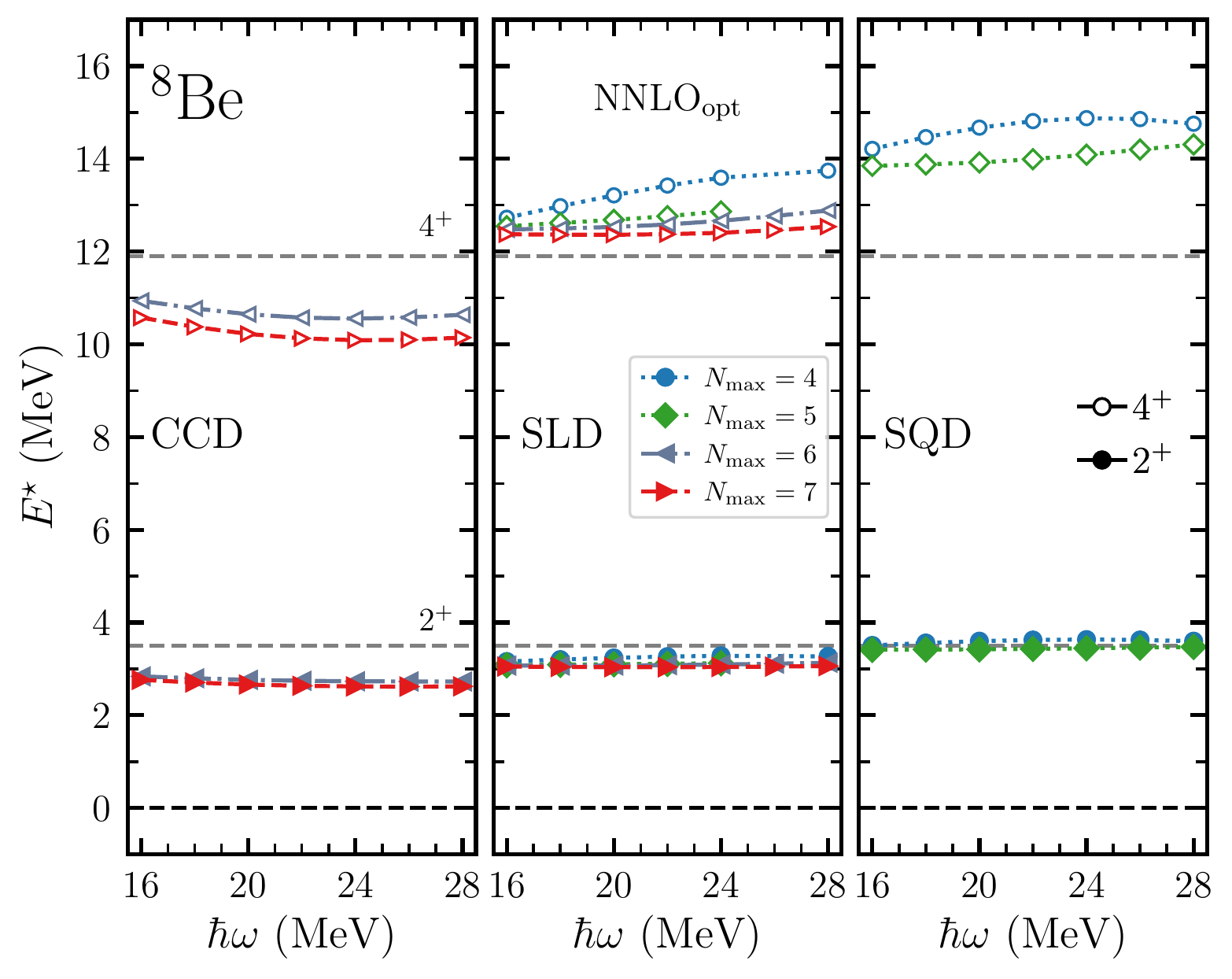}
  \caption{Projected coupled-cluster excitation energies for $^8$Be as a function of oscillator frequency
    obtained from CCD (left panel), SLD (middle panel) and
    SQD approximations (right panel). Model spaces are labeled by
    $N_{\rm max}$. Horizontal dashed grey lines show NCSM results.}
    \label{fig:8BeCompSpec}
\end{figure}

Figure~\ref{fig:8BeCompSpec} displays the ground-state rotational band of $^{8}$Be computed with the CCD (left panel), the SLD (middle panel) and the SQD approximations (right panel). With CCD the   excitation energy of the $J^\pi=2^+$ state is converged for $N_{\rm max}=7$ and practically independent of $\hbar \omega$. The first $4^+$ energy still exhibits some dependence on the basis. The convergence pattern is somewhat different for the SLD and SQD approximations. Here, the the $2^+$ energy starts to converge with $N_{\rm max}=3$ and the $4^+$ energy  at $N_{\rm max}=5$. This is consistent with the discussions below Eqs.~(\ref{schematic2+}) and (\ref{EOM}), because the Fermi surface of $^8$Be is in the $N_{\rm max}=1$ shell.

In the disentangled CCD approximation, the projected energies are lower than the NCSM benchmarks~\cite{caprio2015}, i.e. the  excitation energy of the $J^\pi=2^+$ state is 1~MeV too low whereas the $4^+$ excitation energy differs by about 2~MeV. In the SLD approximation, the $2^+$ energy is close to the NCSM result whereas the converged $4^+$ energy is about $0.5$\,MeV (4\%) too high.  Finally, the $2^+$ energy  from SQD agrees with the benchmark while the the $4^+$ energy is too high.

\begin{figure}[t!]
  \includegraphics[width=1.\columnwidth]{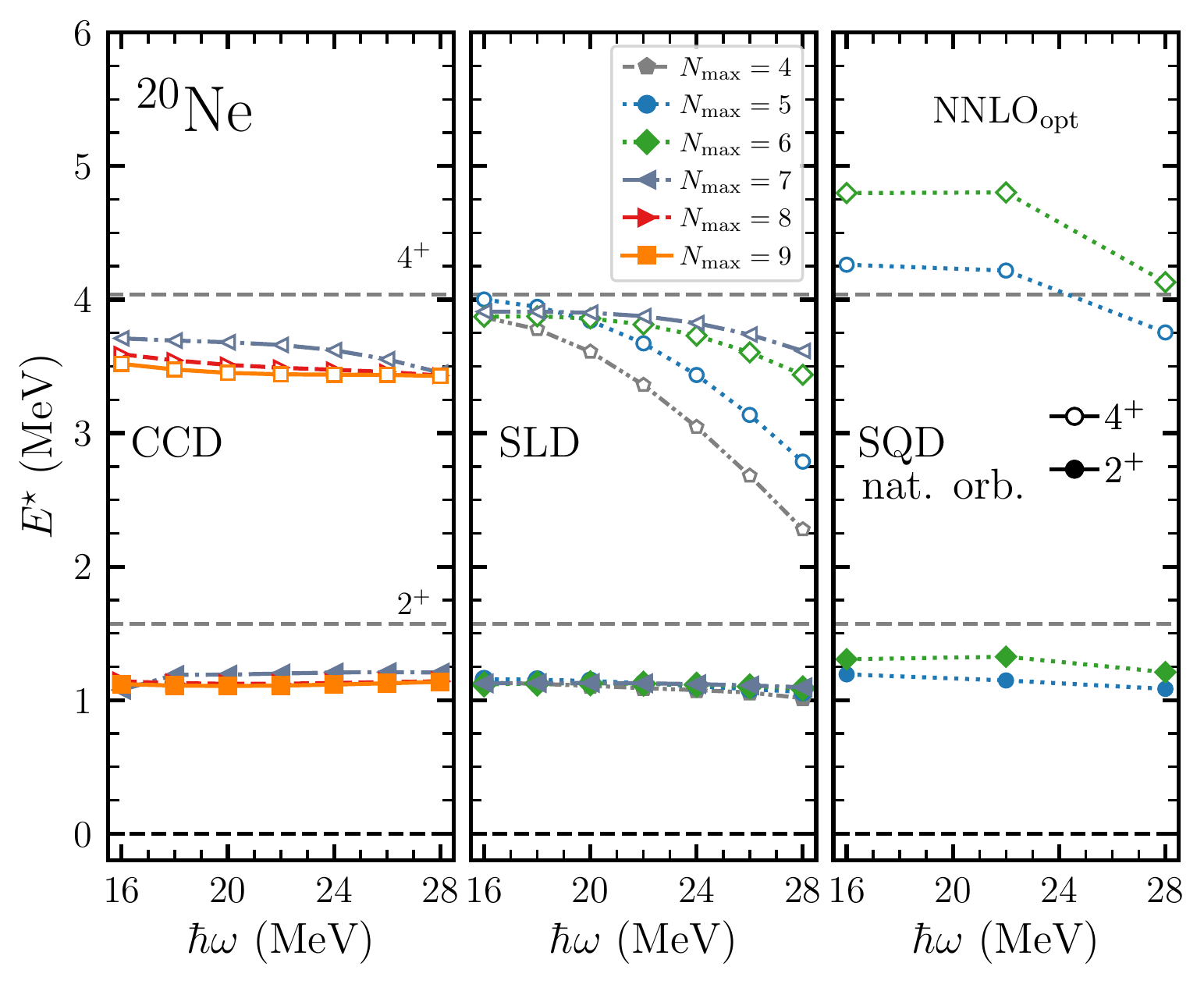}
    \caption{Projected coupled-cluster excitation energies for $^{20}$Ne as a function of oscillator frequency
    obtained from CCD (left panel), SLD (middle panel) and
    SQD approximations (right panel). Model spaces are labeled by
    $N_{\rm max}$. Horizontal dashed grey lines show NCSM results.\label{fig:20NeCompSpec}}
\end{figure}

For $^{20}$Ne we compare with benchmarks from the symmetry adapted NCSM~\cite{dytrych2020} and show results in Fig.~\ref{fig:20NeCompSpec}. For the SLD and SQD approximations the energies of the $J^\pi=2^+$ and $4^+$ states start to converge at $N_{\rm max}=4$ and $N_{\rm max}=4$ and $N_{\rm max}=6$, respectively, because the Fermi surface is in the $N_{\rm max}=2$ shell. The SLD results display an optimal harmonic oscillator frequency around $16$~MeV, and the spectrum is slightly compressed compared to NCSM. For CCD we employed up to 10 major harmonic oscillator shells which is large enough for convergence. As can be seen the spectrum is too compressed and resemble too much that of a rigid-rotor. We also employed the CCSD approximation for the largest model-space and found the $2^+$ energy at 1~MeV and the $4^+$ energy at $\sim 3$~MeV. 

\begin{figure*}[t!]
    \centering
    \includegraphics[width=0.85\textwidth]{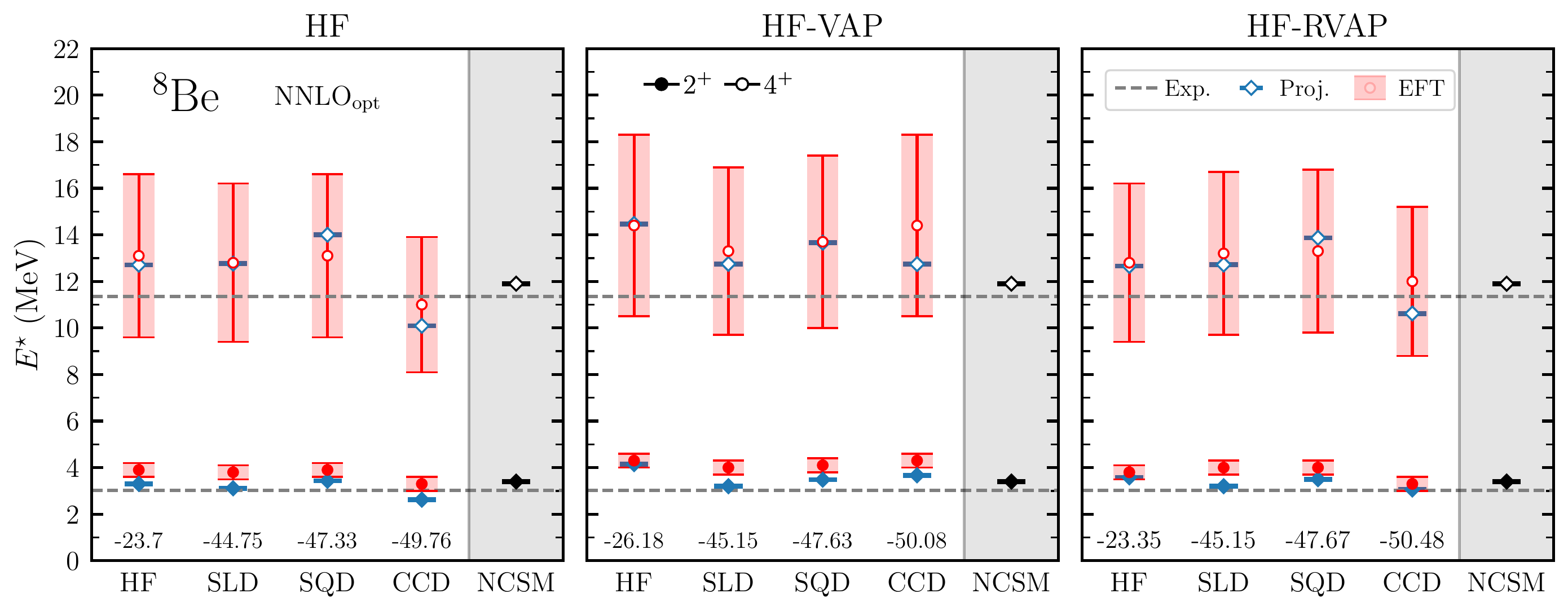}
    \caption{Low-lying excitation energies of $^8$Be from projected Hartree-Fock (HF),  projected coupled-cluster (SLD, SQD, and CCD) computations, and leading-order EFT  using HF, HF-VAP, and HF-RVAP reference states (from left to right). Also shown are NSCM benchmarks and experiment (dashed lines). The numbers at the bottom of the plot are projected ground-state energies (in MeV).}
    \label{fig:eftcomp}
\end{figure*}

Computing rotational spectra via projection requires  much smaller model spaces than required for ground-state energies. This is consistent with our discussion in Subsec.~\ref{subsec:eft} and Ref.~\cite{Frosini:2021ddm}. The symmetry restoration involves a non-perturbative mixing of states close to the Fermi surface such that a comparably small number of harmonic oscillator shells is sufficient to obtain converged results. As observed when going from $^{8}$Be to $^{20}$Ne, the model space requirements increase with increasing mass number because of the reference state.

Let us match the EFT to our results. The leading-order EFT spectrum~(\ref{spec}) depends on two low-energy constants. We use the ground-state energy gained through projection, $\delta E$, and the expectation value $\langle J^2\rangle$ of the unprojected state as input to the EFT and determine the excitation spectrum via Eq.~\eqref{EJ}. Uncertainties are estimated by Eq.~\eqref{uncert}. 
For $^8$Be the breakdown energy $\Lambda_{\rm b}\approx 20 \, \MeV$ is set by the energy of $\alpha$ particle excitations; the corresponding breakdown spin is $J_\text{b}\approx 5$. 
Figure~\ref{fig:eftcomp} shows the EFT spectra (red error bars) based on the input from projected Hartree-Fock and projected coupled-cluster results in Table~\ref{tab:8be} using three different single-particle bases.  Also shown are NCSM benchmarks. About two-thirds of the EFT results agree with the energies of the Hartree-Fock and coupled-cluster calculations. 
For the $J^\pi=2^+$ states the energy ranges from the EFT are systematically higher than the projected energies and the NCSM result. The EFT ranges for the energy of the $4^+$ state agree with the microscopic computations.

The projected Hartree-Fock results are close to the NSCM benchmarks. This is in contrast to the ground-state energies shown at the bottom of the plot (in units of MeV). While the total binding energy requires dynamical correlations, the low-lying rotational excitations are dominated by static correlations and well captured through the symmetry restoration, see also Ref.~\cite{Frosini:2021ddm}. The coupled-cluster computations include dynamical correlations and achieve lower ground-state energies. Clearly, those correlations contribute little to the rotational spectrum.  

Table~\ref{tab:8be} summarizes our results for $^8$Be. We note that $\langle J^2\rangle$ of the unprojected state and energy gained from projection, $\delta E$, decrease a lot from Hartree-Fock to coupled-cluster computations because the latter include much more dynamical correlations. We also note that the projected HF results for the spectrum are more accurate than the projected CCD results in the Hartree-Fock basis.

\begingroup
\renewcommand{\arraystretch}{1.3}
\begin{table*}[ht]
\begin{tabular}{|l|r|r|r|r|r|r|r|r|}\hline\hline
\multicolumn{1}{|c|}{} & \multicolumn{2}{c|}{Unprojected} &
             \multicolumn{4}{c|}{Projected} & \multicolumn{2}{c|}{EFT}\\ \hline
             $^8$Be
             & $E$ & $\langle J^2\rangle$ & $\delta E$  & $E^{(0)}=E+\delta E$  & $E^{(2)}-E^{(0)}$ & $E^{(4)}-E^{(0)}$
             & $E^{(2)}-E^{(0)}$ & $E^{(4)}-E^{(0)}$ \\ \hline
HF     & $-16.304$ & 11.270  & $-7.399$ & $-23.70$  & 3.33 & 12.71 & $3.9\pm 0.3$ & $13.1\pm 3.5$\\
SLD    & $-39.829$ &  7.724  & $-4.923$ & $-44.75$  & 3.11 & 12.76 & $3.8\pm 0.3$ & $12.8\pm 3.4$\\
SQD    & $-43.179$ &  6.311  & $-4.146$ & $-47.33$  & 3.43 & 14.00 & $3.9\pm 0.3$ & $13.1\pm 3.5$\\
CCD    & $-45.907$ &  6.998  & $-3.852$ & $-49.76$  & 2.62 & 10.09 & $3.3\pm 0.3$ & $11.0\pm 2.9$\\
\hline
HF-VAP & $-13.535$ & 17.502  & $-12.640$& $-26.18$  & 4.15 & 14.47 & $4.3\pm 0.3$ & $14.4\pm 3.9$\\
SLD    & $-39.564$ &  8.379  & $-5.588$ & $-45.15$  & 3.20 & 12.75 & $4.0\pm 0.3$ & $13.3\pm 3.6$\\ 
SQD    & $-42.812$ &  7.033  & $-4.819$ & $-47.63$  & 3.48 & 13.66 & $4.1\pm 0.3$ & $13.7\pm 3.7$\\
    CCD    & $-41.961$ & 11.243  & $-8.123$ & $-50.08$  & 3.67 & 12.74 & $4.3\pm 0.3$ & $14.4\pm 3.9$\\
\hline 
HF-RVAP   & $-13.535$ & 15.335  & $-9.815$ & $-23.35$  & 3.58 & 12.66 & $3.8\pm 0.3$ & $12.8\pm 3.4$\\
SLD    & $-39.604$ & 8.418   & $-5.547$ & $-45.15$  & 3.20 & 12.72 & $4.0\pm 0.3$ & $13.2\pm 3.5$\\ 
SQD    & $-43.112$ & 6.844   & $-4.559$ & $-47.67$  & 3.50 & 13.87 & $4.0\pm 0.3$ & $13.3\pm 3.5$\\
CCD    & $-44.979$ & 9.165   & $-5.498$ & $-50.48$  & 3.06 & 10.62 & $3.3\pm 0.3$ & $12.0\pm 3.2$\\
\hline\hline 
\end{tabular}
\caption{Results for $^8$Be. Here, $E$ and $\langle J^2\rangle$ are
  the unprojected ground-state energy and angular momentum expectation
  values, respectively, 
  $\delta E$ the gain of the ground-state energy from projection, and
  $E^{(J)}$ is the energy of the excited state with spin
  $J$. All energies are in MeV. For the projection we employed a
  model-space with $N_{\rm max} = 5$ and $\hbar\omega = 22$~MeV for
  HF, HF-VAP, HF-RVAP, SLD and SQD, while $N_{\rm max} =7$ and $\hbar\omega=24$~MeV for
  CCD.  The SLD and SQD unprojected expectation values are obtained
  using the $\beta=0$ kernels. The CCD unprojected results were obtained using the linear-response approach. 
  The EFT results are obtained by using
  $\delta E$ and $\langle J^2\rangle$ as input; uncertainties are
  based on the assumption that the breakdown scale is at about 20~MeV
  of excitation energy. The NCSM results~\cite{caprio2015} are
  $E^{(2)}-E^{(0)}=3.5$~MeV and $E^{(4)}-E^{(0)}=11.9$~MeV.
\label{tab:8be}}
\end{table*}
\endgroup

\begin{figure}[t!]
    \centering
    \includegraphics[width=0.7\columnwidth]{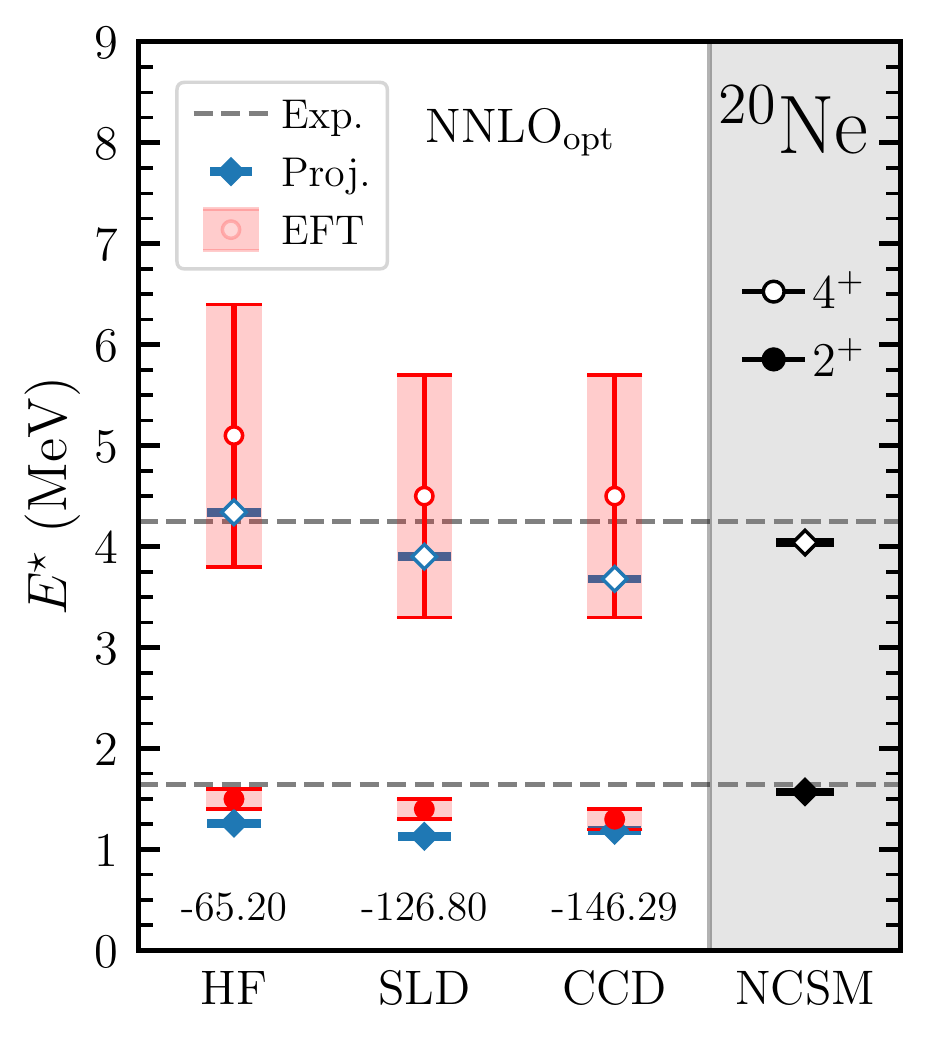}
    \caption{Excitation energies of $^{20}$Ne from projected Hartree-Fock and projected  coupled-cluster (SLD and CCD) computations compared to EFT results, the benchmark from the symmetry-adapted NCSM, and experiment (dashed lines). The numbers at the bottom of the plot are projected ground-state energies (in MeV). For the projection we employed a Hartree-Fock basis in a model-space with $N_{\rm max} = 7$ and with oscillator frequency $\hbar\omega = 20$~MeV.}
    \label{fig:ne20eft}
\end{figure}

For $^{20}$Ne we show results computed with the Hartree-Fock basis in Fig.~\ref{fig:ne20eft} (here the results shown in Table~\ref{tab:20ne} were used as input for the EFT predictions). The EFT uncertainties are based on a breakdown angular momentum $J_{\rm b}\approx 5$ which corresponds to a breakdown energy of about 7~MeV. At this energy, positive-parity states appear that are not part of the ground-state rotational band.  As for $^8$Be, the EFT ranges are above the computed energies for $J^\pi=2^+$ but agree for the $4^+$ levels.

Table~\ref{tab:20ne} summarizes our results for $^{20}$Ne in the Hartree-Fock basis using a
  model-space with $N_{\rm max} = 7$ and with the oscillator frequency
  $\hbar\omega = 20$~MeV. We see in particular that the charge radius $ R_{ch}$ is virtually unchanged by the projection. This validates several calculations of charge radii~\cite{novario2020,koszorus2021}. We also note that the projected HF results are close to the NCSM benchmark.

\begingroup
\renewcommand{\arraystretch}{1.3}

\begin{table*}[ht]
\begin{tabular}{|l|r|r|r|r|r|r|r|r|r|r|}\hline\hline
\multicolumn{1}{|c|}{} & \multicolumn{3}{c|}{Unprojected} &
             \multicolumn{5}{c|}{Projected} & \multicolumn{2}{c|}{EFT}\\ \hline
             $^{20}$Ne
             & $E$ & $\langle J^2\rangle$ & $ R_{\mathrm ch} $ (fm)
             & $\delta E$  & $E^{(0)}=E+\delta E$  & $ R_{\mathrm ch} $ (fm) & $E^{(2)}-E^{(0)}$ & $E^{(4)}-E^{(0)}$
             & $E^{(2)}-E^{(0)}$ & $E^{(4)}-E^{(0)}$ \\ \hline  
HF     & $  -59.442 $ & 22.778  & 2.623   & $-5.760$ & $ -65.202$ & 2.619 & 1.26 & 4.34  & $1.5\pm 0.1$  & $5.1\pm 1.3$ \\
SLD    & $ -122.467 $ & 19.059  & 2.601   & $-4.332$ & $-126.799$ & 2.598 & 1.13 & 3.90  & $1.4\pm 0.1$  & $4.5\pm 1.2$ \\
CCD    & $ -142.666 $ & 16.128  & 2.621   & $-3.627$ & $-146.293$ & 2.620 & 1.19 & 3.68  & $1.3\pm 0.1$  & $4.5\pm 1.2$ \\
\hline\hline
\end{tabular}
\caption{Results for $^{20}$Ne. Here, $E$ and $\langle J^2\rangle$ are
  the unprojected ground-state energy and angular momentum expectation
  values, respectively, $R_{\mathrm ch}$ is the charge radius,
  $\delta E$ the gain of the ground-state energy from projection, and
  $E^{(J)}$ is the energy of the excited state with spin/parity
  $J$. All energies are in MeV. For the projection we employed the Hartree-Fock basis in a
  model-space with $N_{\rm max} = 7$ and with the oscillator frequency
  $\hbar\omega = 20$~MeV. 
  The SLD unprojected expectation values for were obtained
  using the $\beta=0$ kernels. The CCD unprojected expectation value for $ R_{\mathrm ch} $ were obtained using the $\beta=0$ kernel while for $\langle J^2 \rangle$ we used the linear response approach. 
  The EFT results are
  obtained by using $\delta E$ and $\langle J^2\rangle$ as input;
  uncertainties are based on the assumption that the breakdown spin is $J_{\rm b}\approx 5$. The NCSM benchmarks are $E^{(2)}-E^{(0)}=1.64$~MeV and $E^{(4)}-E^{(0)}=4.25$~MeV~\cite{dytrych2020}.  The calculated charge radii are smaller than the experimental charge radius, $R_{\mathrm ch} = 3.0055(21)$~fm~\cite{angeli2013}.
\label{tab:20ne}}
\end{table*}
\endgroup

\subsection{$^{34}$Mg}

Neutron-rich magnesium isotopes have long been in the focus of
experiment~\cite{motobayashi1995,baumann2007,crawford2019,yordanov2012}
and theory~\cite{fossez2016,tsunoda2017,miyagi2020}, because they are located in the
``island of inversion'', where deformed ground states emerge within the shell-model from the presence of intruder orbits~\cite{warburton1990}. The experiments of Refs.~\cite{iwasaki2001,church2005,elekes2006,michimasa2014} observed a ground-state rotational band in
$^{34}$Mg. This provides us with an opportunity to test our projection methods using the chiral interaction NNLO$_{\rm opt}$ (which is fairly accurate for $^8$Be and $^{20}$Ne).

\begin{figure}[t!]
  \includegraphics[width=1.\columnwidth]{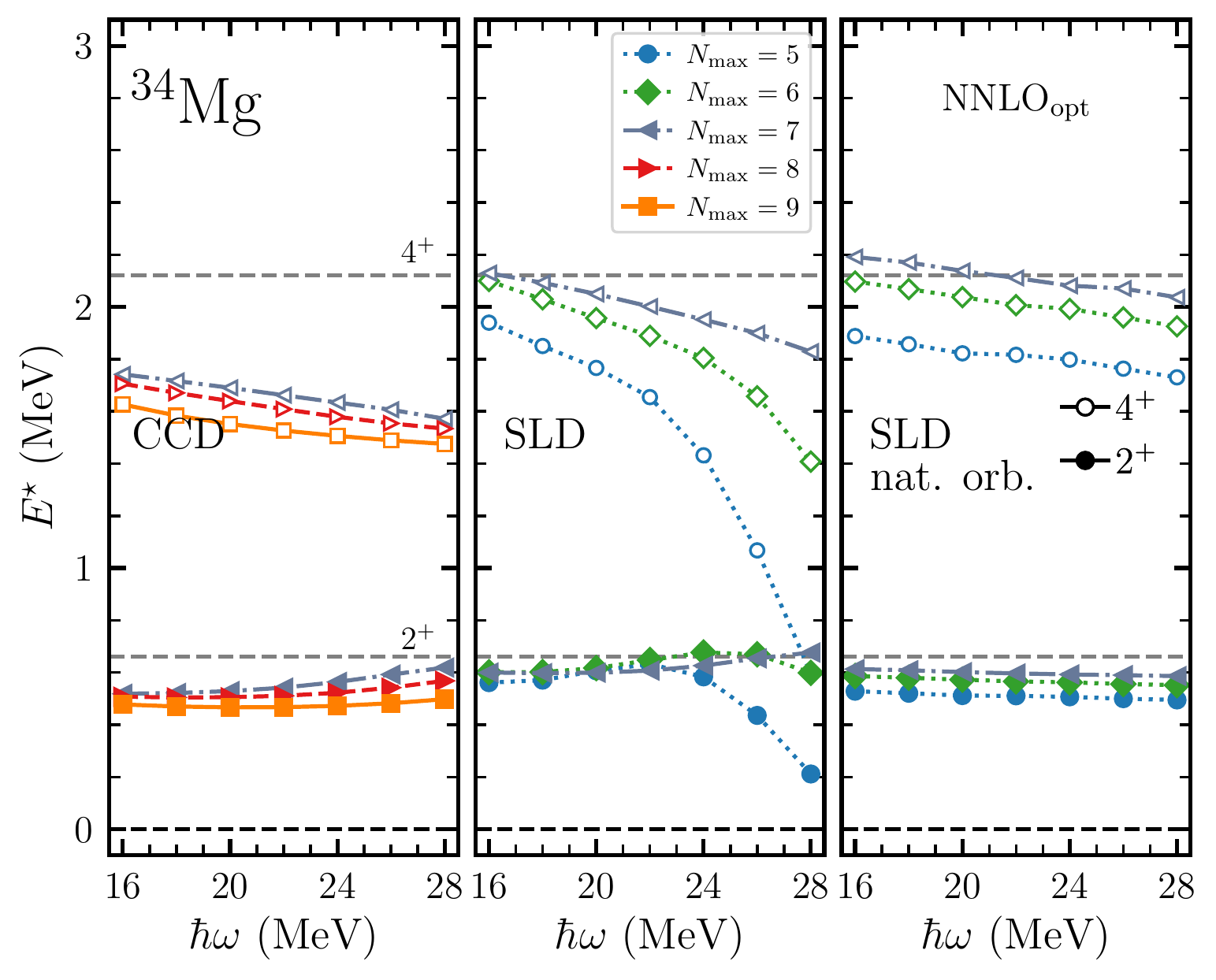}
    \caption{Projected coupled-cluster excitation energies for $^{34}$Mg as a function of oscillator frequency
    obtained from CCD (left panel), SLD (middle panel) and
    SLD approximations with natural orbitals (right panel). Model spaces are labeled by
    $N_{\rm max}$. Horizontal dashed grey lines show experimental data.\label{fig:34MgCompSpec}}
\end{figure}

To facilitate the convergence with respect to the size of the underlying spherical harmonic oscillator basis, we also use natural orbitals for SLD approach.  Those are computed from a one-body density matrix following Refs.~\cite{strayer1973,tichai2019,hoppe2021} in a large basis with $N_{\rm max} =
12$. We truncate the set of natural orbitals selected based on their large occupations following the procedure described in Ref.~\cite{hoppe2021}. The CCD and SLD results for the low-lying spectrum of $^{34}$Mg is shown in Fig.~\ref{fig:34MgCompSpec} for different $N_{\rm max}$ and oscillator frequencies in the range $\hbar\omega = 16,\ldots, 28$~MeV. The SQD results would begin to converge at $N_{\rm max}=7$ which is computationally too expensive for this nucleus. As can be seen the SLD results are close to data while the spectrum obtained with CCD is too compressed. We also here employed the CCSD approximation in the largest model-space and obtained an even more compressed spectrum with the $2^+$ energy at 0.4~MeV and the $4^+$ energy at $\sim 1.3$~MeV.

\begin{figure}[t!]
    \centering
    \includegraphics[width=0.7\columnwidth]{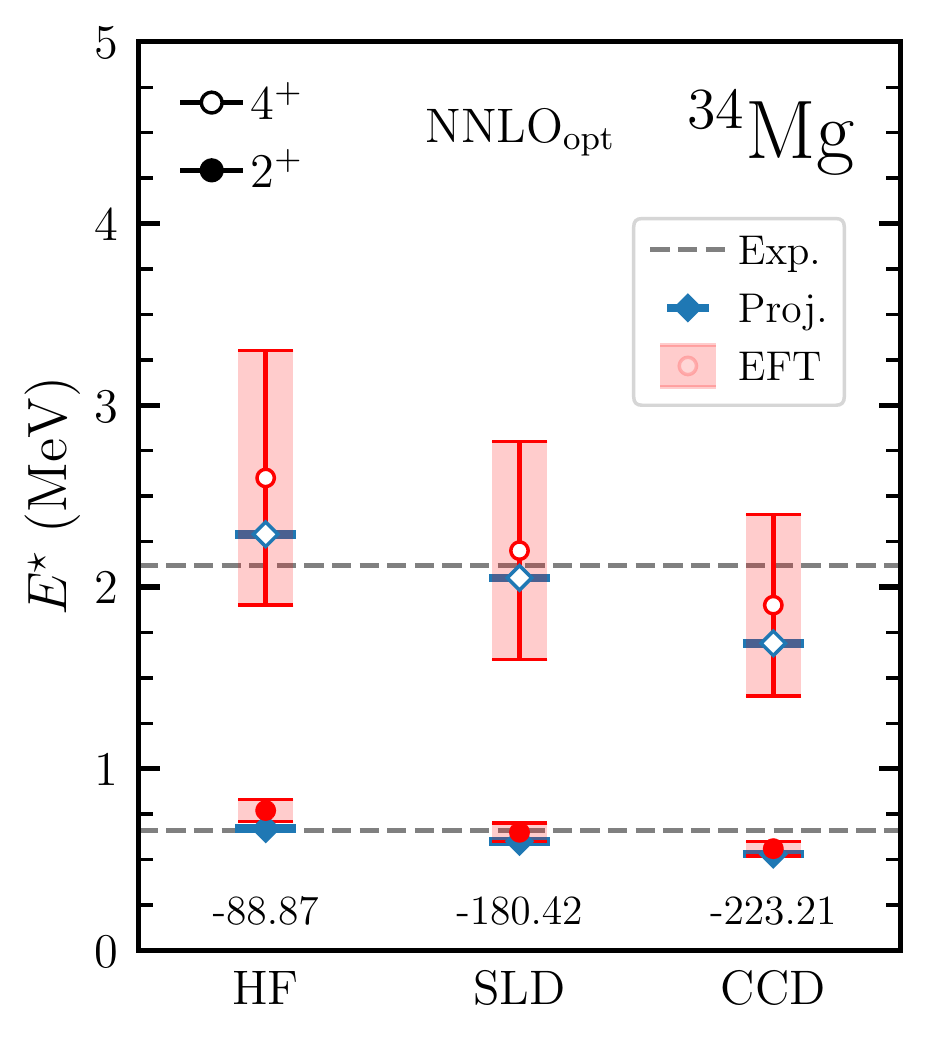}
    \caption{Excitation energies of $^{34}$Mg from projected Hartree-Fock and projected  coupled-cluster (SLD and CCD) computations compared to EFT results and experiment (dashed lines), using the results shown in Table~\ref{tab:34mg}. The numbers at the bottom of the plot are projected ground-state energies (in MeV).}
    \label{fig:mg34eft}
\end{figure}

Figure~\ref{fig:mg34eft} shows the EFT predictions for the low-lying spectrum of $^{34}$Mg obtained from projected calculations using Hartree Fock, SLD, and CCD (taken from Table~\ref{tab:34mg}). The EFT results include uncertainty estimates that are based on 
a breakdown scale of $3.2\,\MeV$. At this energy there is a level which is outside the ground-state rotational band~\cite{michimasa2014}. The corresponding breakdown spin is $J_\text{b}\approx 5$. The EFT results are consistent with the microscopic computations from angular momentum projection. They also agree with data.   

The recent computations by \textcite{miyagi2020} failed to reproduce the collective behavior of $^{34}$Mg.  Calculated  energies of the $J^\pi=2^+$ states are about a factor two too large and $B(E2)$ values are  too small.  Those calculations were based on a multi-shell approach including orbitals up to the $1p0f$ shell. Such a model space is probably too small to capture quadrupole collectivity because the mass quadrupole operator couples neutrons in the $1p0f$ shell to the $2p1f0h$ shell.

Our results for the Hartree-Fock basis and a model-space with $N_{\rm max} = 7$ and with the oscillator frequency $\hbar\omega = 20$~MeV are summarized in Table~\ref{tab:34mg}.  Again we see that charge radii only change at the per-mill level under projection. 

\begingroup
\renewcommand{\arraystretch}{1.3}
\begin{table*}[ht]

\begin{tabular}{|l|r|r|r|r|r|r|r|r|r|r|}\hline\hline
\multicolumn{1}{|c|}{} & \multicolumn{3}{c|}{Unprojected} &
             \multicolumn{5}{c|}{Projected} & \multicolumn{2}{c|}{EFT}\\ \hline
             $^{34}$Mg
             & $E$ & $\langle J^2\rangle$ & $R_{\mathrm ch}$ (fm)
             & $\delta E$  & $E^{(0)}=E+\delta E$  & $ R_{\mathrm ch}$ (fm) & $E^{(2)}-E^{(0)}$ & $E^{(4)}-E^{(0)}$
             & $E^{(2)}-E^{(0)}$ & $E^{(4)}-E^{(0)}$ \\ \hline
HF     & $-85.687$ & 24.740 & 2.727 & $-3.184$ & $-88.87$  & 2.724 & 0.67 & 2.29 & $0.77\pm 0.06$ & $2.6\pm 0.7$\\
SLD    & $-177.938$& 22.790 & 2.707 & $-2.479$ & $-180.42$ & 2.704 & 0.60 & 2.05 & $0.65\pm 0.05$ & $2.2\pm 0.6$\\
CCD    & $-221.315$& 20.213 & 2.725 & $-1.893$ & $-223.21$ & 2.722 & 0.53 & 1.69 & $0.56\pm 0.04$ & $1.9\pm 0.5$\\
\hline\hline
\end{tabular}
\caption{Results for $^{34}$Mg. Here, $E$ and $\langle J^2\rangle$ are
  the unprojected ground-state energy and angular momentum expectation
  values, respectively, $R_p^2$ is the point-proton radius squared (including spin-orbit contributions),
  $\delta E$ the gain of the ground-state energy from projection, and
  $E^{(J)}$ is the energy of the excited state with spin
  $J$. All energies are in MeV. For the projection we employed a
  model-space with $N_{\rm max} = 7$ and with the oscillator frequency
  $\hbar\omega = 20$~MeV. The SLD unprojected expectation values for were obtained
  using the $\beta=0$ kernels. The CCD unprojected expectation value for $\langle R_p^2 \rangle$ were obtained using the $\beta=0$ kernel while for $\langle J^2 \rangle$ we used the linear response approach. 
The EFT results are obtained by using $\delta E$ and $\langle J^2\rangle$ as input;
  uncertainties are based on the assumption that the breakdown scale
  is at about 3.2~MeV, i.e. the breakdown spin is $J_{\rm b}=5$.
\label{tab:34mg}}
\end{table*}
\endgroup

\subsection{Ground-state properties}

\begin{figure}[htb]
    \centering
    \includegraphics[width=.49\textwidth]{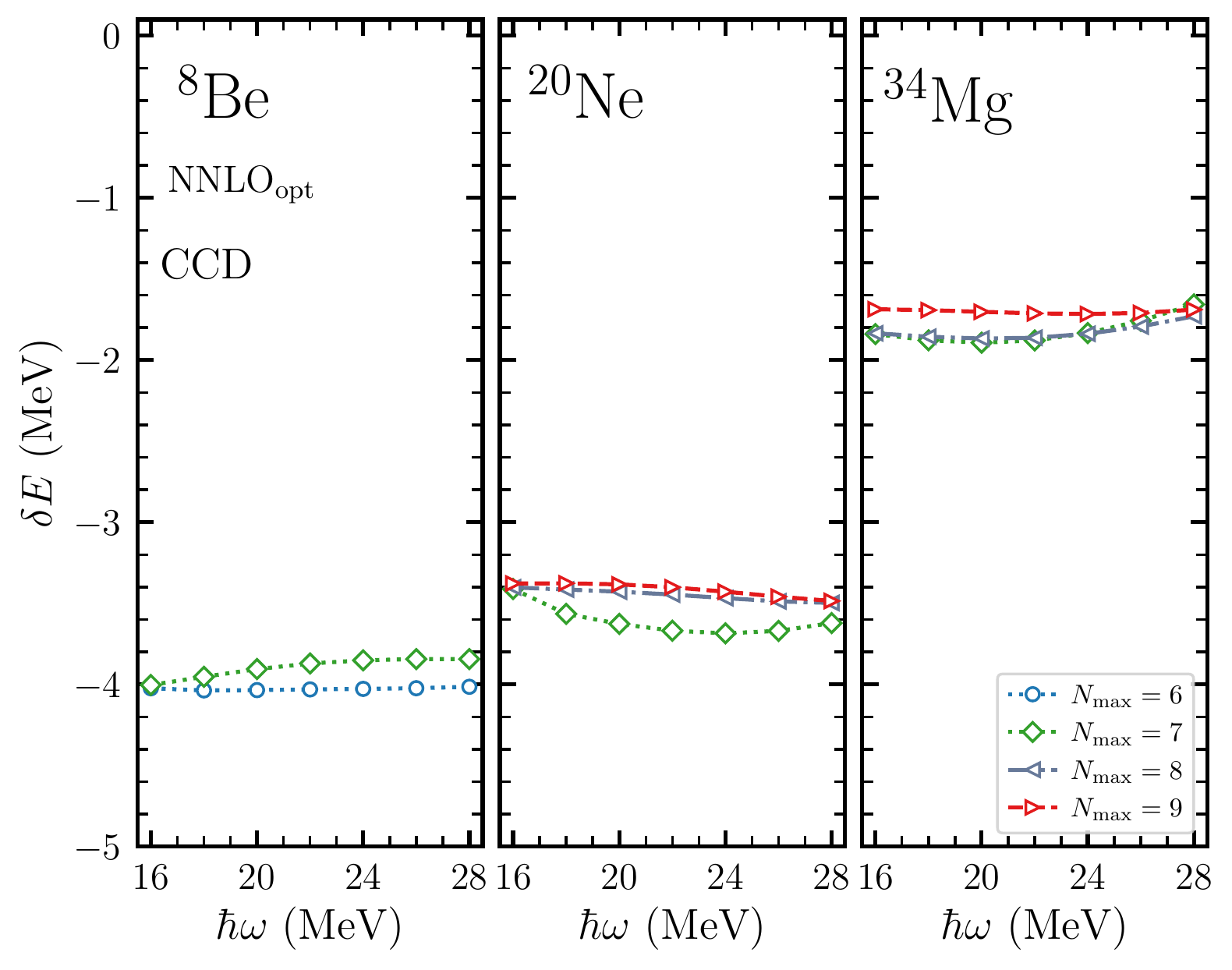}
    \caption{Gain in ground-state energy from angular-momentum projection for $^{8}$Be, $^{20}$Ne and $^{34}$Mg (from left to right) as a function of the harmonic oscillator spacing for model-space sizes as indicated by $N_{\rm max}$. Calculations are performed in the CCD approximation.}
    \label{fig:gain}
\end{figure}

Figure~\ref{fig:gain} shows the energy gain $\delta E$ from angular-momentum projection within CCD for the nuclei $^8$Be, $^{20}$Ne, and $^{34}$Mg. We see that $\delta E$ is sufficiently converged in the model spaces we employed and only find a small residual dependence on the oscillator spacing. The gain decreases with increasing mass number and is not an extensive quantity. 

\begin{figure}[htb]
    \centering
    \includegraphics[width=.49\textwidth]{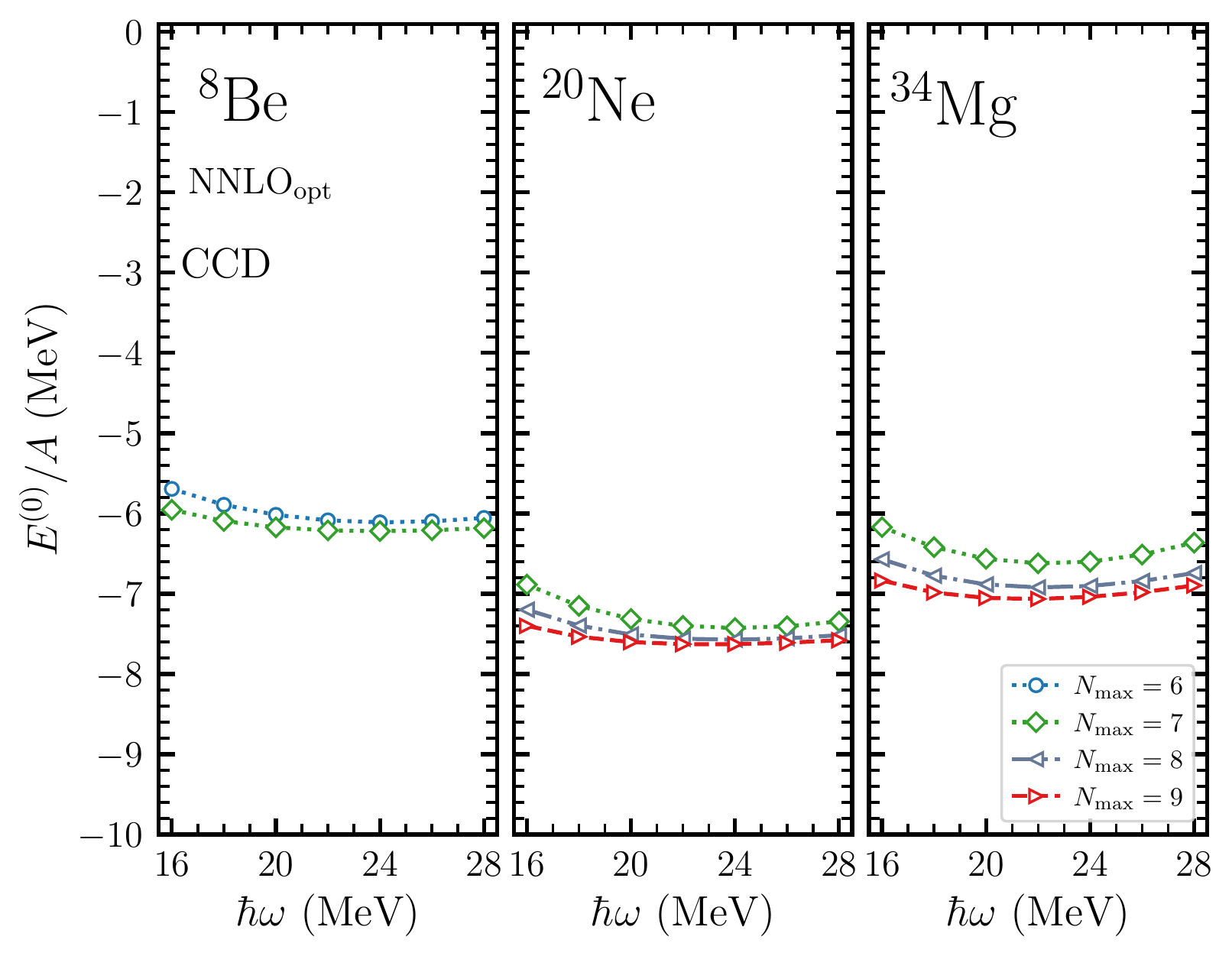}
    \caption{Projected ground-state energy per nucleon for $^{8}$Be, $^{20}$Ne and $^{34}$Mg (from left to right) as a function of the harmonic oscillator spacing for model-space sizes as indicated by $N_{\rm max}$. Calculations are performed in the CCD approximation.}
    \label{fig:BEA}
\end{figure}

Figure~\ref{fig:BEA} shows the ground-state energy per nucleon from angular-momentum projected CCD for the nuclei $^8$Be, $^{20}$Ne, and $^{34}$Mg. For comparison, projected CCSD for $N_{\rm max} = 9$ gives a total of $\sim 100$~keV and $\sim 400$~keV additional binding compared to CCD at the optimal frequency for $^{20}$Ne and $^{34}$Mg, respectively. Again, the results are sufficiently converged with respect to the employed model spaces. We note that the projection only decreases the ground state energy per particle by about 0.5, 0.17, and 0.08~MeV for $^8$Be, $^{20}$Ne, and $^{34}$Mg, respectively.

More accurate ground-state energies require us to include triples cluster excitations. We perform unprojected CCSD and CCSDT-1
computations~\cite{watts1995} starting from Hartree-Fock computations in a large single-particle basis with $N_{\rm
  max}=12$. Using the Hartree-Fock basis we then compute natural-orbitals from second-order perturbation theory and truncate the basis to a size corresponding to  $N_{\rm max}=10$ based on the occupation numbers with respect to the Fermi surface. For the CCSDT-1 calculations we reduce the number of  the \ph{3} amplitudes by an additional cut on occupation numbers, see
Refs.~\cite{tichai2019,novario2020} for details. These calculations yield the
correlation energies $\Delta E_{\rm CCSD}$ and $\Delta E_{\rm
  CCSDT-1}$ along with the ground-state expectation values
$\langle J^2\rangle_{\rm CCSD}$ and $\langle J^2\rangle_{\rm
  CCSDT-1}$.  Projected ground-state energies can be decomposed as
\be
\label{Etot}
E=E_{\rm ref} + \Delta E_{\rm CCSD} + \Delta E_{\rm CCSDT-1} +\delta E \, , 
\ee
where $\delta E$ denotes the gain of the ground-state energy from  angular-momentum projection. As the projection is presently limited to smaller bases, and we have not performed projection of triples results, we use Eq.~(\ref{estimate}) to estimate $\delta E$ based on $\langle J^2\rangle_{\rm CCSDT-1}$ and the energy spacing $E^{(2)}-E^{(0)}$ based on either SQD or SLD calculations; this estimate is denoted as $\delta E_{\rm est}$ in what follows.

\begin{figure*}[t!]
    \centering
    \includegraphics[width=.9\textwidth]{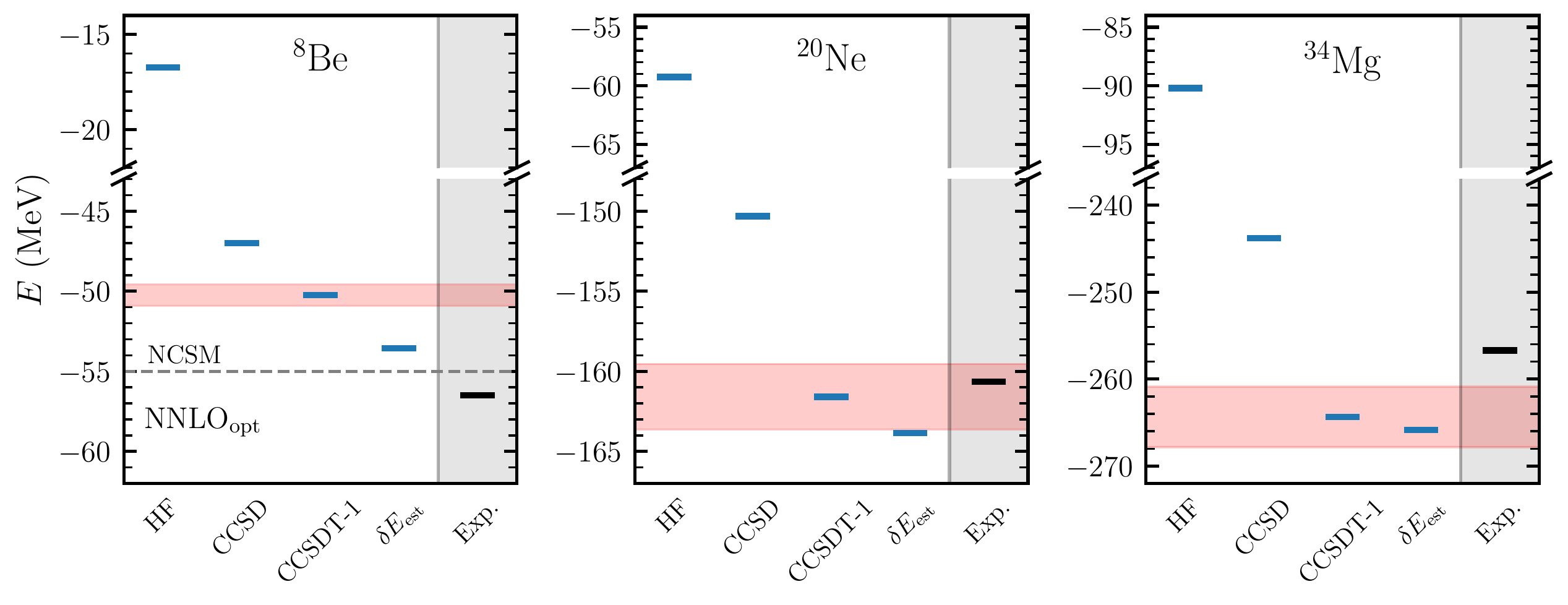}
    \caption{Ground-state energies of $^{8}$Be, $^{20}$Ne and $^{34}$Mg shown (from left to right) with unprojected Hartree Fock, unprojected CCSD, and unprojected CCSDT-1 truncations. The estimated energy gain from angular-momentum projection, $\delta E_{\rm est}$, is based on the leading-order EFT result. The light  red  band  shows  the  estimated  uncertainty  at  the CCSDT-1 truncation. The NCSM result (dashed line in $^8$Be) is from an extrapolation to an infinite model space~\cite{caprio2015}. The right column shows experiment.}
    \label{fig:gse}
\end{figure*}

Figure~\ref{fig:gse} shows how the ground-state energies of $^8$Be, $^{20}$Ne and $^{34}$Mg decrease with increasing 
sophistication of the unprojected many-body computation (from left to right), and with addition of the estimate $\delta E_{\rm est}$ from angular-momentum projection. 
The light red band shows the estimated uncertainty at the CCSDT-1 truncation (which is about 2\% of the total correlation energy). For $^8$Be, the estimated energy gain from projection is clearly outside the uncertainty estimate from the CCSDT-1 truncation. This is possibly due to the strong $\alpha$ correlations in this unbound nucleus. The ground-state energy of the $\alpha$ particle is $-27.76$~MeV with the NNLO$_{\rm opt}$ potential~\cite{ekstrom2013}, and our result (as well as the extrapolated NCSM energy of about $-55.0$~MeV~\cite{caprio2015}) are thus above the $\alpha-\alpha$ threshold.  
We note that the estimated energy gain from projection decreases with increasing mass number.  
The comparison with experiment (shown as black bars) shows 
that the NNLO$_{\rm opt}$ interaction overbinds $^{20}$Ne and $^{34}$Mg. 

The computation of $\langle Q_{20}\rangle$ and $R_{ch}^2$ in the unprojected states allows us to compute the nuclear deformation parameter $\beta$ via Eq.~\eqref{beta2}.
We observe no significant  difference between CCSD and  CCSDT-1 results. 
For $^{20}$Ne we find $\beta_2\approx 0.70$, larger than the value $\beta_2=0.47$ deduced from experimental data~\cite{swiniarski1969}. For $^{34}$Mg, we find $\beta_2\approx 0.57$ and this agrees within experimental uncertainties with the values $\beta_2=0.62(6)$ and $\beta_2=0.68(16)$ extracted~\cite{pritychenko2016} from experimental data in Refs.~\cite{michimasa2014} and~\cite{elekes2006}, respectively. 
Our results for the ground-state properties of $^{8}$Be, $^{20}$Ne and $^{34}$Mg are summarized in Table~\ref{tab:final_results}.

\begingroup
\renewcommand{\arraystretch}{1.3}
\begin{table*}[ht]
\begin{tabular}{|l|r|r|r|r|c|c|c|c|c|r|r|r|}\hline\hline
             & $E_{\rm ref}$ & $\langle J^2\rangle_{\rm ref}$& $\langle Q_2\rangle_{\rm ref}$& $\Delta E_{\rm SD}$ 
             & $\langle J^2\rangle_{\rm SD}$ & $\langle Q_2\rangle_{\rm SD}$& $\Delta E_{\rm
                                                 SDT-1}$ & $\langle
                                                             J^2\rangle_{\rm
                                                             SDT-1}$& $\langle Q_2\rangle_{\rm SDT-1}$&
                                                                        $\delta
                                                                        E_{\rm
                                                                        est}$
  & $E$
             & $E_{\rm Exp}$ \\ \hline
$^8$Be     & $-16.74$ & $11.17$ &$19.46$&$-30.26$ & $6.69$ &$19.64$ &$-3.24$ & $5.82$& $18.86$&$-3.33$ & $-53.58$ & $-56.50$ \\
$^{20}$Ne & $-59.62$ &  $21.26$ & $35.84$& $-91.06$ &$14.71$ &$36.34$& $-11.27$ & $12.09$ &$35.71$& $-2.26$ & $-164.21$ & $-160.64$ \\
$^{34}$Mg & $-90.21$ & $22.62$ & $38.56$& $-153.57$ &$18.40$ & $38.38$& $-20.56$ & $15.03$&$36.97$& $-1.50$ & $-265.84$  & $-256.71$ \\
\hline\hline      
\end{tabular}
\caption{Summary of results for $^8$Be, $^{20}$Ne and
  $^{34}$Mg. Calculations started from a natural orbital basis
  constructed in a $N_{\rm{max}} =12$ model-space and truncated to
  $N_{\rm{max}}^{\rm{nat}} =10 $ for the coupled-cluster
  calculations. We used the oscillator frequencies $\hbar\omega =
  24,22$ and $\hbar\omega = 18$~MeV for $^8$Be, $^{20}$Ne and
  $^{34}$Mg, respectively.
  \label{tab:final_results}}
\end{table*}
\endgroup

\section{Shell-model Hamiltonians}
\label{sec:pf}

We also compared projected coupled-cluster computations with full configuration interaction (FCI) results for the traditional shell-model.
We employ the KB3G interaction for $pf$-shell nuclei~\cite{poves1981}. 
The model space is relatively small and consists of four spherical orbitals above the frozen core of $^{40}$Ca.
In the nuclear shell model, many-body wave functions typically exhibit strong correlations, and truncated coupled-cluster wave functions do not yield accurate total energies~\cite{horoi2007}, while spectra are more accurate~\cite{novario2021}.

\begin{figure}[t!]
  \includegraphics[width=1.0\columnwidth]{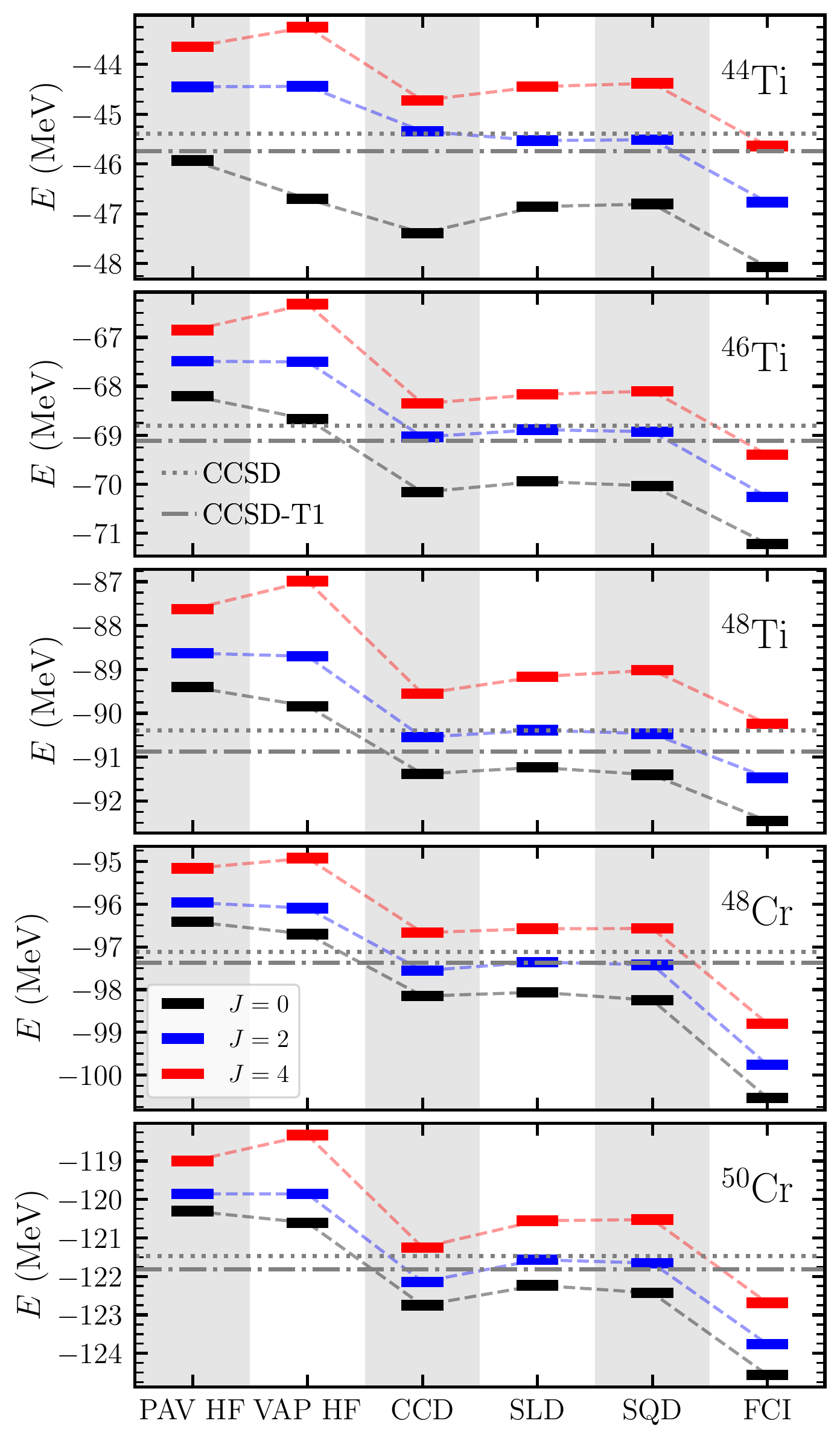}
  \caption{Low-lying states with spin $J$ of selected $pf$-shell nuclei computed with 
    projection-after-variation Hartree-Fock (PAV HF), variation-after-projection Hartree-Fock (VAP HF), and projected CCD, SLD, and SQD methods, and compared to FCI results. Horizontal
    lines shown are the ground-state energies from unprojected
    coupled-cluster computations.\label{fig:pfCombo}}Massachusetts Institute of Technology

\end{figure}

Figure~\ref{fig:pfCombo} compares various projection
techniques with exact FCI results
for the nuclei $^{44}$Ti, $^{46}$Ti, $^{48}$Ti, $^{48}$Cr, and $^{50}$Cr.  
The Hartree-Fock energies lack several MeV of
binding, and the VAP results for the spectrum are not more accurate
than those from PAV. Including correlations beyond Hartree-Fock
significantly lowers the ground-state energies, as shown by the CCD, SLD, and SQD results, and the spectra also improve.  The spacings of the
SQD spectra are close to the FCI results.  The dotted and
dashed-dotted horizontal lines show the results from single-reference
CCSD and CCSDT-1, respectively. We see that projected SLD and SQD
results improve on these ground-state energies, and CCD results gain the most energy. We attribute the
difference between the projected coupled-cluster and FCI results to lacking many-particle--many-hole excitations.

\section{Summary}
\label{sec:sum}

We performed angular-momentum projection after variation within coupled-cluster theory using two different approaches. The first is based on the non-Hermitian energy functional of coupled-cluster theory and  projects within the disentangled formalism via ordinary differential equations. This approach scales favorably and yields somewhat too compressed spectra when compared with benchmarks. We expect that more accurate projections could be achieved within a bi-variational energy functional, i.e. using a more sophisticated parameterization of the bra state. Furthermore, we also expect improvements from inclusion of $W_3$ in the differential equations. The second approach employs truncated coupled-cluster wave functions within a Hermitian energy functional. We studied a linear and a quadratic truncation of the doubles cluster operator. The latter is numerically very  expensive. Spectra are more accurate than in the non-Hermitian approach. Benchmark calculations for $^8$Be and $^{20}$Ne based on a chiral nucleon-nucleon potential show that the angular momentum projections are fairly accurate. Comparing with traditional shell-model calculation of $pf$-shell nuclei revealed the more accurate reproduction of spectra within the Hermitian approach and the larger ground-state energy gain within the non-Hermitian approach. The computed ground-state rotational band of the exotic nucleus $^{34}$Mg agrees with experimental data. This opens the avenue to perform \ai{} predictions of rotational properties in medium-mass exotic nuclei. 

We used our microscopic results to compute the low-energy constants of the rigid-rotor model (which is the leading order Hamiltonian within an effective theory of nuclear rotation) and estimated uncertainties from omitted higher-order corrections. The effective theory accurately reproduces rotational bands. 

The ground-state energy gained from angular-momentum projection decreases with increasing mass number, and this is understood within the effective theory. Thus, such corrections can be computed using methods that are not size extensive. Computations of charge radii revealed that projections have only little impact on this observable.

\allowdisplaybreaks

\begin{acknowledgments}
We thank Thomas~M.~Henderson for many useful discussions and for
sharing his numerical programs with us. We also thank Titus~D.~Morris for
many useful discussions, and Mark Caprio and Kristina Launey for sharing
their results with us. This work was supported by the U.S. Department
of Energy, Office of Science, Office of Nuclear Physics, under Award
Nos.~DE-FG02-96ER40963 and DE-SC0018223.  
A.T. was supported in part by the  Deutsche  Forschungsgemeinschaft  (DFG,  German Research Foundation) -- Projektnummer 279384907 -- SFB 1245 and by the European Research Council (ERC) under the European Union's Horizon 2020 research and innovation programme (Grant Agreement No.~101020842).
Computer time was provided
by the Innovative and Novel Computational Impact on Theory and
Experiment (INCITE) programme. This research used resources of the Oak
Ridge Leadership Computing Facility located at Oak Ridge National
Laboratory, which is supported by the Office of Science of the
Department of Energy under contract No. DE-AC05-00OR22725.
\end{acknowledgments}

\appendix

\section{Transformed operators}

\label{sub-sim}

\subsection{Computation of $\breve{O}$}
\label{subsec:breveO}
Considering an arbitrary operator $O$, one is presently interested in computing its similarity-transformed partner
\ba
\breve{O} &\equiv&  e^{-Z} O e^{Z} \, , \label{simtransoapp}
\ea
where $Z$ is a one-body operator $Z\equiv \sum_{pq} z_{pq} c^\dagger_p c_q$. Because of the one-body nature of $Z$, the Baker-Campbell-Hausdorff expansion is simple and one finds 
\begin{subequations}
\label{simtransfo}
\ba
e^{-Z} c_p e^{Z} &=& \sum_q \left[e^Z\right]_{pq}c_q  \ , \\
e^{-Z} c^\dagger_p e^{Z} &=& \sum_q c^\dagger_q \left[e^{-Z}\right]_{qp}  \ .
\ea
\end{subequations}
Here and in what follows, the matrix of a single-particle operator is denoted using brackets. 

The transformation is thus represented by a matrix exponential that takes a particularly simple form for $Z\equiv T_1$. Let us first provide the matrix representation of $T_1$
\ba
\left[T_1\right] = 
\left(
\begin{array}{cc}
    t_{hh} & t_{hp}\\
    t_{ph} & t_{pp}    
\end{array}
\right)
\equiv \left(\begin{array}{cc}
  0 & 0 \\
  t & 0 \end{array}\right) \, , \label{T1matrix}
\ea
where the one-body basis has been separated into occupied (hole) and unoccupied (particle) states in $| \Phi \rangle$. Because $T_1$ (Eq.~\ref{clusteroperators}) is a pure excitation operator, only the particle-hole block $t$ differs from zero. 

Given Eq.~\eqref{T1matrix}, the matrix exponential that transforms one-body annihilation operators reads as
\ba
\left[e^{T_1}\right] = \left[1+T_1\right]
= \left(\begin{array}{cc}
  1 & 0 \\
  t & 1 \end{array}\right) \ ,  \label{simT1transfo1}
\ea
whereas the one transforming one-body creation operators is
\ba
\left[e^{-T_1}\right] = \left[1-T_1\right]
= \left(\begin{array}{cc}
  1 & 0 \\
  -t & 1 \end{array}\right) \ . \label{simT1transfo2}
\ea
In order to illustrate how Eq.~\eqref{simtransoapp} eventually operates, let us consider a one-body operator  
\ba
O \equiv \sum_{pq} o_{pq} c^\dagger_p c_q \ ,
\ea
as an example. Given Eq.~\eqref{simtransfo}, the similarity-transformed operator is itself a one-body operator
\ba
\breve{O} &\equiv& \sum_{pq} \breve{o}_{pq} c^\dagger_p c_q \ ,
\ea
whose matrix elements read as
\ba
\breve{o}_{pq} &\equiv& \left[\breve{O}\right]_{pq}  \nonumber \\
&=& \sum_{rs} \left[e^{-T_1}\right]_{pr} \left[O\right]_{rs} \left[e^{T_1}\right]_{sq} \ ,
\ea
i.e. to each ket and bra index of the initial operator's matrix elements we associate a right and left multiplication with matrix \eqref{simT1transfo1} and \eqref{simT1transfo2}, respectively.

\subsection{Computation of $\tilde{O}$}

We consider an arbitrary operator $O$ and want to compute 
\ba
\tilde{O} &\equiv&  e^V {\cal R}^{-1}(\beta) O {\cal R}(\beta) e^{-V}\, , \label{tildeop}
\ea
where the modified rotation operator is given by
\be
{\cal R}(\beta) \equiv e^{T_1^\dagger} R(\beta) e^{T_1} \ .
\ee
This similarity transformation was used in Eq.~(\ref{STROTU}). 
The operator $\tilde{O}$ thus results from four successive elementary transformations as described in Subsect.~\ref{subsec:breveO}, taking successively $Z\equiv T_1^{\dagger}$, $Z\equiv-i\beta J_y$, $Z\equiv T_1$, and $Z\equiv -V$. Each of these transformations corresponds to a change of the single-particle basis. The third transformation corresponds to the case worked out in Subsect.~{subsec:breveO}.  The first one is deduced from it by taking the Hermitian conjugate of matrices \eqref{simT1transfo1} and \eqref{simT1transfo2}. The second transformation has  the matrix elements $\left[R(\beta)\right]_{pr}$ of $R(\beta)$ in the Hartree-Fock basis. These matrix elements can be obtained from the initial matrix elements of $R(\beta)$ in the spherical single-particle basis [Eq.~\eqref{reducedwigner}] via the Hartree-Fock transformation. Thus, the first three transformations multiply each ket index of the original matrix elements from the right by the matrix 
\ba
\left[{\cal R}(\beta)\right] = 
\left(
\begin{array}{cc}
  1 & t^{\dagger} \\
  0 & 1 
\end{array}
\right) 
[R(\beta)]
\left(
\begin{array}{cc}
  1 & 0 \\
  t & 1 
\end{array}
\right) 
\ ,  \label{Rcal}
\ea
and each bra index from the left by the matrix
\ba
\left[{\cal R}^{-1}(\beta)\right] = 
\left(
\begin{array}{cc}
  1 & 0 \\
  -t & 1 
\end{array}
\right) 
[R(-\beta)]
\left(
\begin{array}{cc}
  1 & -t^{\dagger} \\
  0 & 1 
\end{array}
\right) 
\, .  \label{Rcal-1}
\ea

Finally, we need to work out the fourth transformation associated with $Z\equiv -V$. Decomposing Eq.~\eqref{Rcal} according to
\be
\left[{\cal R}(\beta)\right] = 
\left(
\begin{array}{cc}
    {\cal R}_{hh} & {\cal R}_{hp}\\
    {\cal R}_{ph} & {\cal R}_{pp}    
\end{array}
\right) \, ,
\ee
the matrix associated with the operator $V$ reads~\cite{qiu2017}
\ba
\left[V\right] \equiv \left(\begin{array}{cc}
  0 & v_{hp} \\
  0 & 0 \end{array}\right) \, 
\ea
with
\ba
v_{hp} = ({\cal R}_{hh})^{-1} {\cal R}_{hp} \equiv v\ .
\ea
Thus, the similarity transformation is worked out such that to each ket index of the initial matrix elements is associated a right multiplication with the matrix 
\ba
\left(\begin{array}{cc}
  1 & -v \\
  0 & 1 
  \end{array}\right) \, ,
\ea
and to each bra index is associated a left multiplication with the matrix
\ba
\left(\begin{array}{cc}
  1 & v \\
  0 & 1 \end{array}\right) \, .
\ea

\section{Many-body matrix elements}

\subsection{Operator $\overline{U}$}
\label{MEDEXCOP}

In Sec.~\ref{algexpr}, the Hamiltonian and norm kernels where expressed in terms of the matrix elements of the excitation part of the similarity-transformed disconnected excitation operator $\overline{U}$
\begin{align}
 u^{a_1 a_2 ...}_{i_1 i_2...} =  \langle\Phi^{a_1 a_2...}_{i_1 i_2 ...}\vert \overline{U}\vert \Phi\rangle \, .
\end{align}
The expressions of these matrix elements up to \ph{4} are 
\begin{subequations}
 \begin{align}
  \overline{U}^{(0)}  & = 1 + \overline{T}_2^{(0)} \nonumber \\
   & + {1\over 2} \left( \sum_{ia}
           \tau^i_a  \tau^a_i + {1\over 4} \sum_{ijab}  \overline{t}^{ij}_{ab}
           \overline{t}^{ab}_{ij} \right) \, , \\
 \overline{u}^a_i  & =  \left( 1 + {1\over 2}\overline{T}_2^{(0)} \right)
          \tau^a_i \nonumber \\
   & + {1\over2} \left( \sum_{c}
           \tau^a_c  \tau^c_i  
          -\sum_{k}\tau^a_k  \tau^k_i
          \right. \nonumber \\
   & + \sum_{kc}\overline{t}^{ac}_{ik}  \tau^k_c  +
     \sum_{kc}\overline{t}^{ak}_{ic}  \tau^c_k\nonumber \\
   & + \left. {1\over 2} \sum_{cdk}\overline{t}^{ak}_{cd}  \overline{t}^{cd}_{ik} -
   {1\over 2}\sum_{ckl}\overline{t}^{ac}_{kl}
     \overline{t}^{kl}_{ic} \right) \, , \\
  \overline{u}^{ab}_{ij}  & =  \left( 1 + {1\over 2}\overline{T}_2^{(0)} \right) \overline{t}^{ab}_{ij} + 
                P(ab)\tau^a_i \tau^b_j  \nonumber \\
   & + {1\over 2} \left( P(ab) \sum_{c} \overline{t}^{ac}_{ij}\tau^b_c - P(ij)\sum_k
     \overline{t}^{ab}_{ik}\tau^k_j \right. \nonumber \\
     &  + P(ij) \sum_c \overline{t}^{ab}_{ic} \tau^c_j - P(ab)\sum_k
       \overline{t}^{ak}_{ij}\tau^b_k  \nonumber \\
   & + P(ab)P(ij)\sum_{kc} \overline{t}^{ac}_{ik}\overline{t}^{kb}_{cj} \nonumber \\
   & \left. + {1\over
     2}\sum_{kl}\overline{t}^{ab}_{kl}\overline{t}^{kl}_{ij} + {1\over 2}
     \sum_{cd}\overline{t}^{ab}_{cd}\overline{t}^{cd}_{ij} \right) \, ,\\
   \label{3p3h}
\overline{u}^{abc}_{ijk}  & =  P(ab/c)P(ij/k) \overline{t}^{ab}_{ij}
                 \tau^c_k \nonumber \\
   & + {1\over 2} P(ab/c)P(ij/k) \left(
                 \sum_{d} \overline{t}^{cd}_{ij}  \overline{t}^{ab}_{dk} -  \sum_{l}
                 \overline{t}^{ab}_{kl}  \overline{t}^{lc}_{ij}
     \right) \nonumber \\
        & =  P(ab/c)P(ij/k) \left( \overline{t}^{ab}_{ij}
          \tau^c_k  -  \sum_{l}
                 \overline{t}^{ab}_{kl}  \overline{t}^{lc}_{ij}
          \right) \, , \\
     \label{4p4h}
   \overline{u}^{abcd}_{ijkl}  & =  {1\over 2}P(ab/cd)P(ij/kl) \overline{t}^{ab}_{ij}
                 \overline{t}^{cd}_{kl}  \, . 
\end{align}
\end{subequations}
Here $P$ denotes the permutation operator~\cite{shavittbartlett2009}. In Eq.~\eqref{3p3h} the sum over $d$ is the negative of the sum over $l$ after
anti-symmetrization, as can be shown by writing out the explicit
similarity transformations of $T_2$. 

Employing the SLD approximation leads to $U=(1+T_2)$ such that the
matrix elements beyond rank two vanish. In this case, the above expressions reduce to
$\overline{U}^{(0)}=1 + \overline{T}_2^{(0)}$, $\overline{u}_i^a=\tau_i^a$,
and $\overline{u}_{ij}^{ab}=\overline{t}_{ij}^{ab}$.  

\subsection{Operator $\overline{H}\overline{U}$}
\label{MESTHU}

Similarly, the matrix elements
\begin{align}
 \overline{x}^{a_1 a_2 ...}_{i_1 i_2...} =  \langle\Phi^{a_1 a_2...}_{i_1 i_2 ...}\vert \overline{H}\overline{U}\vert \Phi\rangle \, ,
\end{align}
are needed up to \ph{4} excitation level. These are  
\begin{subequations}
\begin{align}
\overline{X}^{(0)} &= \overline{H}^{(0)} \overline{U}^{(0)} \nonumber \\
   & + \sum_{ia} \overline{h}^i_a
\overline{u}^a_i + {1\over 4}\sum_{ijab} \overline{h}^{ij}_{ab} \overline{u}^{ab}_{ij} \, , \\
 \overline{x}^a_i  & =  \overline{H}^{(0)} \overline{u}^a_i + \overline{h}^a_i \overline{U}^{(0)} +
          \sum_{bj} \overline{h}^{aj}_{ib} \overline{u}^b_j + \sum_b
          \overline{h}^a_b \overline{u}^b_i
       \nonumber \\
   & + \sum_{bj}\overline{h}^{aj}_{ib}\overline{u}^b_j + {1\over 2}\sum_{bcj}
     \overline{h}^{aj}_{bc} \overline{u}^{bc}_{ij} - {1\over
     2}\sum_{bjk}\overline{h}^{jk}_{ib}\overline{u}^{ab}_{jk}  \nonumber  \\
   & + \sum_{bj}\overline{h}^j_b \overline{u}^{ab}_{ij} + {1\over 4} \sum_{bcjk}
     \overline{h}^{bc}_{jk} \overline{u}^{abc}_{ijk} \, , \\
  \overline{x}^{ab}_{ij} & =  \overline{H}^{(0)} \overline{u}^{ab}_{ij} +
  \overline{h}^{ab}_{ij} \overline{U}^{(0)} + P(ab)P(ij)  \overline{h}^{a}_{i}
                \overline{u}^b_j \nonumber \\
  & + P(ab)\sum_c \overline{h}^b_c \overline{u}^{ac}_{ij} - P(ij)\sum_k \overline{h}^k_j
    \overline{u}^{ab}_{ik} + {1\over2}\sum_{cd}\overline{h}^{ab}_{cd} \overline{u}^{cd}_{ij}
    \nonumber \\
  & {1\over2}\sum_{kl}\overline{h}^{kl}_{ij} \overline{u}^{ab}_{kl} +
    P(ab)P(ij)\sum_{ck} \overline{h}^{kb}_{cj} \overline{u}^{ac}_{ik} +
    P(ij)\sum_{c}\overline{h}^{ab}_{cj} \overline{u}^c_i \nonumber \\
  & -P(ab) \sum_{k}\overline{h}^{kb}_{ij} \overline{u}^a_k +
    \sum_{ck}\overline{h}^k_c\overline{u}^{abc}_{ijk}  + P(ab){1\over
    2}\sum_{cdk}\overline{h}^{bk}_{cd}\overline{u}^{acd}_{ijk} \nonumber \\
 &-  P(ij) {1\over 2}\sum_{ckl}\overline{h}^{kl}_{jc}\overline{u}^{abc}_{ikl} +
   {1\over 4} \sum_{cdkl}\overline{h}^{kl}_{cd}\overline{u}^{abcd}_{ijkl} \, , \\
  \overline{x}^{abc}_{ijk} & =  \overline{H}^{(0)} \overline{u}^{abc}_{ijk} + P(ab/c)P(ij/k)
                  \left( \overline{h}^{ab}_{ij}
                \overline{u}^c_k + \overline{u}^{ab}_{ij}
                  \overline{h}^c_k \right) \nonumber \\
  & + P(ab/c)\sum_d \overline{u}^{abd}_{ijk} \overline{h}^c_d  - P(ij/k)\sum_l \overline{u}^{abc}_{ijl}
    \overline{h}^l_k \nonumber \\
  &  + {1\over 2}P(ab/c)\sum_{de} \overline{u}^{dec}_{ijk} \overline{h}^{ab}_{de}
    + {1\over 2}P(ij/k)\sum_{lm} \overline{u}^{abc}_{lmk} \overline{h}^{lm}_{ij}
    \nonumber \\
  & +P(ij/k) P(ab/c)\left( \sum_{dl} \overline{u}^{abd}_{ijl}
    \overline{h}^{lc}_{dk} + \sum_{d} \overline{u}^{cd}_{ij}
    \overline{h}^{ab}_{dk} \right. \nonumber \\
  & - \left. \sum_{l} \overline{u}^{ab}_{kl}  \overline{h}^{lc}_{ij} \right) +
    \sum_{dl} \overline{u}^{abcd}_{ijkl}\overline{h}^k_d + {1\over
    2}P(ab/c)\sum_{del} \overline{u}^{abde}_{ijkl}\overline{h}^{cl}_{de}
    \nonumber \\
 & - {1\over
   2}P(ij/k)\sum_{dlm} \overline{u}^{abcd}_{ijlm}\overline{h}^{lm}_{kd} \, , \\
  \overline{x}^{abcd}_{ijkl} & = \overline{H}^{(0)} \overline{u}^{abcd}_{ijkl} + P(ab/cd)P(ij/kl)
                    \overline{h}^{ab}_{ij} \overline{u}^{cd}_{kl} \nonumber \\
        & +  P(abc/d)P(ijk/l) \left( \overline{h}^{d}_{l} \overline{u}^{abc}_{ijk}
          + \sum_{em}  \overline{u}^{abce}_{ijkm}\overline{h}^{md}_{el} \right)
          \nonumber \\ 
  & +  P(abc/d)\sum_e\overline{h}^{d}_{e} \overline{u}^{abce}_{ijkl} 
    -  P(ijk/l)\sum_m\overline{h}^{m}_{l} \overline{u}^{abcd}_{ijkm} \nonumber \\
  & + {1\over
    2} P(ab/cd) \sum_{ef}\overline{u}^{abef}_{ijkl} \overline{h}^{cd}_{ef}
     + {1\over
    2} P(ij/kl) \sum_{mn}\overline{u}^{abcd}_{ijmn} \overline{h}^{mn}_{kl}
    \nonumber \\
  & +P(ab/cd)P(ijk/l) \sum_{e}   \overline{u}^{abe}_{ijk} \overline{h}^{cd}_{el}
    \nonumber \\
   &   - P(abc/d)P(ij/kl) \sum_{m}   \overline{u}^{abc}_{ijm}
     \overline{h}^{md}_{kl} \, .
\end{align}
\end{subequations}
Here $\overline{H}^{(0)}$, $\overline{h}^p_q$ and $\overline{h}^{pq}_{rs}$ denote the matrix elements of the normal-ordered zero-, one-, and two-body parts of
$\overline{H}$, respectively.


\begin{thebibliography}{132}%
\makeatletter
\providecommand \@ifxundefined [1]{%
 \@ifx{#1\undefined}
}%
\providecommand \@ifnum [1]{%
 \ifnum #1\expandafter \@firstoftwo
 \else \expandafter \@secondoftwo
 \fi
}%
\providecommand \@ifx [1]{%
 \ifx #1\expandafter \@firstoftwo
 \else \expandafter \@secondoftwo
 \fi
}%
\providecommand \natexlab [1]{#1}%
\providecommand \enquote  [1]{``#1''}%
\providecommand \bibnamefont  [1]{#1}%
\providecommand \bibfnamefont [1]{#1}%
\providecommand \citenamefont [1]{#1}%
\providecommand \href@noop [0]{\@secondoftwo}%
\providecommand \href [0]{\begingroup \@sanitize@url \@href}%
\providecommand \@href[1]{\@@startlink{#1}\@@href}%
\providecommand \@@href[1]{\endgroup#1\@@endlink}%
\providecommand \@sanitize@url [0]{\catcode `\\12\catcode `\$12\catcode
  `\&12\catcode `\#12\catcode `\^12\catcode `\_12\catcode `\%12\relax}%
\providecommand \@@startlink[1]{}%
\providecommand \@@endlink[0]{}%
\providecommand \url  [0]{\begingroup\@sanitize@url \@url }%
\providecommand \@url [1]{\endgroup\@href {#1}{\urlprefix }}%
\providecommand \urlprefix  [0]{URL }%
\providecommand \Eprint [0]{\href }%
\providecommand \doibase [0]{http://dx.doi.org/}%
\providecommand \selectlanguage [0]{\@gobble}%
\providecommand \bibinfo  [0]{\@secondoftwo}%
\providecommand \bibfield  [0]{\@secondoftwo}%
\providecommand \translation [1]{[#1]}%
\providecommand \BibitemOpen [0]{}%
\providecommand \bibitemStop [0]{}%
\providecommand \bibitemNoStop [0]{.\EOS\space}%
\providecommand \EOS [0]{\spacefactor3000\relax}%
\providecommand \BibitemShut  [1]{\csname bibitem#1\endcsname}%
\let\auto@bib@innerbib\@empty
\bibitem [{\citenamefont {Schmid}\ \emph {et~al.}(1989)\citenamefont {Schmid},
  \citenamefont {Ren-Rong}, \citenamefont {Gr{\"u}mmer},\ and\ \citenamefont
  {Faessler}}]{schmidt1989}%
  \BibitemOpen
  \bibfield  {author} {\bibinfo {author} {\bibfnamefont {K.~W.}\ \bibnamefont
  {Schmid}}, \bibinfo {author} {\bibfnamefont {Zheng}\ \bibnamefont
  {Ren-Rong}}, \bibinfo {author} {\bibfnamefont {F.}~\bibnamefont
  {Gr{\"u}mmer}}, \ and\ \bibinfo {author} {\bibfnamefont {Amand}\ \bibnamefont
  {Faessler}},\ }\bibfield  {title} {\enquote {\bibinfo {title} {Beyond
  symmetry-projected quasi-particle mean fields: A new variational procedure
  for nuclear structure calculations},}\ }\href {\doibase
  10.1016/0375-9474(89)90269-8} {\bibfield  {journal} {\bibinfo  {journal}
  {Nucl. Phys. A}\ }\textbf {\bibinfo {volume} {499}},\ \bibinfo {pages} {63 --
  92} (\bibinfo {year} {1989})}\BibitemShut {NoStop}%
\bibitem [{\citenamefont {Bender}\ \emph {et~al.}(2003)\citenamefont {Bender},
  \citenamefont {Heenen},\ and\ \citenamefont {Reinhard}}]{bender2003}%
  \BibitemOpen
  \bibfield  {author} {\bibinfo {author} {\bibfnamefont {Michael}\ \bibnamefont
  {Bender}}, \bibinfo {author} {\bibfnamefont {Paul-Henri}\ \bibnamefont
  {Heenen}}, \ and\ \bibinfo {author} {\bibfnamefont {Paul-Gerhard}\
  \bibnamefont {Reinhard}},\ }\bibfield  {title} {\enquote {\bibinfo {title}
  {Self-consistent mean-field models for nuclear structure},}\ }\href {\doibase
  10.1103/RevModPhys.75.121} {\bibfield  {journal} {\bibinfo  {journal} {Rev.
  Mod. Phys.}\ }\textbf {\bibinfo {volume} {75}},\ \bibinfo {pages} {121--180}
  (\bibinfo {year} {2003})}\BibitemShut {NoStop}%
\bibitem [{\citenamefont {Sheikh}\ \emph {et~al.}(2021)\citenamefont {Sheikh},
  \citenamefont {Dobaczewski}, \citenamefont {Ring}, \citenamefont {Robledo},\
  and\ \citenamefont {Yannouleas}}]{sheikh2021}%
  \BibitemOpen
  \bibfield  {author} {\bibinfo {author} {\bibfnamefont {Javid~A.}\
  \bibnamefont {Sheikh}}, \bibinfo {author} {\bibfnamefont {Jacek~Jan}\
  \bibnamefont {Dobaczewski}}, \bibinfo {author} {\bibfnamefont {Peter}\
  \bibnamefont {Ring}}, \bibinfo {author} {\bibfnamefont {Luis~Miguel}\
  \bibnamefont {Robledo}}, \ and\ \bibinfo {author} {\bibfnamefont
  {Constantine}\ \bibnamefont {Yannouleas}},\ }\bibfield  {title} {\enquote
  {\bibinfo {title} {Symmetry restoration in mean-field approaches},}\ }\href
  {\doibase 10.1088/1361-6471/ac288a} {\bibfield  {journal} {\bibinfo
  {journal} {J. Phys. G.}\ }\textbf {\bibinfo {volume} {48}},\ \bibinfo {pages}
  {123001} (\bibinfo {year} {2021})}\BibitemShut {NoStop}%
\bibitem [{\citenamefont {K{\"u}mmel}\ \emph {et~al.}(1978)\citenamefont
  {K{\"u}mmel}, \citenamefont {L{\"u}hrmann},\ and\ \citenamefont
  {Zabolitzky}}]{kuemmel1978}%
  \BibitemOpen
  \bibfield  {author} {\bibinfo {author} {\bibfnamefont {H.}~\bibnamefont
  {K{\"u}mmel}}, \bibinfo {author} {\bibfnamefont {K.~H.}\ \bibnamefont
  {L{\"u}hrmann}}, \ and\ \bibinfo {author} {\bibfnamefont {J.~G.}\
  \bibnamefont {Zabolitzky}},\ }\bibfield  {title} {\enquote {\bibinfo {title}
  {{Many-fermion theory in expS- (or coupled cluster) form}},}\ }\href
  {\doibase 10.1016/0370-1573(78)90081-9} {\bibfield  {journal} {\bibinfo
  {journal} {Phys. Rep.}\ }\textbf {\bibinfo {volume} {36}},\ \bibinfo {pages}
  {1 -- 63} (\bibinfo {year} {1978})}\BibitemShut {NoStop}%
\bibitem [{\citenamefont {Hagen}\ \emph {et~al.}(2014)\citenamefont {Hagen},
  \citenamefont {Papenbrock}, \citenamefont {Hjorth-Jensen},\ and\
  \citenamefont {Dean}}]{hagen2014}%
  \BibitemOpen
  \bibfield  {author} {\bibinfo {author} {\bibfnamefont {G.}~\bibnamefont
  {Hagen}}, \bibinfo {author} {\bibfnamefont {T.}~\bibnamefont {Papenbrock}},
  \bibinfo {author} {\bibfnamefont {M.}~\bibnamefont {Hjorth-Jensen}}, \ and\
  \bibinfo {author} {\bibfnamefont {D.~J.}\ \bibnamefont {Dean}},\ }\bibfield
  {title} {\enquote {\bibinfo {title} {Coupled-cluster computations of atomic
  nuclei},}\ }\href {\doibase 10.1088/0034-4885/77/9/096302} {\bibfield
  {journal} {\bibinfo  {journal} {Rep. Prog. Phys.}\ }\textbf {\bibinfo
  {volume} {77}},\ \bibinfo {pages} {096302} (\bibinfo {year}
  {2014})}\BibitemShut {NoStop}%
\bibitem [{\citenamefont {Tsukiyama}\ \emph {et~al.}(2011)\citenamefont
  {Tsukiyama}, \citenamefont {Bogner},\ and\ \citenamefont
  {Schwenk}}]{tsukiyama2011}%
  \BibitemOpen
  \bibfield  {author} {\bibinfo {author} {\bibfnamefont {K.}~\bibnamefont
  {Tsukiyama}}, \bibinfo {author} {\bibfnamefont {S.~K.}\ \bibnamefont
  {Bogner}}, \ and\ \bibinfo {author} {\bibfnamefont {A.}~\bibnamefont
  {Schwenk}},\ }\bibfield  {title} {\enquote {\bibinfo {title} {{In-Medium
  Similarity Renormalization Group For Nuclei}},}\ }\href {\doibase
  10.1103/PhysRevLett.106.222502} {\bibfield  {journal} {\bibinfo  {journal}
  {Phys. Rev. Lett.}\ }\textbf {\bibinfo {volume} {106}},\ \bibinfo {pages}
  {222502} (\bibinfo {year} {2011})}\BibitemShut {NoStop}%
\bibitem [{\citenamefont {Hergert}\ \emph {et~al.}(2016)\citenamefont
  {Hergert}, \citenamefont {Bogner}, \citenamefont {Morris}, \citenamefont
  {Schwenk},\ and\ \citenamefont {Tsukiyama}}]{hergert2016}%
  \BibitemOpen
  \bibfield  {author} {\bibinfo {author} {\bibfnamefont {H.}~\bibnamefont
  {Hergert}}, \bibinfo {author} {\bibfnamefont {S.~K.}\ \bibnamefont {Bogner}},
  \bibinfo {author} {\bibfnamefont {T.~D.}\ \bibnamefont {Morris}}, \bibinfo
  {author} {\bibfnamefont {A.}~\bibnamefont {Schwenk}}, \ and\ \bibinfo
  {author} {\bibfnamefont {K.}~\bibnamefont {Tsukiyama}},\ }\bibfield  {title}
  {\enquote {\bibinfo {title} {The in-medium similarity renormalization group:
  A novel ab initio method for nuclei},}\ }\href {\doibase
  10.1016/j.physrep.2015.12.007} {\bibfield  {journal} {\bibinfo  {journal}
  {Phys. Rep.}\ }\textbf {\bibinfo {volume} {621}},\ \bibinfo {pages} {165 --
  222} (\bibinfo {year} {2016})}\BibitemShut {NoStop}%
\bibitem [{\citenamefont {Stroberg}\ \emph {et~al.}(2017)\citenamefont
  {Stroberg}, \citenamefont {Calci}, \citenamefont {Hergert}, \citenamefont
  {Holt}, \citenamefont {Bogner}, \citenamefont {Roth},\ and\ \citenamefont
  {Schwenk}}]{stroberg2017}%
  \BibitemOpen
  \bibfield  {author} {\bibinfo {author} {\bibfnamefont {S.~R.}\ \bibnamefont
  {Stroberg}}, \bibinfo {author} {\bibfnamefont {A.}~\bibnamefont {Calci}},
  \bibinfo {author} {\bibfnamefont {H.}~\bibnamefont {Hergert}}, \bibinfo
  {author} {\bibfnamefont {J.~D.}\ \bibnamefont {Holt}}, \bibinfo {author}
  {\bibfnamefont {S.~K.}\ \bibnamefont {Bogner}}, \bibinfo {author}
  {\bibfnamefont {R.}~\bibnamefont {Roth}}, \ and\ \bibinfo {author}
  {\bibfnamefont {A.}~\bibnamefont {Schwenk}},\ }\bibfield  {title} {\enquote
  {\bibinfo {title} {Nucleus-dependent valence-space approach to nuclear
  structure},}\ }\href {\doibase 10.1103/PhysRevLett.118.032502} {\bibfield
  {journal} {\bibinfo  {journal} {Phys. Rev. Lett.}\ }\textbf {\bibinfo
  {volume} {118}},\ \bibinfo {pages} {032502} (\bibinfo {year}
  {2017})}\BibitemShut {NoStop}%
\bibitem [{\citenamefont {Stroberg}\ \emph {et~al.}(2019)\citenamefont
  {Stroberg}, \citenamefont {Hergert}, \citenamefont {Bogner},\ and\
  \citenamefont {Holt}}]{stroberg2019}%
  \BibitemOpen
  \bibfield  {author} {\bibinfo {author} {\bibfnamefont {S.~R.}\ \bibnamefont
  {Stroberg}}, \bibinfo {author} {\bibfnamefont {H.}~\bibnamefont {Hergert}},
  \bibinfo {author} {\bibfnamefont {S.~K.}\ \bibnamefont {Bogner}}, \ and\
  \bibinfo {author} {\bibfnamefont {J.~D.}\ \bibnamefont {Holt}},\ }\bibfield
  {title} {\enquote {\bibinfo {title} {{Nonempirical Interactions for the
  Nuclear Shell Model: An Update}},}\ }\href {\doibase
  10.1146/annurev-nucl-101917-021120} {\bibfield  {journal} {\bibinfo
  {journal} {Annu. Rev. Nucl. Part. Sci.}\ }\textbf {\bibinfo {volume} {69}},\
  \bibinfo {pages} {307} (\bibinfo {year} {2019})}\BibitemShut {NoStop}%
\bibitem [{\citenamefont {Stroberg}\ \emph
  {et~al.}(2021{\natexlab{a}})\citenamefont {Stroberg}, \citenamefont {Holt},
  \citenamefont {Schwenk},\ and\ \citenamefont {Simonis}}]{stroberg2021}%
  \BibitemOpen
  \bibfield  {author} {\bibinfo {author} {\bibfnamefont {S.~R.}\ \bibnamefont
  {Stroberg}}, \bibinfo {author} {\bibfnamefont {J.~D.}\ \bibnamefont {Holt}},
  \bibinfo {author} {\bibfnamefont {A.}~\bibnamefont {Schwenk}}, \ and\
  \bibinfo {author} {\bibfnamefont {J.}~\bibnamefont {Simonis}},\ }\bibfield
  {title} {\enquote {\bibinfo {title} {Ab initio limits of atomic nuclei},}\
  }\href {\doibase 10.1103/PhysRevLett.126.022501} {\bibfield  {journal}
  {\bibinfo  {journal} {Phys. Rev. Lett.}\ }\textbf {\bibinfo {volume} {126}},\
  \bibinfo {pages} {022501} (\bibinfo {year} {2021}{\natexlab{a}})}\BibitemShut
  {NoStop}%
\bibitem [{\citenamefont {Heinz}\ \emph {et~al.}(2021)\citenamefont {Heinz},
  \citenamefont {Tichai}, \citenamefont {Hoppe}, \citenamefont {Hebeler},\ and\
  \citenamefont {Schwenk}}]{heinz2021}%
  \BibitemOpen
  \bibfield  {author} {\bibinfo {author} {\bibfnamefont {M.}~\bibnamefont
  {Heinz}}, \bibinfo {author} {\bibfnamefont {A.}~\bibnamefont {Tichai}},
  \bibinfo {author} {\bibfnamefont {J.}~\bibnamefont {Hoppe}}, \bibinfo
  {author} {\bibfnamefont {K.}~\bibnamefont {Hebeler}}, \ and\ \bibinfo
  {author} {\bibfnamefont {A.}~\bibnamefont {Schwenk}},\ }\bibfield  {title}
  {\enquote {\bibinfo {title} {In-medium similarity renormalization group with
  three-body operators},}\ }\href {\doibase 10.1103/PhysRevC.103.044318}
  {\bibfield  {journal} {\bibinfo  {journal} {Phys. Rev. C}\ }\textbf {\bibinfo
  {volume} {103}},\ \bibinfo {pages} {044318} (\bibinfo {year}
  {2021})}\BibitemShut {NoStop}%
\bibitem [{\citenamefont {Dickhoff}\ and\ \citenamefont
  {Barbieri}(2004)}]{dickhoff2004}%
  \BibitemOpen
  \bibfield  {author} {\bibinfo {author} {\bibfnamefont {W.H.}\ \bibnamefont
  {Dickhoff}}\ and\ \bibinfo {author} {\bibfnamefont {C.}~\bibnamefont
  {Barbieri}},\ }\bibfield  {title} {\enquote {\bibinfo {title}
  {Self-consistent green's function method for nuclei and nuclear matter},}\
  }\href {\doibase 10.1016/j.ppnp.2004.02.038} {\bibfield  {journal} {\bibinfo
  {journal} {Prog. Part. Nucl. Phys.}\ }\textbf {\bibinfo {volume} {52}},\
  \bibinfo {pages} {377 -- 496} (\bibinfo {year} {2004})}\BibitemShut {NoStop}%
\bibitem [{\citenamefont {Som\`a}\ \emph {et~al.}(2013)\citenamefont {Som\`a},
  \citenamefont {Barbieri},\ and\ \citenamefont {Duguet}}]{soma2013}%
  \BibitemOpen
  \bibfield  {author} {\bibinfo {author} {\bibfnamefont {V.}~\bibnamefont
  {Som\`a}}, \bibinfo {author} {\bibfnamefont {C.}~\bibnamefont {Barbieri}}, \
  and\ \bibinfo {author} {\bibfnamefont {T.}~\bibnamefont {Duguet}},\
  }\bibfield  {title} {\enquote {\bibinfo {title} {\textit{Ab initio}
  gorkov-green's function calculations of open-shell nuclei},}\ }\href
  {\doibase 10.1103/PhysRevC.87.011303} {\bibfield  {journal} {\bibinfo
  {journal} {Phys. Rev. C}\ }\textbf {\bibinfo {volume} {87}},\ \bibinfo
  {pages} {011303} (\bibinfo {year} {2013})}\BibitemShut {NoStop}%
\bibitem [{\citenamefont {Holt}\ \emph {et~al.}(2014)\citenamefont {Holt},
  \citenamefont {Men{\'e}ndez}, \citenamefont {Simonis},\ and\ \citenamefont
  {Schwenk}}]{holt2014}%
  \BibitemOpen
  \bibfield  {author} {\bibinfo {author} {\bibfnamefont {J.~D.}\ \bibnamefont
  {Holt}}, \bibinfo {author} {\bibfnamefont {J.}~\bibnamefont {Men{\'e}ndez}},
  \bibinfo {author} {\bibfnamefont {J.}~\bibnamefont {Simonis}}, \ and\
  \bibinfo {author} {\bibfnamefont {A.}~\bibnamefont {Schwenk}},\ }\bibfield
  {title} {\enquote {\bibinfo {title} {{Three-nucleon forces and spectroscopy
  of neutron-rich calcium isotopes}},}\ }\href {\doibase
  10.1103/PhysRevC.90.024312} {\bibfield  {journal} {\bibinfo  {journal} {Phys.
  Rev. C}\ }\textbf {\bibinfo {volume} {90}},\ \bibinfo {pages} {024312}
  (\bibinfo {year} {2014})}\BibitemShut {NoStop}%
\bibitem [{\citenamefont {Tichai}\ \emph {et~al.}(2016)\citenamefont {Tichai},
  \citenamefont {Langhammer}, \citenamefont {Binder},\ and\ \citenamefont
  {Roth}}]{tichai2016}%
  \BibitemOpen
  \bibfield  {author} {\bibinfo {author} {\bibfnamefont {A.}~\bibnamefont
  {Tichai}}, \bibinfo {author} {\bibfnamefont {J.}~\bibnamefont {Langhammer}},
  \bibinfo {author} {\bibfnamefont {S.}~\bibnamefont {Binder}}, \ and\ \bibinfo
  {author} {\bibfnamefont {R.}~\bibnamefont {Roth}},\ }\bibfield  {title}
  {\enquote {\bibinfo {title} {{Hartree-Fock many-body perturbation theory for
  nuclear ground-states}},}\ }\href {\doibase
  https://doi.org/10.1016/j.physletb.2016.03.029} {\bibfield  {journal}
  {\bibinfo  {journal} {Phys. Lett. B}\ }\textbf {\bibinfo {volume} {756}},\
  \bibinfo {pages} {283--288} (\bibinfo {year} {2016})}\BibitemShut {NoStop}%
\bibitem [{\citenamefont {Hu}\ \emph {et~al.}(2016)\citenamefont {Hu},
  \citenamefont {Xu}, \citenamefont {Sun}, \citenamefont {Vary},\ and\
  \citenamefont {Li}}]{hu2016}%
  \BibitemOpen
  \bibfield  {author} {\bibinfo {author} {\bibfnamefont {B.~S.}\ \bibnamefont
  {Hu}}, \bibinfo {author} {\bibfnamefont {F.~R.}\ \bibnamefont {Xu}}, \bibinfo
  {author} {\bibfnamefont {Z.~H.}\ \bibnamefont {Sun}}, \bibinfo {author}
  {\bibfnamefont {J.~P.}\ \bibnamefont {Vary}}, \ and\ \bibinfo {author}
  {\bibfnamefont {T.}~\bibnamefont {Li}},\ }\bibfield  {title} {\enquote
  {\bibinfo {title} {Ab initio nuclear many-body perturbation calculations in
  the hartree-fock basis},}\ }\href {\doibase 10.1103/PhysRevC.94.014303}
  {\bibfield  {journal} {\bibinfo  {journal} {Phys. Rev. C}\ }\textbf {\bibinfo
  {volume} {94}},\ \bibinfo {pages} {014303} (\bibinfo {year}
  {2016})}\BibitemShut {NoStop}%
\bibitem [{\citenamefont {Tichai}\ \emph
  {et~al.}(2018{\natexlab{a}})\citenamefont {Tichai}, \citenamefont {Arthuis},
  \citenamefont {Duguet}, \citenamefont {Hergert}, \citenamefont {Som{\`a}},\
  and\ \citenamefont {Roth}}]{tichai2018}%
  \BibitemOpen
  \bibfield  {author} {\bibinfo {author} {\bibfnamefont {A.}~\bibnamefont
  {Tichai}}, \bibinfo {author} {\bibfnamefont {P.}~\bibnamefont {Arthuis}},
  \bibinfo {author} {\bibfnamefont {T.}~\bibnamefont {Duguet}}, \bibinfo
  {author} {\bibfnamefont {H.}~\bibnamefont {Hergert}}, \bibinfo {author}
  {\bibfnamefont {V.}~\bibnamefont {Som{\`a}}}, \ and\ \bibinfo {author}
  {\bibfnamefont {R.}~\bibnamefont {Roth}},\ }\bibfield  {title} {\enquote
  {\bibinfo {title} {Bogoliubov many-body perturbation theory for open-shell
  nuclei},}\ }\href {\doibase 10.1016/j.physletb.2018.09.044} {\bibfield
  {journal} {\bibinfo  {journal} {Phys. Lett. B}\ }\textbf {\bibinfo {volume}
  {786}},\ \bibinfo {pages} {195--200} (\bibinfo {year}
  {2018}{\natexlab{a}})}\BibitemShut {NoStop}%
\bibitem [{\citenamefont {Tichai}\ \emph
  {et~al.}(2018{\natexlab{b}})\citenamefont {Tichai}, \citenamefont
  {Gebrerufael}, \citenamefont {Vobig},\ and\ \citenamefont
  {Roth}}]{tichai2018ncsmpt}%
  \BibitemOpen
  \bibfield  {author} {\bibinfo {author} {\bibfnamefont {A.}~\bibnamefont
  {Tichai}}, \bibinfo {author} {\bibfnamefont {E.}~\bibnamefont {Gebrerufael}},
  \bibinfo {author} {\bibfnamefont {K.}~\bibnamefont {Vobig}}, \ and\ \bibinfo
  {author} {\bibfnamefont {R.}~\bibnamefont {Roth}},\ }\bibfield  {title}
  {\enquote {\bibinfo {title} {Open-shell nuclei from no-core shell model with
  perturbative improvement},}\ }\href {\doibase
  https://doi.org/10.1016/j.physletb.2018.10.029} {\bibfield  {journal}
  {\bibinfo  {journal} {Phys. Lett. B}\ }\textbf {\bibinfo {volume} {786}},\
  \bibinfo {pages} {448--452} (\bibinfo {year}
  {2018}{\natexlab{b}})}\BibitemShut {NoStop}%
\bibitem [{\citenamefont {Tichai}\ \emph {et~al.}(2020)\citenamefont {Tichai},
  \citenamefont {Roth},\ and\ \citenamefont {Duguet}}]{tichai2020}%
  \BibitemOpen
  \bibfield  {author} {\bibinfo {author} {\bibfnamefont {A.}~\bibnamefont
  {Tichai}}, \bibinfo {author} {\bibfnamefont {R.}~\bibnamefont {Roth}}, \ and\
  \bibinfo {author} {\bibfnamefont {T.}~\bibnamefont {Duguet}},\ }\bibfield
  {title} {\enquote {\bibinfo {title} {Many-body perturbation theories for
  finite nuclei},}\ }\href {\doibase 10.3389/fphy.2020.00164} {\bibfield
  {journal} {\bibinfo  {journal} {Front. Phys.}\ }\textbf {\bibinfo {volume}
  {8}},\ \bibinfo {pages} {164} (\bibinfo {year} {2020})}\BibitemShut {NoStop}%
\bibitem [{\citenamefont {Novario}\ \emph {et~al.}(2020)\citenamefont
  {Novario}, \citenamefont {Hagen}, \citenamefont {Jansen},\ and\ \citenamefont
  {Papenbrock}}]{novario2020}%
  \BibitemOpen
  \bibfield  {author} {\bibinfo {author} {\bibfnamefont {S.~J.}\ \bibnamefont
  {Novario}}, \bibinfo {author} {\bibfnamefont {G.}~\bibnamefont {Hagen}},
  \bibinfo {author} {\bibfnamefont {G.~R.}\ \bibnamefont {Jansen}}, \ and\
  \bibinfo {author} {\bibfnamefont {T.}~\bibnamefont {Papenbrock}},\ }\bibfield
   {title} {\enquote {\bibinfo {title} {Charge radii of exotic neon and
  magnesium isotopes},}\ }\href {\doibase 10.1103/PhysRevC.102.051303}
  {\bibfield  {journal} {\bibinfo  {journal} {Phys. Rev. C}\ }\textbf {\bibinfo
  {volume} {102}},\ \bibinfo {pages} {051303} (\bibinfo {year}
  {2020})}\BibitemShut {NoStop}%
\bibitem [{\citenamefont {{Som{\`a}}}\ \emph {et~al.}(2021)\citenamefont
  {{Som{\`a}}}, \citenamefont {{Barbieri}}, \citenamefont {{Duguet}},\ and\
  \citenamefont {{Navr{\'a}til}}}]{soma2021}%
  \BibitemOpen
  \bibfield  {author} {\bibinfo {author} {\bibfnamefont {V.}~\bibnamefont
  {{Som{\`a}}}}, \bibinfo {author} {\bibfnamefont {C.}~\bibnamefont
  {{Barbieri}}}, \bibinfo {author} {\bibfnamefont {T.}~\bibnamefont
  {{Duguet}}}, \ and\ \bibinfo {author} {\bibfnamefont {P.}~\bibnamefont
  {{Navr{\'a}til}}},\ }\bibfield  {title} {\enquote {\bibinfo {title} {{Moving
  away from singly-magic nuclei with Gorkov Green's function theory}},}\ }\href
  {\doibase 10.1140/epja/s10050-021-00437-4} {\bibfield  {journal} {\bibinfo
  {journal} {Eur. Phys. J. A}\ }\textbf {\bibinfo {volume} {57}},\ \bibinfo
  {pages} {135} (\bibinfo {year} {2021})}\BibitemShut {NoStop}%
\bibitem [{\citenamefont {Caurier}\ \emph {et~al.}(2005)\citenamefont
  {Caurier}, \citenamefont {Mart{\'i}nez-Pinedo}, \citenamefont {Nowacki},
  \citenamefont {Poves},\ and\ \citenamefont {Zuker}}]{caurier2005}%
  \BibitemOpen
  \bibfield  {author} {\bibinfo {author} {\bibfnamefont {E.}~\bibnamefont
  {Caurier}}, \bibinfo {author} {\bibfnamefont {G.}~\bibnamefont
  {Mart{\'i}nez-Pinedo}}, \bibinfo {author} {\bibfnamefont {F.}~\bibnamefont
  {Nowacki}}, \bibinfo {author} {\bibfnamefont {A.}~\bibnamefont {Poves}}, \
  and\ \bibinfo {author} {\bibfnamefont {A.~P.}\ \bibnamefont {Zuker}},\
  }\bibfield  {title} {\enquote {\bibinfo {title} {The shell model as a unified
  view of nuclear structure},}\ }\href {\doibase 10.1103/RevModPhys.77.427}
  {\bibfield  {journal} {\bibinfo  {journal} {Rev. Mod. Phys.}\ }\textbf
  {\bibinfo {volume} {77}},\ \bibinfo {pages} {427--488} (\bibinfo {year}
  {2005})}\BibitemShut {NoStop}%
\bibitem [{\citenamefont {Caurier}\ \emph {et~al.}(2007)\citenamefont
  {Caurier}, \citenamefont {Men\'endez}, \citenamefont {Nowacki},\ and\
  \citenamefont {Poves}}]{caurier2007}%
  \BibitemOpen
  \bibfield  {author} {\bibinfo {author} {\bibfnamefont {E.}~\bibnamefont
  {Caurier}}, \bibinfo {author} {\bibfnamefont {J.}~\bibnamefont {Men\'endez}},
  \bibinfo {author} {\bibfnamefont {F.}~\bibnamefont {Nowacki}}, \ and\
  \bibinfo {author} {\bibfnamefont {A.}~\bibnamefont {Poves}},\ }\bibfield
  {title} {\enquote {\bibinfo {title} {Coexistence of spherical states with
  deformed and superdeformed bands in doubly magic $^{40}\mathrm{Ca}$: A
  shell-model challenge},}\ }\href {\doibase 10.1103/PhysRevC.75.054317}
  {\bibfield  {journal} {\bibinfo  {journal} {Phys. Rev. C}\ }\textbf {\bibinfo
  {volume} {75}},\ \bibinfo {pages} {054317} (\bibinfo {year}
  {2007})}\BibitemShut {NoStop}%
\bibitem [{\citenamefont {Caprio}\ \emph {et~al.}(2013)\citenamefont {Caprio},
  \citenamefont {Maris},\ and\ \citenamefont {Vary}}]{caprio2013}%
  \BibitemOpen
  \bibfield  {author} {\bibinfo {author} {\bibfnamefont {M.~A.}\ \bibnamefont
  {Caprio}}, \bibinfo {author} {\bibfnamefont {P.}~\bibnamefont {Maris}}, \
  and\ \bibinfo {author} {\bibfnamefont {J.~P.}\ \bibnamefont {Vary}},\
  }\bibfield  {title} {\enquote {\bibinfo {title} {Emergence of rotational
  bands in ab initio no-core configuration interaction calculations of light
  nuclei},}\ }\href {\doibase 10.1016/j.physletb.2012.12.064} {\bibfield
  {journal} {\bibinfo  {journal} {Phys. Lett. B}\ }\textbf {\bibinfo {volume}
  {719}},\ \bibinfo {pages} {179 -- 184} (\bibinfo {year} {2013})}\BibitemShut
  {NoStop}%
\bibitem [{\citenamefont {Jansen}\ \emph {et~al.}(2014)\citenamefont {Jansen},
  \citenamefont {Engel}, \citenamefont {Hagen}, \citenamefont {Navratil},\ and\
  \citenamefont {Signoracci}}]{jansen2014}%
  \BibitemOpen
  \bibfield  {author} {\bibinfo {author} {\bibfnamefont {G.~R.}\ \bibnamefont
  {Jansen}}, \bibinfo {author} {\bibfnamefont {J.}~\bibnamefont {Engel}},
  \bibinfo {author} {\bibfnamefont {G.}~\bibnamefont {Hagen}}, \bibinfo
  {author} {\bibfnamefont {P.}~\bibnamefont {Navratil}}, \ and\ \bibinfo
  {author} {\bibfnamefont {A.}~\bibnamefont {Signoracci}},\ }\bibfield  {title}
  {\enquote {\bibinfo {title} {\textit{Ab Initio}~coupled-cluster effective
  interactions for the shell model: Application to neutron-rich oxygen and
  carbon isotopes},}\ }\href {\doibase 10.1103/PhysRevLett.113.142502}
  {\bibfield  {journal} {\bibinfo  {journal} {Phys. Rev. Lett.}\ }\textbf
  {\bibinfo {volume} {113}},\ \bibinfo {pages} {142502} (\bibinfo {year}
  {2014})}\BibitemShut {NoStop}%
\bibitem [{\citenamefont {Bogner}\ \emph {et~al.}(2014)\citenamefont {Bogner},
  \citenamefont {Hergert}, \citenamefont {Holt}, \citenamefont {Schwenk},
  \citenamefont {Binder}, \citenamefont {Calci}, \citenamefont {Langhammer},\
  and\ \citenamefont {Roth}}]{bogner2014}%
  \BibitemOpen
  \bibfield  {author} {\bibinfo {author} {\bibfnamefont {S.~K.}\ \bibnamefont
  {Bogner}}, \bibinfo {author} {\bibfnamefont {H.}~\bibnamefont {Hergert}},
  \bibinfo {author} {\bibfnamefont {J.~D.}\ \bibnamefont {Holt}}, \bibinfo
  {author} {\bibfnamefont {A.}~\bibnamefont {Schwenk}}, \bibinfo {author}
  {\bibfnamefont {S.}~\bibnamefont {Binder}}, \bibinfo {author} {\bibfnamefont
  {A.}~\bibnamefont {Calci}}, \bibinfo {author} {\bibfnamefont
  {J.}~\bibnamefont {Langhammer}}, \ and\ \bibinfo {author} {\bibfnamefont
  {R.}~\bibnamefont {Roth}},\ }\bibfield  {title} {\enquote {\bibinfo {title}
  {Nonperturbative shell-model interactions from the in-medium similarity
  renormalization group},}\ }\href {\doibase 10.1103/PhysRevLett.113.142501}
  {\bibfield  {journal} {\bibinfo  {journal} {Phys. Rev. Lett.}\ }\textbf
  {\bibinfo {volume} {113}},\ \bibinfo {pages} {142501} (\bibinfo {year}
  {2014})}\BibitemShut {NoStop}%
\bibitem [{\citenamefont {Caprio}\ \emph {et~al.}(2015)\citenamefont {Caprio},
  \citenamefont {Maris}, \citenamefont {Vary},\ and\ \citenamefont
  {Smith}}]{caprio2015}%
  \BibitemOpen
  \bibfield  {author} {\bibinfo {author} {\bibfnamefont {M.~A.}\ \bibnamefont
  {Caprio}}, \bibinfo {author} {\bibfnamefont {P.}~\bibnamefont {Maris}},
  \bibinfo {author} {\bibfnamefont {J.~P.}\ \bibnamefont {Vary}}, \ and\
  \bibinfo {author} {\bibfnamefont {R.}~\bibnamefont {Smith}},\ }\bibfield
  {title} {\enquote {\bibinfo {title} {Collective rotation from ab initio
  theory},}\ }\href {\doibase 10.1142/S0218301315410025} {\bibfield  {journal}
  {\bibinfo  {journal} {Int. J. Mod. Phys. E}\ }\textbf {\bibinfo {volume}
  {24}},\ \bibinfo {pages} {1541002} (\bibinfo {year} {2015})}\BibitemShut
  {NoStop}%
\bibitem [{\citenamefont {Maris}\ \emph {et~al.}(2015)\citenamefont {Maris},
  \citenamefont {Caprio},\ and\ \citenamefont {Vary}}]{maris2015}%
  \BibitemOpen
  \bibfield  {author} {\bibinfo {author} {\bibfnamefont {P.}~\bibnamefont
  {Maris}}, \bibinfo {author} {\bibfnamefont {M.~A.}\ \bibnamefont {Caprio}}, \
  and\ \bibinfo {author} {\bibfnamefont {J.~P.}\ \bibnamefont {Vary}},\
  }\bibfield  {title} {\enquote {\bibinfo {title} {Emergence of rotational
  bands in \textit{ab initio} no-core configuration interaction calculations of
  the be isotopes},}\ }\href {\doibase 10.1103/PhysRevC.91.014310} {\bibfield
  {journal} {\bibinfo  {journal} {Phys. Rev. C}\ }\textbf {\bibinfo {volume}
  {91}},\ \bibinfo {pages} {014310} (\bibinfo {year} {2015})}\BibitemShut
  {NoStop}%
\bibitem [{\citenamefont {Shimizu}\ \emph {et~al.}(2012)\citenamefont
  {Shimizu}, \citenamefont {Abe}, \citenamefont {Tsunoda}, \citenamefont
  {Utsuno}, \citenamefont {Yoshida}, \citenamefont {Mizusaki}, \citenamefont
  {Honma},\ and\ \citenamefont {Otsuka}}]{shimizu2012}%
  \BibitemOpen
  \bibfield  {author} {\bibinfo {author} {\bibfnamefont {Noritaka}\
  \bibnamefont {Shimizu}}, \bibinfo {author} {\bibfnamefont {Takashi}\
  \bibnamefont {Abe}}, \bibinfo {author} {\bibfnamefont {Yusuke}\ \bibnamefont
  {Tsunoda}}, \bibinfo {author} {\bibfnamefont {Yutaka}\ \bibnamefont
  {Utsuno}}, \bibinfo {author} {\bibfnamefont {Tooru}\ \bibnamefont {Yoshida}},
  \bibinfo {author} {\bibfnamefont {Takahiro}\ \bibnamefont {Mizusaki}},
  \bibinfo {author} {\bibfnamefont {Michio}\ \bibnamefont {Honma}}, \ and\
  \bibinfo {author} {\bibfnamefont {Takaharu}\ \bibnamefont {Otsuka}},\
  }\bibfield  {title} {\enquote {\bibinfo {title} {New-generation monte carlo
  shell model for the k computer era},}\ }\href {\doibase 10.1093/ptep/pts012}
  {\bibfield  {journal} {\bibinfo  {journal} {Progress of Theoretical and
  Experimental Physics}\ }\textbf {\bibinfo {volume} {2012}},\ \bibinfo {pages}
  {01A205} (\bibinfo {year} {2012})}\BibitemShut {NoStop}%
\bibitem [{\citenamefont {Dytrych}\ \emph {et~al.}(2013)\citenamefont
  {Dytrych}, \citenamefont {Launey}, \citenamefont {Draayer}, \citenamefont
  {Maris}, \citenamefont {Vary}, \citenamefont {Saule}, \citenamefont
  {Catalyurek}, \citenamefont {Sosonkina}, \citenamefont {Langr},\ and\
  \citenamefont {Caprio}}]{dytrych2013}%
  \BibitemOpen
  \bibfield  {author} {\bibinfo {author} {\bibfnamefont {T.}~\bibnamefont
  {Dytrych}}, \bibinfo {author} {\bibfnamefont {K.~D.}\ \bibnamefont {Launey}},
  \bibinfo {author} {\bibfnamefont {J.~P.}\ \bibnamefont {Draayer}}, \bibinfo
  {author} {\bibfnamefont {P.}~\bibnamefont {Maris}}, \bibinfo {author}
  {\bibfnamefont {J.~P.}\ \bibnamefont {Vary}}, \bibinfo {author}
  {\bibfnamefont {E.}~\bibnamefont {Saule}}, \bibinfo {author} {\bibfnamefont
  {U.}~\bibnamefont {Catalyurek}}, \bibinfo {author} {\bibfnamefont
  {M.}~\bibnamefont {Sosonkina}}, \bibinfo {author} {\bibfnamefont
  {D.}~\bibnamefont {Langr}}, \ and\ \bibinfo {author} {\bibfnamefont {M.~A.}\
  \bibnamefont {Caprio}},\ }\bibfield  {title} {\enquote {\bibinfo {title}
  {Collective modes in light nuclei from first principles},}\ }\href {\doibase
  10.1103/PhysRevLett.111.252501} {\bibfield  {journal} {\bibinfo  {journal}
  {Phys. Rev. Lett.}\ }\textbf {\bibinfo {volume} {111}},\ \bibinfo {pages}
  {252501} (\bibinfo {year} {2013})}\BibitemShut {NoStop}%
\bibitem [{\citenamefont {Dytrych}\ \emph {et~al.}(2020)\citenamefont
  {Dytrych}, \citenamefont {Launey}, \citenamefont {Draayer}, \citenamefont
  {Rowe}, \citenamefont {Wood}, \citenamefont {Rosensteel}, \citenamefont
  {Bahri}, \citenamefont {Langr},\ and\ \citenamefont {Baker}}]{dytrych2020}%
  \BibitemOpen
  \bibfield  {author} {\bibinfo {author} {\bibfnamefont {T.}~\bibnamefont
  {Dytrych}}, \bibinfo {author} {\bibfnamefont {K.~D.}\ \bibnamefont {Launey}},
  \bibinfo {author} {\bibfnamefont {J.~P.}\ \bibnamefont {Draayer}}, \bibinfo
  {author} {\bibfnamefont {D.~J.}\ \bibnamefont {Rowe}}, \bibinfo {author}
  {\bibfnamefont {J.~L.}\ \bibnamefont {Wood}}, \bibinfo {author}
  {\bibfnamefont {G.}~\bibnamefont {Rosensteel}}, \bibinfo {author}
  {\bibfnamefont {C.}~\bibnamefont {Bahri}}, \bibinfo {author} {\bibfnamefont
  {D.}~\bibnamefont {Langr}}, \ and\ \bibinfo {author} {\bibfnamefont {R.~B.}\
  \bibnamefont {Baker}},\ }\bibfield  {title} {\enquote {\bibinfo {title}
  {Physics of nuclei: Key role of an emergent symmetry},}\ }\href {\doibase
  10.1103/PhysRevLett.124.042501} {\bibfield  {journal} {\bibinfo  {journal}
  {Phys. Rev. Lett.}\ }\textbf {\bibinfo {volume} {124}},\ \bibinfo {pages}
  {042501} (\bibinfo {year} {2020})}\BibitemShut {NoStop}%
\bibitem [{\citenamefont {Papenbrock}(2011)}]{papenbrock2011}%
  \BibitemOpen
  \bibfield  {author} {\bibinfo {author} {\bibfnamefont {T.}~\bibnamefont
  {Papenbrock}},\ }\bibfield  {title} {\enquote {\bibinfo {title} {Effective
  theory for deformed nuclei},}\ }\href {\doibase
  10.1016/j.nuclphysa.2010.12.013} {\bibfield  {journal} {\bibinfo  {journal}
  {Nucl. Phys. A}\ }\textbf {\bibinfo {volume} {852}},\ \bibinfo {pages} {36 --
  60} (\bibinfo {year} {2011})}\BibitemShut {NoStop}%
\bibitem [{\citenamefont {Coello~P\'erez}\ and\ \citenamefont
  {Papenbrock}(2015{\natexlab{a}})}]{coelloperez2015}%
  \BibitemOpen
  \bibfield  {author} {\bibinfo {author} {\bibfnamefont {E.~A.}\ \bibnamefont
  {Coello~P\'erez}}\ and\ \bibinfo {author} {\bibfnamefont {T.}~\bibnamefont
  {Papenbrock}},\ }\bibfield  {title} {\enquote {\bibinfo {title} {Effective
  theory for the nonrigid rotor in an electromagnetic field: Toward accurate
  and precise calculations of $e2$ transitions in deformed nuclei},}\ }\href
  {\doibase 10.1103/PhysRevC.92.014323} {\bibfield  {journal} {\bibinfo
  {journal} {Phys. Rev. C}\ }\textbf {\bibinfo {volume} {92}},\ \bibinfo
  {pages} {014323} (\bibinfo {year} {2015}{\natexlab{a}})}\BibitemShut
  {NoStop}%
\bibitem [{\citenamefont {{Chen}}\ \emph {et~al.}(2017)\citenamefont {{Chen}},
  \citenamefont {{Kaiser}}, \citenamefont {{Mei{\ss}ner}},\ and\ \citenamefont
  {{Meng}}}]{chen2017}%
  \BibitemOpen
  \bibfield  {author} {\bibinfo {author} {\bibfnamefont {Q.~B.}\ \bibnamefont
  {{Chen}}}, \bibinfo {author} {\bibfnamefont {N.}~\bibnamefont {{Kaiser}}},
  \bibinfo {author} {\bibfnamefont {Ulf-G.}\ \bibnamefont {{Mei{\ss}ner}}}, \
  and\ \bibinfo {author} {\bibfnamefont {J.}~\bibnamefont {{Meng}}},\
  }\bibfield  {title} {\enquote {\bibinfo {title} {{Effective field theory for
  triaxially deformed nuclei}},}\ }\href {\doibase 10.1140/epja/i2017-12404-5}
  {\bibfield  {journal} {\bibinfo  {journal} {Eur. Phys. J. A}\ }\textbf
  {\bibinfo {volume} {53}},\ \bibinfo {eid} {204} (\bibinfo {year}
  {2017})}\BibitemShut {NoStop}%
\bibitem [{\citenamefont {Papenbrock}\ and\ \citenamefont
  {Weidenm\"uller}(2020)}]{papenbrock2020}%
  \BibitemOpen
  \bibfield  {author} {\bibinfo {author} {\bibfnamefont {T.}~\bibnamefont
  {Papenbrock}}\ and\ \bibinfo {author} {\bibfnamefont {H.~A.}\ \bibnamefont
  {Weidenm\"uller}},\ }\bibfield  {title} {\enquote {\bibinfo {title}
  {Effective field theory for deformed odd-mass nuclei},}\ }\href {\doibase
  10.1103/PhysRevC.102.044324} {\bibfield  {journal} {\bibinfo  {journal}
  {Phys. Rev. C}\ }\textbf {\bibinfo {volume} {102}},\ \bibinfo {pages}
  {044324} (\bibinfo {year} {2020})}\BibitemShut {NoStop}%
\bibitem [{\citenamefont {Alnamlah}\ \emph {et~al.}(2021)\citenamefont
  {Alnamlah}, \citenamefont {Coello~P\'erez},\ and\ \citenamefont
  {Phillips}}]{alnamlah2021}%
  \BibitemOpen
  \bibfield  {author} {\bibinfo {author} {\bibfnamefont {I.~K.}\ \bibnamefont
  {Alnamlah}}, \bibinfo {author} {\bibfnamefont {E.~A.}\ \bibnamefont
  {Coello~P\'erez}}, \ and\ \bibinfo {author} {\bibfnamefont {D.~R.}\
  \bibnamefont {Phillips}},\ }\bibfield  {title} {\enquote {\bibinfo {title}
  {Effective field theory approach to rotational bands in odd-mass nuclei},}\
  }\href {\doibase 10.1103/PhysRevC.104.064311} {\bibfield  {journal} {\bibinfo
   {journal} {Phys. Rev. C}\ }\textbf {\bibinfo {volume} {104}},\ \bibinfo
  {pages} {064311} (\bibinfo {year} {2021})}\BibitemShut {NoStop}%
\bibitem [{\citenamefont {Paar}\ \emph {et~al.}(2007)\citenamefont {Paar},
  \citenamefont {Vretenar}, \citenamefont {Khan},\ and\ \citenamefont
  {Col{\`o}}}]{paar2007}%
  \BibitemOpen
  \bibfield  {author} {\bibinfo {author} {\bibfnamefont {N.}~\bibnamefont
  {Paar}}, \bibinfo {author} {\bibfnamefont {D.}~\bibnamefont {Vretenar}},
  \bibinfo {author} {\bibfnamefont {E.}~\bibnamefont {Khan}}, \ and\ \bibinfo
  {author} {\bibfnamefont {G.}~\bibnamefont {Col{\`o}}},\ }\bibfield  {title}
  {\enquote {\bibinfo {title} {Exotic modes of excitation in atomic nuclei far
  from stability},}\ }\href {\doibase 10.1088/0034-4885/70/5/R02} {\bibfield
  {journal} {\bibinfo  {journal} {Rep. Prog. Phys.}\ }\textbf {\bibinfo
  {volume} {70}},\ \bibinfo {pages} {691} (\bibinfo {year} {2007})}\BibitemShut
  {NoStop}%
\bibitem [{\citenamefont {Erler}\ \emph {et~al.}(2012)\citenamefont {Erler},
  \citenamefont {Birge}, \citenamefont {Kortelainen}, \citenamefont
  {Nazarewicz}, \citenamefont {Olsen}, \citenamefont {Perhac},\ and\
  \citenamefont {Stoitsov}}]{erler2012}%
  \BibitemOpen
  \bibfield  {author} {\bibinfo {author} {\bibfnamefont {Jochen}\ \bibnamefont
  {Erler}}, \bibinfo {author} {\bibfnamefont {Noah}\ \bibnamefont {Birge}},
  \bibinfo {author} {\bibfnamefont {Markus}\ \bibnamefont {Kortelainen}},
  \bibinfo {author} {\bibfnamefont {Witold}\ \bibnamefont {Nazarewicz}},
  \bibinfo {author} {\bibfnamefont {Erik}\ \bibnamefont {Olsen}}, \bibinfo
  {author} {\bibfnamefont {Alexander~M.}\ \bibnamefont {Perhac}}, \ and\
  \bibinfo {author} {\bibfnamefont {Mario}\ \bibnamefont {Stoitsov}},\
  }\bibfield  {title} {\enquote {\bibinfo {title} {The limits of the nuclear
  landscape},}\ }\href {\doibase 10.1038/nature11188} {\bibfield  {journal}
  {\bibinfo  {journal} {Nature}\ }\textbf {\bibinfo {volume} {486}},\ \bibinfo
  {pages} {509 -- 512} (\bibinfo {year} {2012})}\BibitemShut {NoStop}%
\bibitem [{\citenamefont {Goriely}\ \emph {et~al.}(2009)\citenamefont
  {Goriely}, \citenamefont {Chamel},\ and\ \citenamefont
  {Pearson}}]{goriely2009}%
  \BibitemOpen
  \bibfield  {author} {\bibinfo {author} {\bibfnamefont {S.}~\bibnamefont
  {Goriely}}, \bibinfo {author} {\bibfnamefont {N.}~\bibnamefont {Chamel}}, \
  and\ \bibinfo {author} {\bibfnamefont {J.~M.}\ \bibnamefont {Pearson}},\
  }\bibfield  {title} {\enquote {\bibinfo {title}
  {Skyrme-hartree-fock-bogoliubov nuclear mass formulas: Crossing the 0.6 mev
  accuracy threshold with microscopically deduced pairing},}\ }\href {\doibase
  10.1103/PhysRevLett.102.152503} {\bibfield  {journal} {\bibinfo  {journal}
  {Phys. Rev. Lett.}\ }\textbf {\bibinfo {volume} {102}},\ \bibinfo {pages}
  {152503} (\bibinfo {year} {2009})}\BibitemShut {NoStop}%
\bibitem [{\citenamefont {Nik{\v s}i{\'c}}\ \emph {et~al.}(2011)\citenamefont
  {Nik{\v s}i{\'c}}, \citenamefont {Vretenar},\ and\ \citenamefont
  {Ring}}]{niksic2011}%
  \BibitemOpen
  \bibfield  {author} {\bibinfo {author} {\bibfnamefont {T.}~\bibnamefont
  {Nik{\v s}i{\'c}}}, \bibinfo {author} {\bibfnamefont {D.}~\bibnamefont
  {Vretenar}}, \ and\ \bibinfo {author} {\bibfnamefont {P.}~\bibnamefont
  {Ring}},\ }\bibfield  {title} {\enquote {\bibinfo {title} {Relativistic
  nuclear energy density functionals: Mean-field and beyond},}\ }\href
  {\doibase 10.1016/j.ppnp.2011.01.055} {\bibfield  {journal} {\bibinfo
  {journal} {Prog. Part. Nucl. Phys.}\ }\textbf {\bibinfo {volume} {66}},\
  \bibinfo {pages} {519 -- 548} (\bibinfo {year} {2011})}\BibitemShut {NoStop}%
\bibitem [{\citenamefont {Shen}\ \emph {et~al.}(2019)\citenamefont {Shen},
  \citenamefont {Liang}, \citenamefont {Long}, \citenamefont {Meng},\ and\
  \citenamefont {Ring}}]{shen2019}%
  \BibitemOpen
  \bibfield  {author} {\bibinfo {author} {\bibfnamefont {Shihang}\ \bibnamefont
  {Shen}}, \bibinfo {author} {\bibfnamefont {Haozhao}\ \bibnamefont {Liang}},
  \bibinfo {author} {\bibfnamefont {Wen~Hui}\ \bibnamefont {Long}}, \bibinfo
  {author} {\bibfnamefont {Jie}\ \bibnamefont {Meng}}, \ and\ \bibinfo {author}
  {\bibfnamefont {Peter}\ \bibnamefont {Ring}},\ }\bibfield  {title} {\enquote
  {\bibinfo {title} {Towards an ab initio covariant density functional theory
  for nuclear structure},}\ }\href {\doibase 10.1016/j.ppnp.2019.103713}
  {\bibfield  {journal} {\bibinfo  {journal} {Prog. Part. Nucl. Phys.}\
  }\textbf {\bibinfo {volume} {109}},\ \bibinfo {pages} {103713} (\bibinfo
  {year} {2019})}\BibitemShut {NoStop}%
\bibitem [{\citenamefont {Kolck}(1999)}]{vankolck1999}%
  \BibitemOpen
  \bibfield  {author} {\bibinfo {author} {\bibfnamefont {U.~Van}\ \bibnamefont
  {Kolck}},\ }\bibfield  {title} {\enquote {\bibinfo {title} {Effective field
  theory of nuclear forces},}\ }\href {\doibase 10.1016/S0146-6410(99)00097-6}
  {\bibfield  {journal} {\bibinfo  {journal} {Prog. Part. Nucl. Phys.}\
  }\textbf {\bibinfo {volume} {43}},\ \bibinfo {pages} {337 -- 418} (\bibinfo
  {year} {1999})}\BibitemShut {NoStop}%
\bibitem [{\citenamefont {Epelbaum}\ \emph {et~al.}(2008)\citenamefont
  {Epelbaum}, \citenamefont {Krebs},\ and\ \citenamefont
  {Mei{\ss}ner}}]{epelbaum2008}%
  \BibitemOpen
  \bibfield  {author} {\bibinfo {author} {\bibfnamefont {E.}~\bibnamefont
  {Epelbaum}}, \bibinfo {author} {\bibfnamefont {H.}~\bibnamefont {Krebs}}, \
  and\ \bibinfo {author} {\bibfnamefont {U.-G.}\ \bibnamefont {Mei{\ss}ner}},\
  }\bibfield  {title} {\enquote {\bibinfo {title} {{$\Delta$-excitations and
  the three-nucleon force}},}\ }\href {\doibase
  10.1016/j.nuclphysa.2008.02.305} {\bibfield  {journal} {\bibinfo  {journal}
  {Nucl. Phys. A}\ }\textbf {\bibinfo {volume} {806}},\ \bibinfo {pages} {65 --
  78} (\bibinfo {year} {2008})}\BibitemShut {NoStop}%
\bibitem [{\citenamefont {Machleidt}\ and\ \citenamefont
  {Entem}(2011)}]{machleidt2011}%
  \BibitemOpen
  \bibfield  {author} {\bibinfo {author} {\bibfnamefont {R.}~\bibnamefont
  {Machleidt}}\ and\ \bibinfo {author} {\bibfnamefont {D.R.}\ \bibnamefont
  {Entem}},\ }\bibfield  {title} {\enquote {\bibinfo {title} {Chiral effective
  field theory and nuclear forces},}\ }\href {\doibase
  10.1016/j.physrep.2011.02.001} {\bibfield  {journal} {\bibinfo  {journal}
  {Phys. Rep.}\ }\textbf {\bibinfo {volume} {503}},\ \bibinfo {pages} {1 -- 75}
  (\bibinfo {year} {2011})}\BibitemShut {NoStop}%
\bibitem [{\citenamefont {Binder}\ \emph {et~al.}(2014)\citenamefont {Binder},
  \citenamefont {Langhammer}, \citenamefont {Calci},\ and\ \citenamefont
  {Roth}}]{binder2013b}%
  \BibitemOpen
  \bibfield  {author} {\bibinfo {author} {\bibfnamefont {Sven}\ \bibnamefont
  {Binder}}, \bibinfo {author} {\bibfnamefont {Joachim}\ \bibnamefont
  {Langhammer}}, \bibinfo {author} {\bibfnamefont {Angelo}\ \bibnamefont
  {Calci}}, \ and\ \bibinfo {author} {\bibfnamefont {Robert}\ \bibnamefont
  {Roth}},\ }\bibfield  {title} {\enquote {\bibinfo {title} {Ab initio path to
  heavy nuclei},}\ }\href {\doibase 10.1016/j.physletb.2014.07.010} {\bibfield
  {journal} {\bibinfo  {journal} {Phys. Lett. B}\ }\textbf {\bibinfo {volume}
  {736}},\ \bibinfo {pages} {119 -- 123} (\bibinfo {year} {2014})}\BibitemShut
  {NoStop}%
\bibitem [{\citenamefont {Hagen}\ \emph {et~al.}(2016)\citenamefont {Hagen},
  \citenamefont {Jansen},\ and\ \citenamefont {Papenbrock}}]{hagen2017}%
  \BibitemOpen
  \bibfield  {author} {\bibinfo {author} {\bibfnamefont {G.}~\bibnamefont
  {Hagen}}, \bibinfo {author} {\bibfnamefont {G.~R.}\ \bibnamefont {Jansen}}, \
  and\ \bibinfo {author} {\bibfnamefont {T.}~\bibnamefont {Papenbrock}},\
  }\bibfield  {title} {\enquote {\bibinfo {title} {Structure of
  $^{78}\mathrm{Ni}$ from first-principles computations},}\ }\href {\doibase
  10.1103/PhysRevLett.117.172501} {\bibfield  {journal} {\bibinfo  {journal}
  {Phys. Rev. Lett.}\ }\textbf {\bibinfo {volume} {117}},\ \bibinfo {pages}
  {172501} (\bibinfo {year} {2016})}\BibitemShut {NoStop}%
\bibitem [{\citenamefont {Morris}\ \emph {et~al.}(2018)\citenamefont {Morris},
  \citenamefont {Simonis}, \citenamefont {Stroberg}, \citenamefont {Stumpf},
  \citenamefont {Hagen}, \citenamefont {Holt}, \citenamefont {Jansen},
  \citenamefont {Papenbrock}, \citenamefont {Roth},\ and\ \citenamefont
  {Schwenk}}]{morris2018}%
  \BibitemOpen
  \bibfield  {author} {\bibinfo {author} {\bibfnamefont {T.~D.}\ \bibnamefont
  {Morris}}, \bibinfo {author} {\bibfnamefont {J.}~\bibnamefont {Simonis}},
  \bibinfo {author} {\bibfnamefont {S.~R.}\ \bibnamefont {Stroberg}}, \bibinfo
  {author} {\bibfnamefont {C.}~\bibnamefont {Stumpf}}, \bibinfo {author}
  {\bibfnamefont {G.}~\bibnamefont {Hagen}}, \bibinfo {author} {\bibfnamefont
  {J.~D.}\ \bibnamefont {Holt}}, \bibinfo {author} {\bibfnamefont {G.~R.}\
  \bibnamefont {Jansen}}, \bibinfo {author} {\bibfnamefont {T.}~\bibnamefont
  {Papenbrock}}, \bibinfo {author} {\bibfnamefont {R.}~\bibnamefont {Roth}}, \
  and\ \bibinfo {author} {\bibfnamefont {A.}~\bibnamefont {Schwenk}},\
  }\bibfield  {title} {\enquote {\bibinfo {title} {Structure of the lightest
  tin isotopes},}\ }\href {\doibase 10.1103/PhysRevLett.120.152503} {\bibfield
  {journal} {\bibinfo  {journal} {Phys. Rev. Lett.}\ }\textbf {\bibinfo
  {volume} {120}},\ \bibinfo {pages} {152503} (\bibinfo {year}
  {2018})}\BibitemShut {NoStop}%
\bibitem [{\citenamefont {Stroberg}\ \emph
  {et~al.}(2021{\natexlab{b}})\citenamefont {Stroberg}, \citenamefont {Holt},
  \citenamefont {Schwenk},\ and\ \citenamefont {Simonis}}]{holt2021}%
  \BibitemOpen
  \bibfield  {author} {\bibinfo {author} {\bibfnamefont {S.~R.}\ \bibnamefont
  {Stroberg}}, \bibinfo {author} {\bibfnamefont {J.~D.}\ \bibnamefont {Holt}},
  \bibinfo {author} {\bibfnamefont {A.}~\bibnamefont {Schwenk}}, \ and\
  \bibinfo {author} {\bibfnamefont {J.}~\bibnamefont {Simonis}},\ }\bibfield
  {title} {\enquote {\bibinfo {title} {Ab initio limits of atomic nuclei},}\
  }\href {\doibase 10.1103/PhysRevLett.126.022501} {\bibfield  {journal}
  {\bibinfo  {journal} {Phys. Rev. Lett.}\ }\textbf {\bibinfo {volume} {126}},\
  \bibinfo {pages} {022501} (\bibinfo {year} {2021}{\natexlab{b}})}\BibitemShut
  {NoStop}%
\bibitem [{\citenamefont {{Hu}}\ \emph {et~al.}(2021)\citenamefont {{Hu}},
  \citenamefont {{Jiang}}, \citenamefont {{Miyagi}}, \citenamefont {{Sun}},
  \citenamefont {{Ekstr{\"o}m}}, \citenamefont {{Forss{\'e}n}}, \citenamefont
  {{Hagen}}, \citenamefont {{Holt}}, \citenamefont {{Papenbrock}},
  \citenamefont {{Ragnar Stroberg}},\ and\ \citenamefont {{Vernon}}}]{hu2021}%
  \BibitemOpen
  \bibfield  {author} {\bibinfo {author} {\bibfnamefont {Baishan}\ \bibnamefont
  {{Hu}}}, \bibinfo {author} {\bibfnamefont {Weiguang}\ \bibnamefont
  {{Jiang}}}, \bibinfo {author} {\bibfnamefont {Takayuki}\ \bibnamefont
  {{Miyagi}}}, \bibinfo {author} {\bibfnamefont {Zhonghao}\ \bibnamefont
  {{Sun}}}, \bibinfo {author} {\bibfnamefont {Andreas}\ \bibnamefont
  {{Ekstr{\"o}m}}}, \bibinfo {author} {\bibfnamefont {Christian}\ \bibnamefont
  {{Forss{\'e}n}}}, \bibinfo {author} {\bibfnamefont {Gaute}\ \bibnamefont
  {{Hagen}}}, \bibinfo {author} {\bibfnamefont {Jason~D.}\ \bibnamefont
  {{Holt}}}, \bibinfo {author} {\bibfnamefont {Thomas}\ \bibnamefont
  {{Papenbrock}}}, \bibinfo {author} {\bibfnamefont {S.}~\bibnamefont {{Ragnar
  Stroberg}}}, \ and\ \bibinfo {author} {\bibfnamefont {Ian}\ \bibnamefont
  {{Vernon}}},\ }\bibfield  {title} {\enquote {\bibinfo {title} {{Ab initio
  predictions link the neutron skin of ${}^{208}$Pb to nuclear forces}},}\
  }\href@noop {} {\  (\bibinfo {year} {2021})},\ \Eprint
  {http://arxiv.org/abs/2112.01125} {arXiv:2112.01125 [nucl-th]} \BibitemShut
  {NoStop}%
\bibitem [{\citenamefont {Bartlett}\ and\ \citenamefont
  {Musia\l{}}(2007)}]{bartlett2007}%
  \BibitemOpen
  \bibfield  {author} {\bibinfo {author} {\bibfnamefont {Rodney~J.}\
  \bibnamefont {Bartlett}}\ and\ \bibinfo {author} {\bibfnamefont {Monika}\
  \bibnamefont {Musia\l{}}},\ }\bibfield  {title} {\enquote {\bibinfo {title}
  {Coupled-cluster theory in quantum chemistry},}\ }\href {\doibase
  10.1103/RevModPhys.79.291} {\bibfield  {journal} {\bibinfo  {journal} {Rev.
  Mod. Phys.}\ }\textbf {\bibinfo {volume} {79}},\ \bibinfo {pages} {291--352}
  (\bibinfo {year} {2007})}\BibitemShut {NoStop}%
\bibitem [{\citenamefont {Hergert}\ \emph {et~al.}(2014)\citenamefont
  {Hergert}, \citenamefont {Bogner}, \citenamefont {Morris}, \citenamefont
  {Binder}, \citenamefont {Calci}, \citenamefont {Langhammer},\ and\
  \citenamefont {Roth}}]{Hergert:2014iaa}%
  \BibitemOpen
  \bibfield  {author} {\bibinfo {author} {\bibfnamefont {H.}~\bibnamefont
  {Hergert}}, \bibinfo {author} {\bibfnamefont {S.~K.}\ \bibnamefont {Bogner}},
  \bibinfo {author} {\bibfnamefont {T.~D.}\ \bibnamefont {Morris}}, \bibinfo
  {author} {\bibfnamefont {S.}~\bibnamefont {Binder}}, \bibinfo {author}
  {\bibfnamefont {A.}~\bibnamefont {Calci}}, \bibinfo {author} {\bibfnamefont
  {J.}~\bibnamefont {Langhammer}}, \ and\ \bibinfo {author} {\bibfnamefont
  {R.}~\bibnamefont {Roth}},\ }\bibfield  {title} {\enquote {\bibinfo {title}
  {{Ab initio multireference in-medium similarity renormalization group
  calculations of even calcium and nickel isotopes}},}\ }\href {\doibase
  10.1103/PhysRevC.90.041302} {\bibfield  {journal} {\bibinfo  {journal} {Phys.
  Rev. C}\ }\textbf {\bibinfo {volume} {90}},\ \bibinfo {pages} {041302}
  (\bibinfo {year} {2014})}\BibitemShut {NoStop}%
\bibitem [{\citenamefont {Frosini}\ \emph
  {et~al.}(2021{\natexlab{a}})\citenamefont {Frosini}, \citenamefont {Duguet},
  \citenamefont {Ebran},\ and\ \citenamefont {Som\`a}}]{Frosini:2021fjf}%
  \BibitemOpen
  \bibfield  {author} {\bibinfo {author} {\bibfnamefont {Mikael}\ \bibnamefont
  {Frosini}}, \bibinfo {author} {\bibfnamefont {Thomas}\ \bibnamefont
  {Duguet}}, \bibinfo {author} {\bibfnamefont {Jean-Paul}\ \bibnamefont
  {Ebran}}, \ and\ \bibinfo {author} {\bibfnamefont {Vittorio}\ \bibnamefont
  {Som\`a}},\ }\bibfield  {title} {\enquote {\bibinfo {title} {{Multi-reference
  many-body perturbation theory for nuclei I -- Novel PGCM-PT formalism}},}\
  }\href@noop {} {\  (\bibinfo {year} {2021}{\natexlab{a}})},\ \Eprint
  {http://arxiv.org/abs/2110.15737} {arXiv:2110.15737 [nucl-th]} \BibitemShut
  {NoStop}%
\bibitem [{\citenamefont {Frosini}\ \emph
  {et~al.}(2021{\natexlab{b}})\citenamefont {Frosini}, \citenamefont {Duguet},
  \citenamefont {Ebran}, \citenamefont {Bally}, \citenamefont {Mongelli},
  \citenamefont {Rodr\'\i{}guez}, \citenamefont {Roth},\ and\ \citenamefont
  {Som\`a}}]{Frosini:2021sxj}%
  \BibitemOpen
  \bibfield  {author} {\bibinfo {author} {\bibfnamefont {Mikael}\ \bibnamefont
  {Frosini}}, \bibinfo {author} {\bibfnamefont {Thomas}\ \bibnamefont
  {Duguet}}, \bibinfo {author} {\bibfnamefont {Jean-Paul}\ \bibnamefont
  {Ebran}}, \bibinfo {author} {\bibfnamefont {Benjamin}\ \bibnamefont {Bally}},
  \bibinfo {author} {\bibfnamefont {Tobias}\ \bibnamefont {Mongelli}}, \bibinfo
  {author} {\bibfnamefont {Tom\'as~R.}\ \bibnamefont {Rodr\'\i{}guez}},
  \bibinfo {author} {\bibfnamefont {Robert}\ \bibnamefont {Roth}}, \ and\
  \bibinfo {author} {\bibfnamefont {Vittorio}\ \bibnamefont {Som\`a}},\
  }\bibfield  {title} {\enquote {\bibinfo {title} {{Multi-reference many-body
  perturbation theory for nuclei. II. Ab initio study of neon isotopes via PGCM
  and IM-NCSM calculations}},}\ }\href@noop {} {\  (\bibinfo {year}
  {2021}{\natexlab{b}})},\ \Eprint {http://arxiv.org/abs/2111.00797}
  {arXiv:2111.00797 [nucl-th]} \BibitemShut {NoStop}%
\bibitem [{\citenamefont {Frosini}\ \emph
  {et~al.}(2021{\natexlab{c}})\citenamefont {Frosini}, \citenamefont {Duguet},
  \citenamefont {Ebran}, \citenamefont {Bally}, \citenamefont {Hergert},
  \citenamefont {Rodr\'\i{}guez}, \citenamefont {Roth}, \citenamefont {Yao},\
  and\ \citenamefont {Som\`a}}]{Frosini:2021ddm}%
  \BibitemOpen
  \bibfield  {author} {\bibinfo {author} {\bibfnamefont {Mikael}\ \bibnamefont
  {Frosini}}, \bibinfo {author} {\bibfnamefont {Thomas}\ \bibnamefont
  {Duguet}}, \bibinfo {author} {\bibfnamefont {Jean-Paul}\ \bibnamefont
  {Ebran}}, \bibinfo {author} {\bibfnamefont {Benjamin}\ \bibnamefont {Bally}},
  \bibinfo {author} {\bibfnamefont {Heiko}\ \bibnamefont {Hergert}}, \bibinfo
  {author} {\bibfnamefont {Tom\'as~R.}\ \bibnamefont {Rodr\'\i{}guez}},
  \bibinfo {author} {\bibfnamefont {Robert}\ \bibnamefont {Roth}}, \bibinfo
  {author} {\bibfnamefont {Jiangming}\ \bibnamefont {Yao}}, \ and\ \bibinfo
  {author} {\bibfnamefont {Vittorio}\ \bibnamefont {Som\`a}},\ }\bibfield
  {title} {\enquote {\bibinfo {title} {{Multi-reference many-body perturbation
  theory for nuclei III -- Ab initio calculations at second order in
  PGCM-PT}},}\ }\href@noop {} {\  (\bibinfo {year} {2021}{\natexlab{c}})},\
  \Eprint {http://arxiv.org/abs/2111.01461} {arXiv:2111.01461 [nucl-th]}
  \BibitemShut {NoStop}%
\bibitem [{\citenamefont {Duguet}(2015)}]{duguet2015}%
  \BibitemOpen
  \bibfield  {author} {\bibinfo {author} {\bibfnamefont {T.}~\bibnamefont
  {Duguet}},\ }\bibfield  {title} {\enquote {\bibinfo {title} {Symmetry broken
  and restored coupled-cluster theory: I. rotational symmetry and angular
  momentum},}\ }\href {http://stacks.iop.org/0954-3899/42/i=2/a=025107}
  {\bibfield  {journal} {\bibinfo  {journal} {J. Phys. G.}\ }\textbf {\bibinfo
  {volume} {42}},\ \bibinfo {pages} {025107} (\bibinfo {year}
  {2015})}\BibitemShut {NoStop}%
\bibitem [{\citenamefont {Qiu}\ \emph {et~al.}(2017)\citenamefont {Qiu},
  \citenamefont {Henderson}, \citenamefont {Zhao},\ and\ \citenamefont
  {Scuseria}}]{qiu2017}%
  \BibitemOpen
  \bibfield  {author} {\bibinfo {author} {\bibfnamefont {Yiheng}\ \bibnamefont
  {Qiu}}, \bibinfo {author} {\bibfnamefont {Thomas~M.}\ \bibnamefont
  {Henderson}}, \bibinfo {author} {\bibfnamefont {Jinmo}\ \bibnamefont {Zhao}},
  \ and\ \bibinfo {author} {\bibfnamefont {Gustavo~E.}\ \bibnamefont
  {Scuseria}},\ }\bibfield  {title} {\enquote {\bibinfo {title} {Projected
  coupled cluster theory},}\ }\href {\doibase 10.1063/1.4991020} {\bibfield
  {journal} {\bibinfo  {journal} {J. Chem. Phys.}\ }\textbf {\bibinfo {volume}
  {147}},\ \bibinfo {pages} {064111} (\bibinfo {year} {2017})}\BibitemShut
  {NoStop}%
\bibitem [{\citenamefont {Duguet}\ and\ \citenamefont
  {Signoracci}(2017)}]{Duguet:2015yle}%
  \BibitemOpen
  \bibfield  {author} {\bibinfo {author} {\bibfnamefont {T.}~\bibnamefont
  {Duguet}}\ and\ \bibinfo {author} {\bibfnamefont {A.}~\bibnamefont
  {Signoracci}},\ }\bibfield  {title} {\enquote {\bibinfo {title} {{Symmetry
  broken and restored coupled-cluster theory. II. Global gauge symmetry and
  particle number}},}\ }\href {\doibase 10.1088/0954-3899/44/1/015103}
  {\bibfield  {journal} {\bibinfo  {journal} {J. Phys. G}\ }\textbf {\bibinfo
  {volume} {44}},\ \bibinfo {pages} {015103} (\bibinfo {year} {2017})},\
  \bibinfo {note} {[Erratum: J.Phys.G 44, 049601 (2017)]}\BibitemShut {NoStop}%
\bibitem [{\citenamefont {Qiu}\ \emph {et~al.}(2018)\citenamefont {Qiu},
  \citenamefont {Henderson}, \citenamefont {Zhao},\ and\ \citenamefont
  {Scuseria}}]{qiu2018}%
  \BibitemOpen
  \bibfield  {author} {\bibinfo {author} {\bibfnamefont {Yiheng}\ \bibnamefont
  {Qiu}}, \bibinfo {author} {\bibfnamefont {Thomas~M.}\ \bibnamefont
  {Henderson}}, \bibinfo {author} {\bibfnamefont {Jinmo}\ \bibnamefont {Zhao}},
  \ and\ \bibinfo {author} {\bibfnamefont {Gustavo~E.}\ \bibnamefont
  {Scuseria}},\ }\bibfield  {title} {\enquote {\bibinfo {title} {Projected
  coupled cluster theory: Optimization of cluster amplitudes in the presence of
  symmetry projection},}\ }\href {\doibase 10.1063/1.5053605} {\bibfield
  {journal} {\bibinfo  {journal} {J. Chem. Phys.}\ }\textbf {\bibinfo {volume}
  {149}},\ \bibinfo {pages} {164108} (\bibinfo {year} {2018})}\BibitemShut
  {NoStop}%
\bibitem [{\citenamefont {Qiu}\ \emph {et~al.}(2019)\citenamefont {Qiu},
  \citenamefont {Henderson}, \citenamefont {Duguet},\ and\ \citenamefont
  {Scuseria}}]{qiu2019}%
  \BibitemOpen
  \bibfield  {author} {\bibinfo {author} {\bibfnamefont {Y.}~\bibnamefont
  {Qiu}}, \bibinfo {author} {\bibfnamefont {T.~M.}\ \bibnamefont {Henderson}},
  \bibinfo {author} {\bibfnamefont {T.}~\bibnamefont {Duguet}}, \ and\ \bibinfo
  {author} {\bibfnamefont {G.~E.}\ \bibnamefont {Scuseria}},\ }\bibfield
  {title} {\enquote {\bibinfo {title} {Particle-number projected
  bogoliubov-coupled-cluster theory: Application to the pairing hamiltonian},}\
  }\href {\doibase 10.1103/PhysRevC.99.044301} {\bibfield  {journal} {\bibinfo
  {journal} {Phys. Rev. C}\ }\textbf {\bibinfo {volume} {99}},\ \bibinfo
  {pages} {044301} (\bibinfo {year} {2019})}\BibitemShut {NoStop}%
\bibitem [{\citenamefont {Tsuchimochi}\ and\ \citenamefont
  {Ten-no}(2018)}]{tsuchimochi2018}%
  \BibitemOpen
  \bibfield  {author} {\bibinfo {author} {\bibfnamefont {Takashi}\ \bibnamefont
  {Tsuchimochi}}\ and\ \bibinfo {author} {\bibfnamefont {Seiichiro~L.}\
  \bibnamefont {Ten-no}},\ }\bibfield  {title} {\enquote {\bibinfo {title}
  {Orbital-invariant spin-extended approximate coupled-cluster for
  multi-reference systems},}\ }\href {\doibase 10.1063/1.5036542} {\bibfield
  {journal} {\bibinfo  {journal} {J. Chem. Phys.}\ }\textbf {\bibinfo {volume}
  {149}},\ \bibinfo {pages} {044109} (\bibinfo {year} {2018})}\BibitemShut
  {NoStop}%
\bibitem [{\citenamefont {Mizusaki}\ and\ \citenamefont
  {Schuck}(2021)}]{mizusaki2021}%
  \BibitemOpen
  \bibfield  {author} {\bibinfo {author} {\bibfnamefont {Takahiro}\
  \bibnamefont {Mizusaki}}\ and\ \bibinfo {author} {\bibfnamefont {Peter}\
  \bibnamefont {Schuck}},\ }\bibfield  {title} {\enquote {\bibinfo {title}
  {Symmetry projection to coupled-cluster singles plus doubles wave function
  through the monte carlo method},}\ }\href {\doibase
  10.1103/PhysRevC.104.L031305} {\bibfield  {journal} {\bibinfo  {journal}
  {Phys. Rev. C}\ }\textbf {\bibinfo {volume} {104}},\ \bibinfo {pages}
  {L031305} (\bibinfo {year} {2021})}\BibitemShut {NoStop}%
\bibitem [{\citenamefont {Arponen}(1983)}]{arponen1983}%
  \BibitemOpen
  \bibfield  {author} {\bibinfo {author} {\bibfnamefont {Jouko}\ \bibnamefont
  {Arponen}},\ }\bibfield  {title} {\enquote {\bibinfo {title} {{Variational
  principles and linked-cluster exp S expansions for static and dynamic
  many-body problems}},}\ }\href {\doibase 10.1016/0003-4916(83)90284-1}
  {\bibfield  {journal} {\bibinfo  {journal} {Ann. Phys.}\ }\textbf {\bibinfo
  {volume} {151}},\ \bibinfo {pages} {311 -- 382} (\bibinfo {year}
  {1983})}\BibitemShut {NoStop}%
\bibitem [{\citenamefont {Arponen}\ \emph {et~al.}(1987)\citenamefont
  {Arponen}, \citenamefont {Bishop},\ and\ \citenamefont
  {Pajanne}}]{arponen1987}%
  \BibitemOpen
  \bibfield  {author} {\bibinfo {author} {\bibfnamefont {J.~S.}\ \bibnamefont
  {Arponen}}, \bibinfo {author} {\bibfnamefont {R.~F.}\ \bibnamefont {Bishop}},
  \ and\ \bibinfo {author} {\bibfnamefont {E.}~\bibnamefont {Pajanne}},\
  }\bibfield  {title} {\enquote {\bibinfo {title} {Extended coupled-cluster
  method. i. generalized coherent bosonization as a mapping of quantum theory
  into classical hamiltonian mechanics},}\ }\href {\doibase
  10.1103/PhysRevA.36.2519} {\bibfield  {journal} {\bibinfo  {journal} {Phys.
  Rev. A}\ }\textbf {\bibinfo {volume} {36}},\ \bibinfo {pages} {2519--2538}
  (\bibinfo {year} {1987})}\BibitemShut {NoStop}%
\bibitem [{\citenamefont {{Van Voorhis}}\ and\ \citenamefont
  {Head-Gordon}(2000)}]{vanvoorhis2000}%
  \BibitemOpen
  \bibfield  {author} {\bibinfo {author} {\bibfnamefont {Troy}\ \bibnamefont
  {{Van Voorhis}}}\ and\ \bibinfo {author} {\bibfnamefont {Martin}\
  \bibnamefont {Head-Gordon}},\ }\bibfield  {title} {\enquote {\bibinfo {title}
  {The quadratic coupled cluster doubles model},}\ }\href {\doibase
  10.1016/S0009-2614(00)01137-4} {\bibfield  {journal} {\bibinfo  {journal}
  {Chem. Phys. Lett.}\ }\textbf {\bibinfo {volume} {330}},\ \bibinfo {pages}
  {585--594} (\bibinfo {year} {2000})}\BibitemShut {NoStop}%
\bibitem [{\citenamefont {Byrd}\ \emph {et~al.}(2002)\citenamefont {Byrd},
  \citenamefont {Van~Voorhis},\ and\ \citenamefont {Head-Gordon}}]{byrd2002}%
  \BibitemOpen
  \bibfield  {author} {\bibinfo {author} {\bibfnamefont {Edward F.~C.}\
  \bibnamefont {Byrd}}, \bibinfo {author} {\bibfnamefont {Troy}\ \bibnamefont
  {Van~Voorhis}}, \ and\ \bibinfo {author} {\bibfnamefont {Martin}\
  \bibnamefont {Head-Gordon}},\ }\bibfield  {title} {\enquote {\bibinfo {title}
  {Quadratic coupled-cluster doubles: Implementation and assessment of perfect
  pairing optimized geometries},}\ }\href {\doibase 10.1021/jp020255u}
  {\bibfield  {journal} {\bibinfo  {journal} {J. Phys. Chem. B}\ }\textbf
  {\bibinfo {volume} {106}},\ \bibinfo {pages} {8070--8077} (\bibinfo {year}
  {2002})}\BibitemShut {NoStop}%
\bibitem [{\citenamefont {Bartlett}\ and\ \citenamefont
  {Noga}(1988)}]{bartlett1988}%
  \BibitemOpen
  \bibfield  {author} {\bibinfo {author} {\bibfnamefont {Rodney~J.}\
  \bibnamefont {Bartlett}}\ and\ \bibinfo {author} {\bibfnamefont {Jozef}\
  \bibnamefont {Noga}},\ }\bibfield  {title} {\enquote {\bibinfo {title} {The
  expectation value coupled-cluster method and analytical energy
  derivatives},}\ }\href {\doibase 10.1016/0009-2614(88)80392-0} {\bibfield
  {journal} {\bibinfo  {journal} {Chem. Phys. Lett.}\ }\textbf {\bibinfo
  {volume} {150}},\ \bibinfo {pages} {29--36} (\bibinfo {year}
  {1988})}\BibitemShut {NoStop}%
\bibitem [{\citenamefont {Szalay}\ \emph {et~al.}(1995)\citenamefont {Szalay},
  \citenamefont {Nooijen},\ and\ \citenamefont {Bartlett}}]{szalay1995}%
  \BibitemOpen
  \bibfield  {author} {\bibinfo {author} {\bibfnamefont {P{\'e}ter~G.}\
  \bibnamefont {Szalay}}, \bibinfo {author} {\bibfnamefont {Marcel}\
  \bibnamefont {Nooijen}}, \ and\ \bibinfo {author} {\bibfnamefont {Rodney~J.}\
  \bibnamefont {Bartlett}},\ }\bibfield  {title} {\enquote {\bibinfo {title}
  {Alternative ans{\"a}tze in single reference coupled-cluster theory. iii. a
  critical analysis of different methods},}\ }\href {\doibase 10.1063/1.469641}
  {\bibfield  {journal} {\bibinfo  {journal} {J. Chem. Phys.}\ }\textbf
  {\bibinfo {volume} {103}},\ \bibinfo {pages} {281--298} (\bibinfo {year}
  {1995})}\BibitemShut {NoStop}%
\bibitem [{\citenamefont {Duch}\ and\ \citenamefont
  {Diercksen}(1994)}]{duch1994}%
  \BibitemOpen
  \bibfield  {author} {\bibinfo {author} {\bibfnamefont {W\l{}odzis\l{}aw}\
  \bibnamefont {Duch}}\ and\ \bibinfo {author} {\bibfnamefont {Geerd H.~F.}\
  \bibnamefont {Diercksen}},\ }\bibfield  {title} {\enquote {\bibinfo {title}
  {Size-extensivity corrections in configuration interaction methods},}\ }\href
  {\doibase 10.1063/1.467615} {\bibfield  {journal} {\bibinfo  {journal} {J.
  Chem. Phys.}\ }\textbf {\bibinfo {volume} {101}},\ \bibinfo {pages}
  {3018--3030} (\bibinfo {year} {1994})}\BibitemShut {NoStop}%
\bibitem [{\citenamefont {Ramos-Cordoba}\ \emph {et~al.}(2016)\citenamefont
  {Ramos-Cordoba}, \citenamefont {Salvador},\ and\ \citenamefont
  {Matito}}]{ramoscordoba2016}%
  \BibitemOpen
  \bibfield  {author} {\bibinfo {author} {\bibfnamefont {Eloy}\ \bibnamefont
  {Ramos-Cordoba}}, \bibinfo {author} {\bibfnamefont {Pedro}\ \bibnamefont
  {Salvador}}, \ and\ \bibinfo {author} {\bibfnamefont {Eduard}\ \bibnamefont
  {Matito}},\ }\bibfield  {title} {\enquote {\bibinfo {title} {Separation of
  dynamic and nondynamic correlation},}\ }\href {\doibase 10.1039/C6CP03072F}
  {\bibfield  {journal} {\bibinfo  {journal} {Phys. Chem. Chem. Phys.}\
  }\textbf {\bibinfo {volume} {18}},\ \bibinfo {pages} {24015--24023} (\bibinfo
  {year} {2016})}\BibitemShut {NoStop}%
\bibitem [{\citenamefont {Bertsch}\ \emph {et~al.}(2019)\citenamefont
  {Bertsch}, \citenamefont {Kawano},\ and\ \citenamefont
  {Robledo}}]{bertsch2019}%
  \BibitemOpen
  \bibfield  {author} {\bibinfo {author} {\bibfnamefont {G.~F.}\ \bibnamefont
  {Bertsch}}, \bibinfo {author} {\bibfnamefont {T.}~\bibnamefont {Kawano}}, \
  and\ \bibinfo {author} {\bibfnamefont {L.~M.}\ \bibnamefont {Robledo}},\
  }\bibfield  {title} {\enquote {\bibinfo {title} {Angular momentum of fission
  fragments},}\ }\href {\doibase 10.1103/PhysRevC.99.034603} {\bibfield
  {journal} {\bibinfo  {journal} {Phys. Rev. C}\ }\textbf {\bibinfo {volume}
  {99}},\ \bibinfo {pages} {034603} (\bibinfo {year} {2019})}\BibitemShut
  {NoStop}%
\bibitem [{\citenamefont {Fujita}\ and\ \citenamefont
  {Miyazawa}(1957)}]{fujita1957}%
  \BibitemOpen
  \bibfield  {author} {\bibinfo {author} {\bibfnamefont {{Jun-ichi}}\
  \bibnamefont {Fujita}}\ and\ \bibinfo {author} {\bibfnamefont {Hironari}\
  \bibnamefont {Miyazawa}},\ }\bibfield  {title} {\enquote {\bibinfo {title}
  {Pion theory of three-body forces},}\ }\href {\doibase 10.1143/PTP.17.360}
  {\bibfield  {journal} {\bibinfo  {journal} {Prog. Theor. Phys.}\ }\textbf
  {\bibinfo {volume} {17}},\ \bibinfo {pages} {360--365} (\bibinfo {year}
  {1957})}\BibitemShut {NoStop}%
\bibitem [{\citenamefont {Bedaque}\ \emph {et~al.}(1999)\citenamefont
  {Bedaque}, \citenamefont {Hammer},\ and\ \citenamefont {van
  Kolck}}]{bedaque1999}%
  \BibitemOpen
  \bibfield  {author} {\bibinfo {author} {\bibfnamefont {P.~F.}\ \bibnamefont
  {Bedaque}}, \bibinfo {author} {\bibfnamefont {H.-W.}\ \bibnamefont {Hammer}},
  \ and\ \bibinfo {author} {\bibfnamefont {U.}~\bibnamefont {van Kolck}},\
  }\bibfield  {title} {\enquote {\bibinfo {title} {Renormalization of the
  three-body system with short-range interactions},}\ }\href {\doibase
  10.1103/PhysRevLett.82.463} {\bibfield  {journal} {\bibinfo  {journal} {Phys.
  Rev. Lett.}\ }\textbf {\bibinfo {volume} {82}},\ \bibinfo {pages} {463--467}
  (\bibinfo {year} {1999})}\BibitemShut {NoStop}%
\bibitem [{\citenamefont {Epelbaum}\ \emph {et~al.}(2009)\citenamefont
  {Epelbaum}, \citenamefont {Hammer},\ and\ \citenamefont
  {Mei\ss{}ner}}]{epelbaum2009}%
  \BibitemOpen
  \bibfield  {author} {\bibinfo {author} {\bibfnamefont {E.}~\bibnamefont
  {Epelbaum}}, \bibinfo {author} {\bibfnamefont {H.-W.}\ \bibnamefont
  {Hammer}}, \ and\ \bibinfo {author} {\bibfnamefont {Ulf-G.}\ \bibnamefont
  {Mei\ss{}ner}},\ }\bibfield  {title} {\enquote {\bibinfo {title} {Modern
  theory of nuclear forces},}\ }\href {\doibase 10.1103/RevModPhys.81.1773}
  {\bibfield  {journal} {\bibinfo  {journal} {Rev. Mod. Phys.}\ }\textbf
  {\bibinfo {volume} {81}},\ \bibinfo {pages} {1773--1825} (\bibinfo {year}
  {2009})}\BibitemShut {NoStop}%
\bibitem [{\citenamefont {Hammer}\ \emph {et~al.}(2013)\citenamefont {Hammer},
  \citenamefont {Nogga},\ and\ \citenamefont {Schwenk}}]{hammer2013}%
  \BibitemOpen
  \bibfield  {author} {\bibinfo {author} {\bibfnamefont {Hans-Werner}\
  \bibnamefont {Hammer}}, \bibinfo {author} {\bibfnamefont {Andreas}\
  \bibnamefont {Nogga}}, \ and\ \bibinfo {author} {\bibfnamefont {Achim}\
  \bibnamefont {Schwenk}},\ }\bibfield  {title} {\enquote {\bibinfo {title}
  {\textit{Colloquium} : Three-body forces: From cold atoms to nuclei},}\
  }\href {\doibase 10.1103/RevModPhys.85.197} {\bibfield  {journal} {\bibinfo
  {journal} {Rev. Mod. Phys.}\ }\textbf {\bibinfo {volume} {85}},\ \bibinfo
  {pages} {197--217} (\bibinfo {year} {2013})}\BibitemShut {NoStop}%
\bibitem [{\citenamefont {Bohr}\ and\ \citenamefont
  {Mottelson}(1975)}]{bohr1975}%
  \BibitemOpen
  \bibfield  {author} {\bibinfo {author} {\bibfnamefont {A.}~\bibnamefont
  {Bohr}}\ and\ \bibinfo {author} {\bibfnamefont {B.~R.}\ \bibnamefont
  {Mottelson}},\ }\href@noop {} {\emph {\bibinfo {title} {Nuclear
  Structure}}},\ Vol.\ \bibinfo {volume} {II: Nuclear Deformation}\ (\bibinfo
  {publisher} {W. A. Benjamin},\ \bibinfo {address} {Reading, Massachusetts,
  USA},\ \bibinfo {year} {1975})\BibitemShut {NoStop}%
\bibitem [{\citenamefont {{Staszczak}}\ \emph {et~al.}(2010)\citenamefont
  {{Staszczak}}, \citenamefont {{Stoitsov}}, \citenamefont {{Baran}},\ and\
  \citenamefont {{Nazarewicz}}}]{staszscak2010}%
  \BibitemOpen
  \bibfield  {author} {\bibinfo {author} {\bibfnamefont {A.}~\bibnamefont
  {{Staszczak}}}, \bibinfo {author} {\bibfnamefont {M.}~\bibnamefont
  {{Stoitsov}}}, \bibinfo {author} {\bibfnamefont {A.}~\bibnamefont {{Baran}}},
  \ and\ \bibinfo {author} {\bibfnamefont {W.}~\bibnamefont {{Nazarewicz}}},\
  }\bibfield  {title} {\enquote {\bibinfo {title} {{Augmented Lagrangian method
  for constrained nuclear density functional theory}},}\ }\href {\doibase
  10.1140/epja/i2010-11018-9} {\bibfield  {journal} {\bibinfo  {journal} {Eur.
  Phys. J. A}\ }\textbf {\bibinfo {volume} {46}},\ \bibinfo {pages} {85--90}
  (\bibinfo {year} {2010})}\BibitemShut {NoStop}%
\bibitem [{\citenamefont {Rodr\'iguez}\ \emph {et~al.}(2005)\citenamefont
  {Rodr\'iguez}, \citenamefont {Egido},\ and\ \citenamefont {Robledo}}]{Rod05}%
  \BibitemOpen
  \bibfield  {author} {\bibinfo {author} {\bibfnamefont {Tom\'as~R.}\
  \bibnamefont {Rodr\'iguez}}, \bibinfo {author} {\bibfnamefont {J.~L.}\
  \bibnamefont {Egido}}, \ and\ \bibinfo {author} {\bibfnamefont {L.~M.}\
  \bibnamefont {Robledo}},\ }\href@noop {} {\bibfield  {journal} {\bibinfo
  {journal} {Phys. Rev.}\ }\textbf {\bibinfo {volume} {C72}},\ \bibinfo {pages}
  {064303} (\bibinfo {year} {2005})}\BibitemShut {NoStop}%
\bibitem [{\citenamefont {Ripoche}\ \emph {et~al.}(2018)\citenamefont
  {Ripoche}, \citenamefont {Duguet}, \citenamefont {Ebran},\ and\ \citenamefont
  {Lacroix}}]{Ripoche:2017ydv}%
  \BibitemOpen
  \bibfield  {author} {\bibinfo {author} {\bibfnamefont {Julien}\ \bibnamefont
  {Ripoche}}, \bibinfo {author} {\bibfnamefont {Thomas}\ \bibnamefont
  {Duguet}}, \bibinfo {author} {\bibfnamefont {Jean-Paul}\ \bibnamefont
  {Ebran}}, \ and\ \bibinfo {author} {\bibfnamefont {Denis}\ \bibnamefont
  {Lacroix}},\ }\bibfield  {title} {\enquote {\bibinfo {title} {{Combining
  symmetry breaking and restoration with configuration interaction: extension
  to z-signature symmetry in the case of the Lipkin Model}},}\ }\href {\doibase
  10.1103/PhysRevC.97.064316} {\bibfield  {journal} {\bibinfo  {journal} {Phys.
  Rev. C}\ }\textbf {\bibinfo {volume} {97}},\ \bibinfo {pages} {064316}
  (\bibinfo {year} {2018})}\BibitemShut {NoStop}%
\bibitem [{\citenamefont {Ring}\ and\ \citenamefont
  {Schuck}(1980)}]{ringschuck}%
  \BibitemOpen
  \bibfield  {author} {\bibinfo {author} {\bibfnamefont {P.}~\bibnamefont
  {Ring}}\ and\ \bibinfo {author} {\bibfnamefont {P.}~\bibnamefont {Schuck}},\
  }\href@noop {} {\emph {\bibinfo {title} {The Nuclear Many-Body Problem}}}\
  (\bibinfo  {publisher} {Springer},\ \bibinfo {address} {Heidelberg},\
  \bibinfo {year} {1980})\BibitemShut {NoStop}%
\bibitem [{\citenamefont {Jiménez-Hoyos}\ \emph {et~al.}(2012)\citenamefont
  {Jiménez-Hoyos}, \citenamefont {Henderson}, \citenamefont {Tsuchimochi},\
  and\ \citenamefont {Scuseria}}]{hoyos2012}%
  \BibitemOpen
  \bibfield  {author} {\bibinfo {author} {\bibfnamefont {Carlos~A.}\
  \bibnamefont {Jiménez-Hoyos}}, \bibinfo {author} {\bibfnamefont {Thomas~M.}\
  \bibnamefont {Henderson}}, \bibinfo {author} {\bibfnamefont {Takashi}\
  \bibnamefont {Tsuchimochi}}, \ and\ \bibinfo {author} {\bibfnamefont
  {Gustavo~E.}\ \bibnamefont {Scuseria}},\ }\bibfield  {title} {\enquote
  {\bibinfo {title} {Projected hartree–fock theory},}\ }\href {\doibase
  10.1063/1.4705280} {\bibfield  {journal} {\bibinfo  {journal} {J. Chem.
  Phys.}\ }\textbf {\bibinfo {volume} {136}},\ \bibinfo {pages} {164109}
  (\bibinfo {year} {2012})}\BibitemShut {NoStop}%
\bibitem [{\citenamefont {Ripoche}\ \emph {et~al.}(2020)\citenamefont
  {Ripoche}, \citenamefont {Tichai},\ and\ \citenamefont
  {Duguet}}]{ripoche2020}%
  \BibitemOpen
  \bibfield  {author} {\bibinfo {author} {\bibfnamefont {Julien}\ \bibnamefont
  {Ripoche}}, \bibinfo {author} {\bibfnamefont {Alexander}\ \bibnamefont
  {Tichai}}, \ and\ \bibinfo {author} {\bibfnamefont {Thomas}\ \bibnamefont
  {Duguet}},\ }\bibfield  {title} {\enquote {\bibinfo {title} {{Normal-ordered
  $k$-body approximation in particle-number-breaking theories}},}\ }\href
  {\doibase 10.1140/epja/s10050-020-00045-8} {\bibfield  {journal} {\bibinfo
  {journal} {Eur. Phys. J. A}\ }\textbf {\bibinfo {volume} {56}},\ \bibinfo
  {pages} {40} (\bibinfo {year} {2020})}\BibitemShut {NoStop}%
\bibitem [{\citenamefont {Frosini}\ \emph
  {et~al.}(2021{\natexlab{d}})\citenamefont {Frosini}, \citenamefont {Duguet},
  \citenamefont {Bally}, \citenamefont {Beaujeault-Taudi\`ere}, \citenamefont
  {Ebran},\ and\ \citenamefont {Som\`a}}]{Frosini:2021tuj}%
  \BibitemOpen
  \bibfield  {author} {\bibinfo {author} {\bibfnamefont {M.}~\bibnamefont
  {Frosini}}, \bibinfo {author} {\bibfnamefont {T.}~\bibnamefont {Duguet}},
  \bibinfo {author} {\bibfnamefont {B.}~\bibnamefont {Bally}}, \bibinfo
  {author} {\bibfnamefont {Y.}~\bibnamefont {Beaujeault-Taudi\`ere}}, \bibinfo
  {author} {\bibfnamefont {J.~P.}\ \bibnamefont {Ebran}}, \ and\ \bibinfo
  {author} {\bibfnamefont {V.}~\bibnamefont {Som\`a}},\ }\bibfield  {title}
  {\enquote {\bibinfo {title} {{In-medium $k$-body reduction of $n$-body
  operators: A flexible symmetry-conserving approach based on the sole one-body
  density matrix}},}\ }\href {\doibase 10.1140/epja/s10050-021-00458-z}
  {\bibfield  {journal} {\bibinfo  {journal} {Eur. Phys. J. A}\ }\textbf
  {\bibinfo {volume} {57}},\ \bibinfo {pages} {151} (\bibinfo {year}
  {2021}{\natexlab{d}})}\BibitemShut {NoStop}%
\bibitem [{\citenamefont {Hagen}\ \emph {et~al.}(2007)\citenamefont {Hagen},
  \citenamefont {Papenbrock}, \citenamefont {Dean}, \citenamefont {Schwenk},
  \citenamefont {Nogga}, \citenamefont {W\l{}och},\ and\ \citenamefont
  {Piecuch}}]{hagen2007a}%
  \BibitemOpen
  \bibfield  {author} {\bibinfo {author} {\bibfnamefont {G.}~\bibnamefont
  {Hagen}}, \bibinfo {author} {\bibfnamefont {T.}~\bibnamefont {Papenbrock}},
  \bibinfo {author} {\bibfnamefont {D.~J.}\ \bibnamefont {Dean}}, \bibinfo
  {author} {\bibfnamefont {A.}~\bibnamefont {Schwenk}}, \bibinfo {author}
  {\bibfnamefont {A.}~\bibnamefont {Nogga}}, \bibinfo {author} {\bibfnamefont
  {M.}~\bibnamefont {W\l{}och}}, \ and\ \bibinfo {author} {\bibfnamefont
  {P.}~\bibnamefont {Piecuch}},\ }\bibfield  {title} {\enquote {\bibinfo
  {title} {{Coupled-cluster theory for three-body Hamiltonians}},}\ }\href
  {\doibase 10.1103/PhysRevC.76.034302} {\bibfield  {journal} {\bibinfo
  {journal} {Phys. Rev. C}\ }\textbf {\bibinfo {volume} {76}},\ \bibinfo
  {pages} {034302} (\bibinfo {year} {2007})}\BibitemShut {NoStop}%
\bibitem [{\citenamefont {Roth}\ \emph {et~al.}(2012)\citenamefont {Roth},
  \citenamefont {Binder}, \citenamefont {Vobig}, \citenamefont {Calci},
  \citenamefont {Langhammer},\ and\ \citenamefont {Navr\'atil}}]{roth2012}%
  \BibitemOpen
  \bibfield  {author} {\bibinfo {author} {\bibfnamefont {Robert}\ \bibnamefont
  {Roth}}, \bibinfo {author} {\bibfnamefont {Sven}\ \bibnamefont {Binder}},
  \bibinfo {author} {\bibfnamefont {Klaus}\ \bibnamefont {Vobig}}, \bibinfo
  {author} {\bibfnamefont {Angelo}\ \bibnamefont {Calci}}, \bibinfo {author}
  {\bibfnamefont {Joachim}\ \bibnamefont {Langhammer}}, \ and\ \bibinfo
  {author} {\bibfnamefont {Petr}\ \bibnamefont {Navr\'atil}},\ }\bibfield
  {title} {\enquote {\bibinfo {title} {{Medium-Mass Nuclei with Normal-Ordered
  Chiral $NN\mathbf{+}3N$ Interactions}},}\ }\href {\doibase
  10.1103/PhysRevLett.109.052501} {\bibfield  {journal} {\bibinfo  {journal}
  {Phys. Rev. Lett.}\ }\textbf {\bibinfo {volume} {109}},\ \bibinfo {pages}
  {052501} (\bibinfo {year} {2012})}\BibitemShut {NoStop}%
\bibitem [{\citenamefont {Shavitt}\ and\ \citenamefont
  {Bartlett}(2009)}]{shavittbartlett2009}%
  \BibitemOpen
  \bibfield  {author} {\bibinfo {author} {\bibfnamefont {I.}~\bibnamefont
  {Shavitt}}\ and\ \bibinfo {author} {\bibfnamefont {R.~J.}\ \bibnamefont
  {Bartlett}},\ }\href@noop {} {\emph {\bibinfo {title} {Many-body Methods in
  Chemistry and Physics}}}\ (\bibinfo  {publisher} {Cambridge University
  Press},\ \bibinfo {address} {Cambridge UK},\ \bibinfo {year}
  {2009})\BibitemShut {NoStop}%
\bibitem [{\citenamefont {Tichai}\ \emph {et~al.}(2019)\citenamefont {Tichai},
  \citenamefont {M\"uller}, \citenamefont {Vobig},\ and\ \citenamefont
  {Roth}}]{tichai2019}%
  \BibitemOpen
  \bibfield  {author} {\bibinfo {author} {\bibfnamefont {A.}~\bibnamefont
  {Tichai}}, \bibinfo {author} {\bibfnamefont {J.}~\bibnamefont {M\"uller}},
  \bibinfo {author} {\bibfnamefont {K.}~\bibnamefont {Vobig}}, \ and\ \bibinfo
  {author} {\bibfnamefont {R.}~\bibnamefont {Roth}},\ }\bibfield  {title}
  {\enquote {\bibinfo {title} {Natural orbitals for ab initio no-core shell
  model calculations},}\ }\href {\doibase 10.1103/PhysRevC.99.034321}
  {\bibfield  {journal} {\bibinfo  {journal} {Phys. Rev. C}\ }\textbf {\bibinfo
  {volume} {99}},\ \bibinfo {pages} {034321} (\bibinfo {year}
  {2019})}\BibitemShut {NoStop}%
\bibitem [{\citenamefont {Hoppe}\ \emph {et~al.}(2021)\citenamefont {Hoppe},
  \citenamefont {Tichai}, \citenamefont {Heinz}, \citenamefont {Hebeler},\ and\
  \citenamefont {Schwenk}}]{hoppe2021}%
  \BibitemOpen
  \bibfield  {author} {\bibinfo {author} {\bibfnamefont {J.}~\bibnamefont
  {Hoppe}}, \bibinfo {author} {\bibfnamefont {A.}~\bibnamefont {Tichai}},
  \bibinfo {author} {\bibfnamefont {M.}~\bibnamefont {Heinz}}, \bibinfo
  {author} {\bibfnamefont {K.}~\bibnamefont {Hebeler}}, \ and\ \bibinfo
  {author} {\bibfnamefont {A.}~\bibnamefont {Schwenk}},\ }\bibfield  {title}
  {\enquote {\bibinfo {title} {Natural orbitals for many-body expansion
  methods},}\ }\href {\doibase 10.1103/PhysRevC.103.014321} {\bibfield
  {journal} {\bibinfo  {journal} {Phys. Rev. C}\ }\textbf {\bibinfo {volume}
  {103}},\ \bibinfo {pages} {014321} (\bibinfo {year} {2021})}\BibitemShut
  {NoStop}%
\bibitem [{\citenamefont {Kortelainen}\ \emph {et~al.}(2021)\citenamefont
  {Kortelainen}, \citenamefont {Sun}, \citenamefont {Hagen}, \citenamefont
  {Nazarewicz}, \citenamefont {Papenbrock},\ and\ \citenamefont
  {Reinhard}}]{kortelainen2021}%
  \BibitemOpen
  \bibfield  {author} {\bibinfo {author} {\bibfnamefont {Markus}\ \bibnamefont
  {Kortelainen}}, \bibinfo {author} {\bibfnamefont {Zhonghao}\ \bibnamefont
  {Sun}}, \bibinfo {author} {\bibfnamefont {Gaute}\ \bibnamefont {Hagen}},
  \bibinfo {author} {\bibfnamefont {Witold}\ \bibnamefont {Nazarewicz}},
  \bibinfo {author} {\bibfnamefont {Thomas}\ \bibnamefont {Papenbrock}}, \ and\
  \bibinfo {author} {\bibfnamefont {Paul-Gerhard}\ \bibnamefont {Reinhard}},\
  }\href@noop {} {\enquote {\bibinfo {title} {Universal trend of charge radii
  of even-even ca-zn nuclei},}\ } (\bibinfo {year} {2021}),\ \Eprint
  {http://arxiv.org/abs/2111.12464} {arXiv:2111.12464 [nucl-th]} \BibitemShut
  {NoStop}%
\bibitem [{\citenamefont {Varshalovich}\ \emph {et~al.}(1988)\citenamefont
  {Varshalovich}, \citenamefont {Moskalev},\ and\ \citenamefont
  {Khersonskii}}]{varshalovich88a}%
  \BibitemOpen
  \bibfield  {author} {\bibinfo {author} {\bibfnamefont {D.~A.}\ \bibnamefont
  {Varshalovich}}, \bibinfo {author} {\bibfnamefont {A.~N.}\ \bibnamefont
  {Moskalev}}, \ and\ \bibinfo {author} {\bibfnamefont {V.~K.}\ \bibnamefont
  {Khersonskii}},\ }\href@noop {} {\emph {\bibinfo {title} {Quantum Theory of
  Angular Momentum}}}\ (\bibinfo  {publisher} {World Scientific},\ \bibinfo
  {address} {Singapor},\ \bibinfo {year} {1988})\BibitemShut {NoStop}%
\bibitem [{\citenamefont {{Thouless}}(1960)}]{thouless1960}%
  \BibitemOpen
  \bibfield  {author} {\bibinfo {author} {\bibfnamefont {D.~J.}\ \bibnamefont
  {{Thouless}}},\ }\bibfield  {title} {\enquote {\bibinfo {title} {{Stability
  conditions and nuclear rotations in the Hartree-Fock theory}},}\ }\href
  {\doibase 10.1016/0029-5582(60)90048-1} {\bibfield  {journal} {\bibinfo
  {journal} {Nuclear Physics}\ }\textbf {\bibinfo {volume} {21}},\ \bibinfo
  {pages} {225--232} (\bibinfo {year} {1960})}\BibitemShut {NoStop}%
\bibitem [{\citenamefont {{Wa Wong}}(1975)}]{wawong1975}%
  \BibitemOpen
  \bibfield  {author} {\bibinfo {author} {\bibfnamefont {Chun}\ \bibnamefont
  {{Wa Wong}}},\ }\bibfield  {title} {\enquote {\bibinfo {title}
  {Generator-coordinate methods in nuclear physics},}\ }\href {\doibase
  10.1016/0370-1573(75)90036-8} {\bibfield  {journal} {\bibinfo  {journal}
  {Phys. Rep.}\ }\textbf {\bibinfo {volume} {15}},\ \bibinfo {pages} {283--357}
  (\bibinfo {year} {1975})}\BibitemShut {NoStop}%
\bibitem [{\citenamefont {Broglia}\ \emph {et~al.}(2000)\citenamefont
  {Broglia}, \citenamefont {Terasaki},\ and\ \citenamefont
  {Giovanardi}}]{broglia2000}%
  \BibitemOpen
  \bibfield  {author} {\bibinfo {author} {\bibfnamefont {R.~A.}\ \bibnamefont
  {Broglia}}, \bibinfo {author} {\bibfnamefont {J.}~\bibnamefont {Terasaki}}, \
  and\ \bibinfo {author} {\bibfnamefont {N.}~\bibnamefont {Giovanardi}},\
  }\bibfield  {title} {\enquote {\bibinfo {title} {The
  anderson–goldstone–nambu mode in finite and in infinite systems},}\
  }\href {\doibase 10.1016/S0370-1573(00)00046-6} {\bibfield  {journal}
  {\bibinfo  {journal} {Phys. Rep.}\ }\textbf {\bibinfo {volume} {335}},\
  \bibinfo {pages} {1--18} (\bibinfo {year} {2000})}\BibitemShut {NoStop}%
\bibitem [{\citenamefont {Yannouleas}\ and\ \citenamefont
  {Landman}(2007)}]{yannouleas2007}%
  \BibitemOpen
  \bibfield  {author} {\bibinfo {author} {\bibfnamefont {C.}~\bibnamefont
  {Yannouleas}}\ and\ \bibinfo {author} {\bibfnamefont {U.}~\bibnamefont
  {Landman}},\ }\bibfield  {title} {\enquote {\bibinfo {title} {Symmetry
  breaking and quantum correlations in finite systems: studies of quantum dots
  and ultracold bose gases and related nuclear and chemical methods},}\ }\href
  {\doibase 10.1088/0034-4885/70/12/R02} {\bibfield  {journal} {\bibinfo
  {journal} {Rep. Prog. Phys.}\ }\textbf {\bibinfo {volume} {70}},\ \bibinfo
  {pages} {2067} (\bibinfo {year} {2007})}\BibitemShut {NoStop}%
\bibitem [{\citenamefont {Deserno}(2004)}]{deserno2004}%
  \BibitemOpen
  \bibfield  {author} {\bibinfo {author} {\bibfnamefont {Markus}\ \bibnamefont
  {Deserno}},\ }\href
  {https://www.cmu.edu/biolphys/deserno/pdf/sphere_equi.pdf} {\enquote
  {\bibinfo {title} {How to generate equidistributed points on the surface of a
  sphere},}\ } (\bibinfo {year} {2004})\BibitemShut {NoStop}%
\bibitem [{\citenamefont {Weinberg}(1996)}]{weinberg_v2_1996}%
  \BibitemOpen
  \bibfield  {author} {\bibinfo {author} {\bibfnamefont {S.}~\bibnamefont
  {Weinberg}},\ }\href@noop {} {\emph {\bibinfo {title} {The Quantum Theory of
  Fields}}},\ Vol.~\bibinfo {volume} {II}\ (\bibinfo  {publisher} {Cambridge
  University Press},\ \bibinfo {address} {Cambridge, UK},\ \bibinfo {year}
  {1996})\BibitemShut {NoStop}%
\bibitem [{\citenamefont {Papenbrock}\ and\ \citenamefont
  {Weidenm\"uller}(2014)}]{papenbrock2014}%
  \BibitemOpen
  \bibfield  {author} {\bibinfo {author} {\bibfnamefont {T.}~\bibnamefont
  {Papenbrock}}\ and\ \bibinfo {author} {\bibfnamefont {H.~A.}\ \bibnamefont
  {Weidenm\"uller}},\ }\bibfield  {title} {\enquote {\bibinfo {title}
  {Effective field theory for finite systems with spontaneously broken
  symmetry},}\ }\href {\doibase 10.1103/PhysRevC.89.014334} {\bibfield
  {journal} {\bibinfo  {journal} {Phys. Rev. C}\ }\textbf {\bibinfo {volume}
  {89}},\ \bibinfo {pages} {014334} (\bibinfo {year} {2014})}\BibitemShut
  {NoStop}%
\bibitem [{\citenamefont {Peierls}\ and\ \citenamefont
  {Yoccoz}(1957)}]{peierls1957}%
  \BibitemOpen
  \bibfield  {author} {\bibinfo {author} {\bibfnamefont {R.~E.}\ \bibnamefont
  {Peierls}}\ and\ \bibinfo {author} {\bibfnamefont {J.}~\bibnamefont
  {Yoccoz}},\ }\bibfield  {title} {\enquote {\bibinfo {title} {The collective
  model of nuclear motion},}\ }\href {\doibase 10.1088/0370-1298/70/5/309}
  {\bibfield  {journal} {\bibinfo  {journal} {Proc. Phys. Soc. A}\ }\textbf
  {\bibinfo {volume} {70}},\ \bibinfo {pages} {381--387} (\bibinfo {year}
  {1957})}\BibitemShut {NoStop}%
\bibitem [{\citenamefont {Schindler}\ and\ \citenamefont
  {Phillips}(2009)}]{schindler2009}%
  \BibitemOpen
  \bibfield  {author} {\bibinfo {author} {\bibfnamefont {M.~R.}\ \bibnamefont
  {Schindler}}\ and\ \bibinfo {author} {\bibfnamefont {D.~R.}\ \bibnamefont
  {Phillips}},\ }\bibfield  {title} {\enquote {\bibinfo {title} {Bayesian
  methods for parameter estimation in effective field theories},}\ }\href
  {\doibase 10.1016/j.aop.2008.09.003} {\bibfield  {journal} {\bibinfo
  {journal} {Ann. Phys.}\ }\textbf {\bibinfo {volume} {324}},\ \bibinfo {pages}
  {682 -- 708} (\bibinfo {year} {2009})}\BibitemShut {NoStop}%
\bibitem [{\citenamefont {{Furnstahl}}\ \emph {et~al.}(2015)\citenamefont
  {{Furnstahl}}, \citenamefont {{Phillips}},\ and\ \citenamefont
  {{Wesolowski}}}]{furnstahl2014c}%
  \BibitemOpen
  \bibfield  {author} {\bibinfo {author} {\bibfnamefont {R.~J.}\ \bibnamefont
  {{Furnstahl}}}, \bibinfo {author} {\bibfnamefont {D.~R.}\ \bibnamefont
  {{Phillips}}}, \ and\ \bibinfo {author} {\bibfnamefont {S.}~\bibnamefont
  {{Wesolowski}}},\ }\bibfield  {title} {\enquote {\bibinfo {title} {{A recipe
  for EFT uncertainty quantification in nuclear physics}},}\ }\href
  {http://stacks.iop.org/0954-3899/42/i=3/a=034028} {\bibfield  {journal}
  {\bibinfo  {journal} {J. Phys. G.}\ }\textbf {\bibinfo {volume} {42}},\
  \bibinfo {pages} {034028} (\bibinfo {year} {2015})}\BibitemShut {NoStop}%
\bibitem [{\citenamefont {Coello~P\'erez}\ and\ \citenamefont
  {Papenbrock}(2015{\natexlab{b}})}]{coelloperez2015b}%
  \BibitemOpen
  \bibfield  {author} {\bibinfo {author} {\bibfnamefont {E.~A.}\ \bibnamefont
  {Coello~P\'erez}}\ and\ \bibinfo {author} {\bibfnamefont {T.}~\bibnamefont
  {Papenbrock}},\ }\bibfield  {title} {\enquote {\bibinfo {title} {Effective
  field theory for nuclear vibrations with quantified uncertainties},}\ }\href
  {\doibase 10.1103/PhysRevC.92.064309} {\bibfield  {journal} {\bibinfo
  {journal} {Phys. Rev. C}\ }\textbf {\bibinfo {volume} {92}},\ \bibinfo
  {pages} {064309} (\bibinfo {year} {2015}{\natexlab{b}})}\BibitemShut
  {NoStop}%
\bibitem [{\citenamefont {L\"obner}\ \emph {et~al.}(1970)\citenamefont
  {L\"obner}, \citenamefont {Vetter},\ and\ \citenamefont
  {H\"onig}}]{loebner1970}%
  \BibitemOpen
  \bibfield  {author} {\bibinfo {author} {\bibfnamefont {K.~E.~G.}\
  \bibnamefont {L\"obner}}, \bibinfo {author} {\bibfnamefont {M.}~\bibnamefont
  {Vetter}}, \ and\ \bibinfo {author} {\bibfnamefont {V.}~\bibnamefont
  {H\"onig}},\ }\bibfield  {title} {\enquote {\bibinfo {title} {Nuclear
  intrinsic quadrupole moments and deformation parameters},}\ }\href {\doibase
  10.1016/S0092-640X(18)30059-7} {\bibfield  {journal} {\bibinfo  {journal}
  {At. Data Nucl. Data Tables}\ }\textbf {\bibinfo {volume} {7}},\ \bibinfo
  {pages} {495--564} (\bibinfo {year} {1970})}\BibitemShut {NoStop}%
\bibitem [{\citenamefont {Dytrych}\ \emph {et~al.}(2015)\citenamefont
  {Dytrych}, \citenamefont {Hayes}, \citenamefont {Launey}, \citenamefont
  {Draayer}, \citenamefont {Maris}, \citenamefont {Vary}, \citenamefont
  {Langr},\ and\ \citenamefont {Oberhuber}}]{dytrych2015}%
  \BibitemOpen
  \bibfield  {author} {\bibinfo {author} {\bibfnamefont {T.}~\bibnamefont
  {Dytrych}}, \bibinfo {author} {\bibfnamefont {A.~C.}\ \bibnamefont {Hayes}},
  \bibinfo {author} {\bibfnamefont {K.~D.}\ \bibnamefont {Launey}}, \bibinfo
  {author} {\bibfnamefont {J.~P.}\ \bibnamefont {Draayer}}, \bibinfo {author}
  {\bibfnamefont {P.}~\bibnamefont {Maris}}, \bibinfo {author} {\bibfnamefont
  {J.~P.}\ \bibnamefont {Vary}}, \bibinfo {author} {\bibfnamefont
  {D.}~\bibnamefont {Langr}}, \ and\ \bibinfo {author} {\bibfnamefont
  {T.}~\bibnamefont {Oberhuber}},\ }\bibfield  {title} {\enquote {\bibinfo
  {title} {Electron-scattering form factors for $^{6}\mathrm{Li}$ in the ab
  initio symmetry-guided framework},}\ }\href {\doibase
  10.1103/PhysRevC.91.024326} {\bibfield  {journal} {\bibinfo  {journal} {Phys.
  Rev. C}\ }\textbf {\bibinfo {volume} {91}},\ \bibinfo {pages} {024326}
  (\bibinfo {year} {2015})}\BibitemShut {NoStop}%
\bibitem [{\citenamefont {Kanungo}\ \emph {et~al.}(2016)\citenamefont
  {Kanungo}, \citenamefont {Horiuchi}, \citenamefont {Hagen}, \citenamefont
  {Jansen}, \citenamefont {Navratil}, \citenamefont {Ameil}, \citenamefont
  {Atkinson}, \citenamefont {Ayyad}, \citenamefont {Cortina-Gil}, \citenamefont
  {Dillmann}, \citenamefont {Estrad\'e}, \citenamefont {Evdokimov},
  \citenamefont {Farinon}, \citenamefont {Geissel}, \citenamefont {Guastalla},
  \citenamefont {Janik}, \citenamefont {Kimura}, \citenamefont {Kn\"obel},
  \citenamefont {Kurcewicz}, \citenamefont {Litvinov}, \citenamefont {Marta},
  \citenamefont {Mostazo}, \citenamefont {Mukha}, \citenamefont {Nociforo},
  \citenamefont {Ong}, \citenamefont {Pietri}, \citenamefont {Prochazka},
  \citenamefont {Scheidenberger}, \citenamefont {Sitar}, \citenamefont
  {Strmen}, \citenamefont {Suzuki}, \citenamefont {Takechi}, \citenamefont
  {Tanaka}, \citenamefont {Tanihata}, \citenamefont {Terashima}, \citenamefont
  {Vargas}, \citenamefont {Weick},\ and\ \citenamefont
  {Winfield}}]{kanungo2016}%
  \BibitemOpen
  \bibfield  {author} {\bibinfo {author} {\bibfnamefont {R.}~\bibnamefont
  {Kanungo}}, \bibinfo {author} {\bibfnamefont {W.}~\bibnamefont {Horiuchi}},
  \bibinfo {author} {\bibfnamefont {G.}~\bibnamefont {Hagen}}, \bibinfo
  {author} {\bibfnamefont {G.~R.}\ \bibnamefont {Jansen}}, \bibinfo {author}
  {\bibfnamefont {P.}~\bibnamefont {Navratil}}, \bibinfo {author}
  {\bibfnamefont {F.}~\bibnamefont {Ameil}}, \bibinfo {author} {\bibfnamefont
  {J.}~\bibnamefont {Atkinson}}, \bibinfo {author} {\bibfnamefont
  {Y.}~\bibnamefont {Ayyad}}, \bibinfo {author} {\bibfnamefont
  {D.}~\bibnamefont {Cortina-Gil}}, \bibinfo {author} {\bibfnamefont
  {I.}~\bibnamefont {Dillmann}}, \bibinfo {author} {\bibfnamefont
  {A.}~\bibnamefont {Estrad\'e}}, \bibinfo {author} {\bibfnamefont
  {A.}~\bibnamefont {Evdokimov}}, \bibinfo {author} {\bibfnamefont
  {F.}~\bibnamefont {Farinon}}, \bibinfo {author} {\bibfnamefont
  {H.}~\bibnamefont {Geissel}}, \bibinfo {author} {\bibfnamefont
  {G.}~\bibnamefont {Guastalla}}, \bibinfo {author} {\bibfnamefont
  {R.}~\bibnamefont {Janik}}, \bibinfo {author} {\bibfnamefont
  {M.}~\bibnamefont {Kimura}}, \bibinfo {author} {\bibfnamefont
  {R.}~\bibnamefont {Kn\"obel}}, \bibinfo {author} {\bibfnamefont
  {J.}~\bibnamefont {Kurcewicz}}, \bibinfo {author} {\bibfnamefont {Yu.~A.}\
  \bibnamefont {Litvinov}}, \bibinfo {author} {\bibfnamefont {M.}~\bibnamefont
  {Marta}}, \bibinfo {author} {\bibfnamefont {M.}~\bibnamefont {Mostazo}},
  \bibinfo {author} {\bibfnamefont {I.}~\bibnamefont {Mukha}}, \bibinfo
  {author} {\bibfnamefont {C.}~\bibnamefont {Nociforo}}, \bibinfo {author}
  {\bibfnamefont {H.~J.}\ \bibnamefont {Ong}}, \bibinfo {author} {\bibfnamefont
  {S.}~\bibnamefont {Pietri}}, \bibinfo {author} {\bibfnamefont
  {A.}~\bibnamefont {Prochazka}}, \bibinfo {author} {\bibfnamefont
  {C.}~\bibnamefont {Scheidenberger}}, \bibinfo {author} {\bibfnamefont
  {B.}~\bibnamefont {Sitar}}, \bibinfo {author} {\bibfnamefont
  {P.}~\bibnamefont {Strmen}}, \bibinfo {author} {\bibfnamefont
  {Y.}~\bibnamefont {Suzuki}}, \bibinfo {author} {\bibfnamefont
  {M.}~\bibnamefont {Takechi}}, \bibinfo {author} {\bibfnamefont
  {J.}~\bibnamefont {Tanaka}}, \bibinfo {author} {\bibfnamefont
  {I.}~\bibnamefont {Tanihata}}, \bibinfo {author} {\bibfnamefont
  {S.}~\bibnamefont {Terashima}}, \bibinfo {author} {\bibfnamefont
  {J.}~\bibnamefont {Vargas}}, \bibinfo {author} {\bibfnamefont
  {H.}~\bibnamefont {Weick}}, \ and\ \bibinfo {author} {\bibfnamefont {J.~S.}\
  \bibnamefont {Winfield}},\ }\bibfield  {title} {\enquote {\bibinfo {title}
  {Proton distribution radii of $^{12--19}\mathrm{C}$ illuminate features of
  neutron halos},}\ }\href {\doibase 10.1103/PhysRevLett.117.102501} {\bibfield
   {journal} {\bibinfo  {journal} {Phys. Rev. Lett.}\ }\textbf {\bibinfo
  {volume} {117}},\ \bibinfo {pages} {102501} (\bibinfo {year}
  {2016})}\BibitemShut {NoStop}%
\bibitem [{\citenamefont {Duguet}\ \emph {et~al.}(2017)\citenamefont {Duguet},
  \citenamefont {Som\`a}, \citenamefont {Lecluse}, \citenamefont {Barbieri},\
  and\ \citenamefont {Navr\'atil}}]{duguet2017}%
  \BibitemOpen
  \bibfield  {author} {\bibinfo {author} {\bibfnamefont {T.}~\bibnamefont
  {Duguet}}, \bibinfo {author} {\bibfnamefont {V.}~\bibnamefont {Som\`a}},
  \bibinfo {author} {\bibfnamefont {S.}~\bibnamefont {Lecluse}}, \bibinfo
  {author} {\bibfnamefont {C.}~\bibnamefont {Barbieri}}, \ and\ \bibinfo
  {author} {\bibfnamefont {P.}~\bibnamefont {Navr\'atil}},\ }\bibfield  {title}
  {\enquote {\bibinfo {title} {Ab initio calculation of the potential bubble
  nucleus $^{34}\mathrm{Si}$},}\ }\href {\doibase 10.1103/PhysRevC.95.034319}
  {\bibfield  {journal} {\bibinfo  {journal} {Phys. Rev. C}\ }\textbf {\bibinfo
  {volume} {95}},\ \bibinfo {pages} {034319} (\bibinfo {year}
  {2017})}\BibitemShut {NoStop}%
\bibitem [{\citenamefont {Burrows}\ \emph {et~al.}(2020)\citenamefont
  {Burrows}, \citenamefont {Baker}, \citenamefont {Elster}, \citenamefont
  {Weppner}, \citenamefont {Launey}, \citenamefont {Maris},\ and\ \citenamefont
  {Popa}}]{burrows2020}%
  \BibitemOpen
  \bibfield  {author} {\bibinfo {author} {\bibfnamefont {M.}~\bibnamefont
  {Burrows}}, \bibinfo {author} {\bibfnamefont {R.~B.}\ \bibnamefont {Baker}},
  \bibinfo {author} {\bibfnamefont {Ch.}\ \bibnamefont {Elster}}, \bibinfo
  {author} {\bibfnamefont {S.~P.}\ \bibnamefont {Weppner}}, \bibinfo {author}
  {\bibfnamefont {K.~D.}\ \bibnamefont {Launey}}, \bibinfo {author}
  {\bibfnamefont {P.}~\bibnamefont {Maris}}, \ and\ \bibinfo {author}
  {\bibfnamefont {G.}~\bibnamefont {Popa}},\ }\bibfield  {title} {\enquote
  {\bibinfo {title} {Ab initio leading order effective potentials for elastic
  nucleon-nucleus scattering},}\ }\href {\doibase 10.1103/PhysRevC.102.034606}
  {\bibfield  {journal} {\bibinfo  {journal} {Phys. Rev. C}\ }\textbf {\bibinfo
  {volume} {102}},\ \bibinfo {pages} {034606} (\bibinfo {year}
  {2020})}\BibitemShut {NoStop}%
\bibitem [{\citenamefont {Brink}\ and\ \citenamefont
  {Broglia}(2005)}]{brink-broglia2005}%
  \BibitemOpen
  \bibfield  {author} {\bibinfo {author} {\bibfnamefont {D.~M.}\ \bibnamefont
  {Brink}}\ and\ \bibinfo {author} {\bibfnamefont {R.~A.}\ \bibnamefont
  {Broglia}},\ }\href@noop {} {\emph {\bibinfo {title} {Nuclear
  Superfluidity}}}\ (\bibinfo  {publisher} {Cambridge University Press},\
  \bibinfo {address} {Cambridge, UK},\ \bibinfo {year} {2005})\BibitemShut
  {NoStop}%
\bibitem [{\citenamefont {Ekstr\"om}\ \emph {et~al.}(2013)\citenamefont
  {Ekstr\"om}, \citenamefont {Baardsen}, \citenamefont {Forss\'en},
  \citenamefont {Hagen}, \citenamefont {Hjorth-Jensen}, \citenamefont {Jansen},
  \citenamefont {Machleidt}, \citenamefont {Nazarewicz}, \citenamefont
  {Papenbrock}, \citenamefont {Sarich},\ and\ \citenamefont
  {Wild}}]{ekstrom2013}%
  \BibitemOpen
  \bibfield  {author} {\bibinfo {author} {\bibfnamefont {A.}~\bibnamefont
  {Ekstr\"om}}, \bibinfo {author} {\bibfnamefont {G.}~\bibnamefont {Baardsen}},
  \bibinfo {author} {\bibfnamefont {C.}~\bibnamefont {Forss\'en}}, \bibinfo
  {author} {\bibfnamefont {G.}~\bibnamefont {Hagen}}, \bibinfo {author}
  {\bibfnamefont {M.}~\bibnamefont {Hjorth-Jensen}}, \bibinfo {author}
  {\bibfnamefont {G.~R.}\ \bibnamefont {Jansen}}, \bibinfo {author}
  {\bibfnamefont {R.}~\bibnamefont {Machleidt}}, \bibinfo {author}
  {\bibfnamefont {W.}~\bibnamefont {Nazarewicz}}, \bibinfo {author}
  {\bibfnamefont {T.}~\bibnamefont {Papenbrock}}, \bibinfo {author}
  {\bibfnamefont {J.}~\bibnamefont {Sarich}}, \ and\ \bibinfo {author}
  {\bibfnamefont {S.~M.}\ \bibnamefont {Wild}},\ }\bibfield  {title} {\enquote
  {\bibinfo {title} {Optimized chiral nucleon-nucleon interaction at
  next-to-next-to-leading order},}\ }\href {\doibase
  10.1103/PhysRevLett.110.192502} {\bibfield  {journal} {\bibinfo  {journal}
  {Phys. Rev. Lett.}\ }\textbf {\bibinfo {volume} {110}},\ \bibinfo {pages}
  {192502} (\bibinfo {year} {2013})}\BibitemShut {NoStop}%
\bibitem [{\citenamefont {Hagen}\ \emph {et~al.}(2009)\citenamefont {Hagen},
  \citenamefont {Papenbrock},\ and\ \citenamefont {Dean}}]{hagen2009a}%
  \BibitemOpen
  \bibfield  {author} {\bibinfo {author} {\bibfnamefont {G.}~\bibnamefont
  {Hagen}}, \bibinfo {author} {\bibfnamefont {T.}~\bibnamefont {Papenbrock}}, \
  and\ \bibinfo {author} {\bibfnamefont {D.~J.}\ \bibnamefont {Dean}},\
  }\bibfield  {title} {\enquote {\bibinfo {title} {Solution of the
  center-of-mass problem in nuclear structure calculations},}\ }\href {\doibase
  10.1103/PhysRevLett.103.062503} {\bibfield  {journal} {\bibinfo  {journal}
  {Phys. Rev. Lett.}\ }\textbf {\bibinfo {volume} {103}},\ \bibinfo {pages}
  {062503} (\bibinfo {year} {2009})}\BibitemShut {NoStop}%
\bibitem [{\citenamefont {Hagen}\ \emph {et~al.}(2010)\citenamefont {Hagen},
  \citenamefont {Papenbrock}, \citenamefont {Dean},\ and\ \citenamefont
  {Hjorth-Jensen}}]{hagen2010b}%
  \BibitemOpen
  \bibfield  {author} {\bibinfo {author} {\bibfnamefont {G.}~\bibnamefont
  {Hagen}}, \bibinfo {author} {\bibfnamefont {T.}~\bibnamefont {Papenbrock}},
  \bibinfo {author} {\bibfnamefont {D.~J.}\ \bibnamefont {Dean}}, \ and\
  \bibinfo {author} {\bibfnamefont {M.}~\bibnamefont {Hjorth-Jensen}},\
  }\bibfield  {title} {\enquote {\bibinfo {title} {\textit{Ab initio}
  coupled-cluster approach to nuclear structure with modern nucleon-nucleon
  interactions},}\ }\href {\doibase 10.1103/PhysRevC.82.034330} {\bibfield
  {journal} {\bibinfo  {journal} {Phys. Rev. C}\ }\textbf {\bibinfo {volume}
  {82}},\ \bibinfo {pages} {034330} (\bibinfo {year} {2010})}\BibitemShut
  {NoStop}%
\bibitem [{\citenamefont {Parzuchowski}\ \emph {et~al.}(2017)\citenamefont
  {Parzuchowski}, \citenamefont {Stroberg}, \citenamefont {Navr\'atil},
  \citenamefont {Hergert},\ and\ \citenamefont {Bogner}}]{parzuchowski2017}%
  \BibitemOpen
  \bibfield  {author} {\bibinfo {author} {\bibfnamefont {N.~M.}\ \bibnamefont
  {Parzuchowski}}, \bibinfo {author} {\bibfnamefont {S.~R.}\ \bibnamefont
  {Stroberg}}, \bibinfo {author} {\bibfnamefont {P.}~\bibnamefont
  {Navr\'atil}}, \bibinfo {author} {\bibfnamefont {H.}~\bibnamefont {Hergert}},
  \ and\ \bibinfo {author} {\bibfnamefont {S.~K.}\ \bibnamefont {Bogner}},\
  }\bibfield  {title} {\enquote {\bibinfo {title} {Ab initio electromagnetic
  observables with the in-medium similarity renormalization group},}\ }\href
  {\doibase 10.1103/PhysRevC.96.034324} {\bibfield  {journal} {\bibinfo
  {journal} {Phys. Rev. C}\ }\textbf {\bibinfo {volume} {96}},\ \bibinfo
  {pages} {034324} (\bibinfo {year} {2017})}\BibitemShut {NoStop}%
\bibitem [{\citenamefont {Mayer}\ and\ \citenamefont
  {Jensen}(1955)}]{mayer1955}%
  \BibitemOpen
  \bibfield  {author} {\bibinfo {author} {\bibfnamefont {M.~G.}\ \bibnamefont
  {Mayer}}\ and\ \bibinfo {author} {\bibfnamefont {J.~H.~D.}\ \bibnamefont
  {Jensen}},\ }\href@noop {} {\emph {\bibinfo {title} {Elementary Theory of
  Nuclear Shell Structure}}}\ (\bibinfo  {publisher} {John Wiley \& Sons},\
  \bibinfo {address} {New York},\ \bibinfo {year} {1955})\BibitemShut {NoStop}%
\bibitem [{\citenamefont {{Koszor{\'u}s}}\ \emph {et~al.}(2021)\citenamefont
  {{Koszor{\'u}s}}, \citenamefont {{Yang}}, \citenamefont {{Jiang}},
  \citenamefont {{Novario}}, \citenamefont {{Bai}}, \citenamefont {{Billowes}},
  \citenamefont {{Binnersley}}, \citenamefont {{Bissell}}, \citenamefont
  {{Cocolios}}, \citenamefont {{Cooper}}, \citenamefont {{de Groote}},
  \citenamefont {{Ekstr{\"o}m}}, \citenamefont {{Flanagan}}, \citenamefont
  {{Forss{\'e}n}}, \citenamefont {{Franchoo}}, \citenamefont {{Ruiz}},
  \citenamefont {{Gustafsson}}, \citenamefont {{Hagen}}, \citenamefont
  {{Jansen}}, \citenamefont {{Kanellakopoulos}}, \citenamefont {{Kortelainen}},
  \citenamefont {{Nazarewicz}}, \citenamefont {{Neyens}}, \citenamefont
  {{Papenbrock}}, \citenamefont {{Reinhard}}, \citenamefont {{Ricketts}},
  \citenamefont {{Sahoo}}, \citenamefont {{Vernon}},\ and\ \citenamefont
  {{Wilkins}}}]{koszorus2021}%
  \BibitemOpen
  \bibfield  {author} {\bibinfo {author} {\bibfnamefont {{\'A}.}~\bibnamefont
  {{Koszor{\'u}s}}}, \bibinfo {author} {\bibfnamefont {X.~F.}\ \bibnamefont
  {{Yang}}}, \bibinfo {author} {\bibfnamefont {W.~G.}\ \bibnamefont {{Jiang}}},
  \bibinfo {author} {\bibfnamefont {S.~J.}\ \bibnamefont {{Novario}}}, \bibinfo
  {author} {\bibfnamefont {S.~W.}\ \bibnamefont {{Bai}}}, \bibinfo {author}
  {\bibfnamefont {J.}~\bibnamefont {{Billowes}}}, \bibinfo {author}
  {\bibfnamefont {C.~L.}\ \bibnamefont {{Binnersley}}}, \bibinfo {author}
  {\bibfnamefont {M.~L.}\ \bibnamefont {{Bissell}}}, \bibinfo {author}
  {\bibfnamefont {T.~E.}\ \bibnamefont {{Cocolios}}}, \bibinfo {author}
  {\bibfnamefont {B.~S.}\ \bibnamefont {{Cooper}}}, \bibinfo {author}
  {\bibfnamefont {R.~P.}\ \bibnamefont {{de Groote}}}, \bibinfo {author}
  {\bibfnamefont {A.}~\bibnamefont {{Ekstr{\"o}m}}}, \bibinfo {author}
  {\bibfnamefont {K.~T.}\ \bibnamefont {{Flanagan}}}, \bibinfo {author}
  {\bibfnamefont {C.}~\bibnamefont {{Forss{\'e}n}}}, \bibinfo {author}
  {\bibfnamefont {S.}~\bibnamefont {{Franchoo}}}, \bibinfo {author}
  {\bibfnamefont {R.~F.~Garcia}\ \bibnamefont {{Ruiz}}}, \bibinfo {author}
  {\bibfnamefont {F.~P.}\ \bibnamefont {{Gustafsson}}}, \bibinfo {author}
  {\bibfnamefont {G.}~\bibnamefont {{Hagen}}}, \bibinfo {author} {\bibfnamefont
  {G.~R.}\ \bibnamefont {{Jansen}}}, \bibinfo {author} {\bibfnamefont
  {A.}~\bibnamefont {{Kanellakopoulos}}}, \bibinfo {author} {\bibfnamefont
  {M.}~\bibnamefont {{Kortelainen}}}, \bibinfo {author} {\bibfnamefont
  {W.}~\bibnamefont {{Nazarewicz}}}, \bibinfo {author} {\bibfnamefont
  {G.}~\bibnamefont {{Neyens}}}, \bibinfo {author} {\bibfnamefont
  {T.}~\bibnamefont {{Papenbrock}}}, \bibinfo {author} {\bibfnamefont {P.~G.}\
  \bibnamefont {{Reinhard}}}, \bibinfo {author} {\bibfnamefont {C.~M.}\
  \bibnamefont {{Ricketts}}}, \bibinfo {author} {\bibfnamefont {B.~K.}\
  \bibnamefont {{Sahoo}}}, \bibinfo {author} {\bibfnamefont {A.~R.}\
  \bibnamefont {{Vernon}}}, \ and\ \bibinfo {author} {\bibfnamefont {S.~G.}\
  \bibnamefont {{Wilkins}}},\ }\bibfield  {title} {\enquote {\bibinfo {title}
  {{Charge radii of exotic potassium isotopes challenge nuclear theory and the
  magic character of N = 32}},}\ }\href {\doibase 10.1038/s41567-020-01136-5}
  {\bibfield  {journal} {\bibinfo  {journal} {Nature Physics}\ }\textbf
  {\bibinfo {volume} {17}},\ \bibinfo {pages} {439--443} (\bibinfo {year}
  {2021})}\BibitemShut {NoStop}%
\bibitem [{\citenamefont {Angeli}\ and\ \citenamefont
  {Marinova}(2013)}]{angeli2013}%
  \BibitemOpen
  \bibfield  {author} {\bibinfo {author} {\bibfnamefont {I.}~\bibnamefont
  {Angeli}}\ and\ \bibinfo {author} {\bibfnamefont {K.P.}\ \bibnamefont
  {Marinova}},\ }\bibfield  {title} {\enquote {\bibinfo {title} {Table of
  experimental nuclear ground state charge radii: An update},}\ }\href
  {\doibase 10.1016/j.adt.2011.12.006} {\bibfield  {journal} {\bibinfo
  {journal} {At. Data Nucl. Data Tables}\ }\textbf {\bibinfo {volume} {99}},\
  \bibinfo {pages} {69 -- 95} (\bibinfo {year} {2013})}\BibitemShut {NoStop}%
\bibitem [{\citenamefont {Motobayashi}\ \emph {et~al.}(1995)\citenamefont
  {Motobayashi}, \citenamefont {Ikeda}, \citenamefont {Ieki}, \citenamefont
  {Inoue}, \citenamefont {Iwasa}, \citenamefont {Kikuchi}, \citenamefont
  {Kurokawa}, \citenamefont {Moriya}, \citenamefont {Ogawa}, \citenamefont
  {Murakami}, \citenamefont {Shimoura}, \citenamefont {Yanagisawa},
  \citenamefont {Nakamura}, \citenamefont {Watanabe}, \citenamefont {Ishihara},
  \citenamefont {Teranishi}, \citenamefont {Okuno},\ and\ \citenamefont
  {Casten}}]{motobayashi1995}%
  \BibitemOpen
  \bibfield  {author} {\bibinfo {author} {\bibfnamefont {T.}~\bibnamefont
  {Motobayashi}}, \bibinfo {author} {\bibfnamefont {Y.}~\bibnamefont {Ikeda}},
  \bibinfo {author} {\bibfnamefont {K.}~\bibnamefont {Ieki}}, \bibinfo {author}
  {\bibfnamefont {M.}~\bibnamefont {Inoue}}, \bibinfo {author} {\bibfnamefont
  {N.}~\bibnamefont {Iwasa}}, \bibinfo {author} {\bibfnamefont
  {T.}~\bibnamefont {Kikuchi}}, \bibinfo {author} {\bibfnamefont
  {M.}~\bibnamefont {Kurokawa}}, \bibinfo {author} {\bibfnamefont
  {S.}~\bibnamefont {Moriya}}, \bibinfo {author} {\bibfnamefont
  {S.}~\bibnamefont {Ogawa}}, \bibinfo {author} {\bibfnamefont
  {H.}~\bibnamefont {Murakami}}, \bibinfo {author} {\bibfnamefont
  {S.}~\bibnamefont {Shimoura}}, \bibinfo {author} {\bibfnamefont
  {Y.}~\bibnamefont {Yanagisawa}}, \bibinfo {author} {\bibfnamefont
  {T.}~\bibnamefont {Nakamura}}, \bibinfo {author} {\bibfnamefont
  {Y.}~\bibnamefont {Watanabe}}, \bibinfo {author} {\bibfnamefont
  {M.}~\bibnamefont {Ishihara}}, \bibinfo {author} {\bibfnamefont
  {T.}~\bibnamefont {Teranishi}}, \bibinfo {author} {\bibfnamefont
  {H.}~\bibnamefont {Okuno}}, \ and\ \bibinfo {author} {\bibfnamefont {R.~F.}\
  \bibnamefont {Casten}},\ }\bibfield  {title} {\enquote {\bibinfo {title}
  {Large deformation of the very neutron-rich nucleus 32mg from
  intermediate-energy coulomb excitation},}\ }\href {\doibase
  10.1016/0370-2693(95)00012-A} {\bibfield  {journal} {\bibinfo  {journal}
  {Phys. Lett. B}\ }\textbf {\bibinfo {volume} {346}},\ \bibinfo {pages} {9 --
  14} (\bibinfo {year} {1995})}\BibitemShut {NoStop}%
\bibitem [{\citenamefont {Baumann}\ \emph {et~al.}(2007)\citenamefont
  {Baumann}, \citenamefont {Amthor}, \citenamefont {Bazin}, \citenamefont
  {Brown}, \citenamefont {Folden}, \citenamefont {Gade}, \citenamefont
  {Ginter}, \citenamefont {Hausmann}, \citenamefont {Matos}, \citenamefont
  {Morrissey}, \citenamefont {Portillo}, \citenamefont {Schiller},
  \citenamefont {Sherrill}, \citenamefont {Stolz}, \citenamefont {Tarasov},\
  and\ \citenamefont {Thoennessen}}]{baumann2007}%
  \BibitemOpen
  \bibfield  {author} {\bibinfo {author} {\bibfnamefont {T.}~\bibnamefont
  {Baumann}}, \bibinfo {author} {\bibfnamefont {A.~M.}\ \bibnamefont {Amthor}},
  \bibinfo {author} {\bibfnamefont {D.}~\bibnamefont {Bazin}}, \bibinfo
  {author} {\bibfnamefont {B.~A.}\ \bibnamefont {Brown}}, \bibinfo {author}
  {\bibfnamefont {C.~M.}\ \bibnamefont {Folden}}, \bibinfo {author}
  {\bibfnamefont {A.}~\bibnamefont {Gade}}, \bibinfo {author} {\bibfnamefont
  {T.~N.}\ \bibnamefont {Ginter}}, \bibinfo {author} {\bibfnamefont
  {M.}~\bibnamefont {Hausmann}}, \bibinfo {author} {\bibfnamefont
  {M.}~\bibnamefont {Matos}}, \bibinfo {author} {\bibfnamefont {D.~J.}\
  \bibnamefont {Morrissey}}, \bibinfo {author} {\bibfnamefont {M.}~\bibnamefont
  {Portillo}}, \bibinfo {author} {\bibfnamefont {A.}~\bibnamefont {Schiller}},
  \bibinfo {author} {\bibfnamefont {B.~M.}\ \bibnamefont {Sherrill}}, \bibinfo
  {author} {\bibfnamefont {A.}~\bibnamefont {Stolz}}, \bibinfo {author}
  {\bibfnamefont {O.~B.}\ \bibnamefont {Tarasov}}, \ and\ \bibinfo {author}
  {\bibfnamefont {M.}~\bibnamefont {Thoennessen}},\ }\bibfield  {title}
  {\enquote {\bibinfo {title} {Discovery of 40mg and 42al suggests neutron
  drip-line slant towards heavier isotopes},}\ }\href {\doibase
  10.1038/nature06213} {\bibfield  {journal} {\bibinfo  {journal} {Nature}\
  }\textbf {\bibinfo {volume} {449}},\ \bibinfo {pages} {1022 -- 1024}
  (\bibinfo {year} {2007})}\BibitemShut {NoStop}%
\bibitem [{\citenamefont {Crawford}\ \emph {et~al.}(2019)\citenamefont
  {Crawford}, \citenamefont {Fallon}, \citenamefont {Macchiavelli},
  \citenamefont {Doornenbal}, \citenamefont {Aoi}, \citenamefont {Browne},
  \citenamefont {Campbell}, \citenamefont {Chen}, \citenamefont {Clark},
  \citenamefont {Cort\'es}, \citenamefont {Cromaz}, \citenamefont {Ideguchi},
  \citenamefont {Jones}, \citenamefont {Kanungo}, \citenamefont {MacCormick},
  \citenamefont {Momiyama}, \citenamefont {Murray}, \citenamefont {Niikura},
  \citenamefont {Paschalis}, \citenamefont {Petri}, \citenamefont {Sakurai},
  \citenamefont {Salathe}, \citenamefont {Schrock}, \citenamefont
  {Steppenbeck}, \citenamefont {Takeuchi}, \citenamefont {Tanaka},
  \citenamefont {Taniuchi}, \citenamefont {Wang},\ and\ \citenamefont
  {Wimmer}}]{crawford2019}%
  \BibitemOpen
  \bibfield  {author} {\bibinfo {author} {\bibfnamefont {H.~L.}\ \bibnamefont
  {Crawford}}, \bibinfo {author} {\bibfnamefont {P.}~\bibnamefont {Fallon}},
  \bibinfo {author} {\bibfnamefont {A.~O.}\ \bibnamefont {Macchiavelli}},
  \bibinfo {author} {\bibfnamefont {P.}~\bibnamefont {Doornenbal}}, \bibinfo
  {author} {\bibfnamefont {N.}~\bibnamefont {Aoi}}, \bibinfo {author}
  {\bibfnamefont {F.}~\bibnamefont {Browne}}, \bibinfo {author} {\bibfnamefont
  {C.~M.}\ \bibnamefont {Campbell}}, \bibinfo {author} {\bibfnamefont
  {S.}~\bibnamefont {Chen}}, \bibinfo {author} {\bibfnamefont {R.~M.}\
  \bibnamefont {Clark}}, \bibinfo {author} {\bibfnamefont {M.~L.}\ \bibnamefont
  {Cort\'es}}, \bibinfo {author} {\bibfnamefont {M.}~\bibnamefont {Cromaz}},
  \bibinfo {author} {\bibfnamefont {E.}~\bibnamefont {Ideguchi}}, \bibinfo
  {author} {\bibfnamefont {M.~D.}\ \bibnamefont {Jones}}, \bibinfo {author}
  {\bibfnamefont {R.}~\bibnamefont {Kanungo}}, \bibinfo {author} {\bibfnamefont
  {M.}~\bibnamefont {MacCormick}}, \bibinfo {author} {\bibfnamefont
  {S.}~\bibnamefont {Momiyama}}, \bibinfo {author} {\bibfnamefont
  {I.}~\bibnamefont {Murray}}, \bibinfo {author} {\bibfnamefont
  {M.}~\bibnamefont {Niikura}}, \bibinfo {author} {\bibfnamefont
  {S.}~\bibnamefont {Paschalis}}, \bibinfo {author} {\bibfnamefont
  {M.}~\bibnamefont {Petri}}, \bibinfo {author} {\bibfnamefont
  {H.}~\bibnamefont {Sakurai}}, \bibinfo {author} {\bibfnamefont
  {M.}~\bibnamefont {Salathe}}, \bibinfo {author} {\bibfnamefont
  {P.}~\bibnamefont {Schrock}}, \bibinfo {author} {\bibfnamefont
  {D.}~\bibnamefont {Steppenbeck}}, \bibinfo {author} {\bibfnamefont
  {S.}~\bibnamefont {Takeuchi}}, \bibinfo {author} {\bibfnamefont {Y.~K.}\
  \bibnamefont {Tanaka}}, \bibinfo {author} {\bibfnamefont {R.}~\bibnamefont
  {Taniuchi}}, \bibinfo {author} {\bibfnamefont {H.}~\bibnamefont {Wang}}, \
  and\ \bibinfo {author} {\bibfnamefont {K.}~\bibnamefont {Wimmer}},\
  }\bibfield  {title} {\enquote {\bibinfo {title} {First spectroscopy of the
  near drip-line nucleus $^{40}\mathrm{Mg}$},}\ }\href {\doibase
  10.1103/PhysRevLett.122.052501} {\bibfield  {journal} {\bibinfo  {journal}
  {Phys. Rev. Lett.}\ }\textbf {\bibinfo {volume} {122}},\ \bibinfo {pages}
  {052501} (\bibinfo {year} {2019})}\BibitemShut {NoStop}%
\bibitem [{\citenamefont {Yordanov}\ \emph {et~al.}(2012)\citenamefont
  {Yordanov}, \citenamefont {Bissell}, \citenamefont {Blaum}, \citenamefont
  {De~Rydt}, \citenamefont {Geppert}, \citenamefont {Kowalska}, \citenamefont
  {Kr\"amer}, \citenamefont {Kreim}, \citenamefont {Krieger}, \citenamefont
  {Lievens}, \citenamefont {Neff}, \citenamefont {Neugart}, \citenamefont
  {Neyens}, \citenamefont {N\"ortersh\"auser}, \citenamefont {S\'anchez},\ and\
  \citenamefont {Vingerhoets}}]{yordanov2012}%
  \BibitemOpen
  \bibfield  {author} {\bibinfo {author} {\bibfnamefont {D.~T.}\ \bibnamefont
  {Yordanov}}, \bibinfo {author} {\bibfnamefont {M.~L.}\ \bibnamefont
  {Bissell}}, \bibinfo {author} {\bibfnamefont {K.}~\bibnamefont {Blaum}},
  \bibinfo {author} {\bibfnamefont {M.}~\bibnamefont {De~Rydt}}, \bibinfo
  {author} {\bibfnamefont {Ch.}\ \bibnamefont {Geppert}}, \bibinfo {author}
  {\bibfnamefont {M.}~\bibnamefont {Kowalska}}, \bibinfo {author}
  {\bibfnamefont {J.}~\bibnamefont {Kr\"amer}}, \bibinfo {author}
  {\bibfnamefont {K.}~\bibnamefont {Kreim}}, \bibinfo {author} {\bibfnamefont
  {A.}~\bibnamefont {Krieger}}, \bibinfo {author} {\bibfnamefont
  {P.}~\bibnamefont {Lievens}}, \bibinfo {author} {\bibfnamefont
  {T.}~\bibnamefont {Neff}}, \bibinfo {author} {\bibfnamefont {R.}~\bibnamefont
  {Neugart}}, \bibinfo {author} {\bibfnamefont {G.}~\bibnamefont {Neyens}},
  \bibinfo {author} {\bibfnamefont {W.}~\bibnamefont {N\"ortersh\"auser}},
  \bibinfo {author} {\bibfnamefont {R.}~\bibnamefont {S\'anchez}}, \ and\
  \bibinfo {author} {\bibfnamefont {P.}~\bibnamefont {Vingerhoets}},\
  }\bibfield  {title} {\enquote {\bibinfo {title} {Nuclear charge radii of
  $^{21\mathrm{\text{\ensuremath{-}}}32}\mathrm{Mg}$},}\ }\href {\doibase
  10.1103/PhysRevLett.108.042504} {\bibfield  {journal} {\bibinfo  {journal}
  {Phys. Rev. Lett.}\ }\textbf {\bibinfo {volume} {108}},\ \bibinfo {pages}
  {042504} (\bibinfo {year} {2012})}\BibitemShut {NoStop}%
\bibitem [{\citenamefont {Fossez}\ \emph {et~al.}(2016)\citenamefont {Fossez},
  \citenamefont {Rotureau}, \citenamefont {Michel}, \citenamefont {Liu},\ and\
  \citenamefont {Nazarewicz}}]{fossez2016}%
  \BibitemOpen
  \bibfield  {author} {\bibinfo {author} {\bibfnamefont {K.}~\bibnamefont
  {Fossez}}, \bibinfo {author} {\bibfnamefont {J.}~\bibnamefont {Rotureau}},
  \bibinfo {author} {\bibfnamefont {N.}~\bibnamefont {Michel}}, \bibinfo
  {author} {\bibfnamefont {Quan}\ \bibnamefont {Liu}}, \ and\ \bibinfo {author}
  {\bibfnamefont {W.}~\bibnamefont {Nazarewicz}},\ }\bibfield  {title}
  {\enquote {\bibinfo {title} {Single-particle and collective motion in unbound
  deformed $^{39}\mathrm{Mg}$},}\ }\href {\doibase 10.1103/PhysRevC.94.054302}
  {\bibfield  {journal} {\bibinfo  {journal} {Phys. Rev. C}\ }\textbf {\bibinfo
  {volume} {94}},\ \bibinfo {pages} {054302} (\bibinfo {year}
  {2016})}\BibitemShut {NoStop}%
\bibitem [{\citenamefont {Tsunoda}\ \emph {et~al.}(2017)\citenamefont
  {Tsunoda}, \citenamefont {Otsuka}, \citenamefont {Shimizu}, \citenamefont
  {Hjorth-Jensen}, \citenamefont {Takayanagi},\ and\ \citenamefont
  {Suzuki}}]{tsunoda2017}%
  \BibitemOpen
  \bibfield  {author} {\bibinfo {author} {\bibfnamefont {Naofumi}\ \bibnamefont
  {Tsunoda}}, \bibinfo {author} {\bibfnamefont {Takaharu}\ \bibnamefont
  {Otsuka}}, \bibinfo {author} {\bibfnamefont {Noritaka}\ \bibnamefont
  {Shimizu}}, \bibinfo {author} {\bibfnamefont {Morten}\ \bibnamefont
  {Hjorth-Jensen}}, \bibinfo {author} {\bibfnamefont {Kazuo}\ \bibnamefont
  {Takayanagi}}, \ and\ \bibinfo {author} {\bibfnamefont {Toshio}\ \bibnamefont
  {Suzuki}},\ }\bibfield  {title} {\enquote {\bibinfo {title} {Exotic
  neutron-rich medium-mass nuclei with realistic nuclear forces},}\ }\href
  {\doibase 10.1103/PhysRevC.95.021304} {\bibfield  {journal} {\bibinfo
  {journal} {Phys. Rev. C}\ }\textbf {\bibinfo {volume} {95}},\ \bibinfo
  {pages} {021304} (\bibinfo {year} {2017})}\BibitemShut {NoStop}%
\bibitem [{\citenamefont {Miyagi}\ \emph {et~al.}(2020)\citenamefont {Miyagi},
  \citenamefont {Stroberg}, \citenamefont {Holt},\ and\ \citenamefont
  {Shimizu}}]{miyagi2020}%
  \BibitemOpen
  \bibfield  {author} {\bibinfo {author} {\bibfnamefont {T.}~\bibnamefont
  {Miyagi}}, \bibinfo {author} {\bibfnamefont {S.~R.}\ \bibnamefont
  {Stroberg}}, \bibinfo {author} {\bibfnamefont {J.~D.}\ \bibnamefont {Holt}},
  \ and\ \bibinfo {author} {\bibfnamefont {N.}~\bibnamefont {Shimizu}},\
  }\bibfield  {title} {\enquote {\bibinfo {title} {Ab initio multishell
  valence-space hamiltonians and the island of inversion},}\ }\href {\doibase
  10.1103/PhysRevC.102.034320} {\bibfield  {journal} {\bibinfo  {journal}
  {Phys. Rev. C}\ }\textbf {\bibinfo {volume} {102}},\ \bibinfo {pages}
  {034320} (\bibinfo {year} {2020})}\BibitemShut {NoStop}%
\bibitem [{\citenamefont {Warburton}\ \emph {et~al.}(1990)\citenamefont
  {Warburton}, \citenamefont {Becker},\ and\ \citenamefont
  {Brown}}]{warburton1990}%
  \BibitemOpen
  \bibfield  {author} {\bibinfo {author} {\bibfnamefont {E.~K.}\ \bibnamefont
  {Warburton}}, \bibinfo {author} {\bibfnamefont {J.~A.}\ \bibnamefont
  {Becker}}, \ and\ \bibinfo {author} {\bibfnamefont {B.~A.}\ \bibnamefont
  {Brown}},\ }\bibfield  {title} {\enquote {\bibinfo {title} {Mass systematics
  for a=29--44 nuclei: The deformed a\ensuremath{\sim}32 region},}\ }\href
  {\doibase 10.1103/PhysRevC.41.1147} {\bibfield  {journal} {\bibinfo
  {journal} {Phys. Rev. C}\ }\textbf {\bibinfo {volume} {41}},\ \bibinfo
  {pages} {1147--1166} (\bibinfo {year} {1990})}\BibitemShut {NoStop}%
\bibitem [{\citenamefont {Iwasaki}\ \emph {et~al.}(2001)\citenamefont
  {Iwasaki}, \citenamefont {Motobayashi}, \citenamefont {Sakurai},
  \citenamefont {Yoneda}, \citenamefont {Gomi}, \citenamefont {Aoi},
  \citenamefont {Fukuda}, \citenamefont {F\"ul\"op}, \citenamefont {Futakami},
  \citenamefont {Gacsi}, \citenamefont {Higurashi}, \citenamefont {Imai},
  \citenamefont {Iwasa}, \citenamefont {Kubo}, \citenamefont {Kunibu},
  \citenamefont {Kurokawa}, \citenamefont {Liu}, \citenamefont {Minemura},
  \citenamefont {Saito}, \citenamefont {Serata}, \citenamefont {Shimoura},
  \citenamefont {Takeuchi}, \citenamefont {Watanabe}, \citenamefont {Yamada},
  \citenamefont {Yanagisawa},\ and\ \citenamefont {Ishihara}}]{iwasaki2001}%
  \BibitemOpen
  \bibfield  {author} {\bibinfo {author} {\bibfnamefont {H.}~\bibnamefont
  {Iwasaki}}, \bibinfo {author} {\bibfnamefont {T.}~\bibnamefont
  {Motobayashi}}, \bibinfo {author} {\bibfnamefont {H.}~\bibnamefont
  {Sakurai}}, \bibinfo {author} {\bibfnamefont {K.}~\bibnamefont {Yoneda}},
  \bibinfo {author} {\bibfnamefont {T.}~\bibnamefont {Gomi}}, \bibinfo {author}
  {\bibfnamefont {N.}~\bibnamefont {Aoi}}, \bibinfo {author} {\bibfnamefont
  {N.}~\bibnamefont {Fukuda}}, \bibinfo {author} {\bibfnamefont
  {Z.}~\bibnamefont {F\"ul\"op}}, \bibinfo {author} {\bibfnamefont
  {U.}~\bibnamefont {Futakami}}, \bibinfo {author} {\bibfnamefont
  {Z.}~\bibnamefont {Gacsi}}, \bibinfo {author} {\bibfnamefont
  {Y.}~\bibnamefont {Higurashi}}, \bibinfo {author} {\bibfnamefont
  {N.}~\bibnamefont {Imai}}, \bibinfo {author} {\bibfnamefont {N.}~\bibnamefont
  {Iwasa}}, \bibinfo {author} {\bibfnamefont {T.}~\bibnamefont {Kubo}},
  \bibinfo {author} {\bibfnamefont {M.}~\bibnamefont {Kunibu}}, \bibinfo
  {author} {\bibfnamefont {M.}~\bibnamefont {Kurokawa}}, \bibinfo {author}
  {\bibfnamefont {Z.}~\bibnamefont {Liu}}, \bibinfo {author} {\bibfnamefont
  {T.}~\bibnamefont {Minemura}}, \bibinfo {author} {\bibfnamefont
  {A.}~\bibnamefont {Saito}}, \bibinfo {author} {\bibfnamefont
  {M.}~\bibnamefont {Serata}}, \bibinfo {author} {\bibfnamefont
  {S.}~\bibnamefont {Shimoura}}, \bibinfo {author} {\bibfnamefont
  {S.}~\bibnamefont {Takeuchi}}, \bibinfo {author} {\bibfnamefont {Y.~X.}\
  \bibnamefont {Watanabe}}, \bibinfo {author} {\bibfnamefont {K.}~\bibnamefont
  {Yamada}}, \bibinfo {author} {\bibfnamefont {Y.}~\bibnamefont {Yanagisawa}},
  \ and\ \bibinfo {author} {\bibfnamefont {M.}~\bibnamefont {Ishihara}},\
  }\bibfield  {title} {\enquote {\bibinfo {title} {Large collectivity of
  34mg},}\ }\href {\doibase 10.1016/S0370-2693(01)01244-8} {\bibfield
  {journal} {\bibinfo  {journal} {Phys. Lett. B}\ }\textbf {\bibinfo {volume}
  {522}},\ \bibinfo {pages} {227--232} (\bibinfo {year} {2001})}\BibitemShut
  {NoStop}%
\bibitem [{\citenamefont {Church}\ \emph {et~al.}(2005)\citenamefont {Church},
  \citenamefont {Campbell}, \citenamefont {Dinca}, \citenamefont {Enders},
  \citenamefont {Gade}, \citenamefont {Glasmacher}, \citenamefont {Hu},
  \citenamefont {Janssens}, \citenamefont {Mueller}, \citenamefont {Olliver},
  \citenamefont {Perry}, \citenamefont {Riley},\ and\ \citenamefont
  {Yurkewicz}}]{church2005}%
  \BibitemOpen
  \bibfield  {author} {\bibinfo {author} {\bibfnamefont {J.~A.}\ \bibnamefont
  {Church}}, \bibinfo {author} {\bibfnamefont {C.~M.}\ \bibnamefont
  {Campbell}}, \bibinfo {author} {\bibfnamefont {D.-C.}\ \bibnamefont {Dinca}},
  \bibinfo {author} {\bibfnamefont {J.}~\bibnamefont {Enders}}, \bibinfo
  {author} {\bibfnamefont {A.}~\bibnamefont {Gade}}, \bibinfo {author}
  {\bibfnamefont {T.}~\bibnamefont {Glasmacher}}, \bibinfo {author}
  {\bibfnamefont {Z.}~\bibnamefont {Hu}}, \bibinfo {author} {\bibfnamefont
  {R.~V.~F.}\ \bibnamefont {Janssens}}, \bibinfo {author} {\bibfnamefont
  {W.~F.}\ \bibnamefont {Mueller}}, \bibinfo {author} {\bibfnamefont
  {H.}~\bibnamefont {Olliver}}, \bibinfo {author} {\bibfnamefont {B.~C.}\
  \bibnamefont {Perry}}, \bibinfo {author} {\bibfnamefont {L.~A.}\ \bibnamefont
  {Riley}}, \ and\ \bibinfo {author} {\bibfnamefont {K.~L.}\ \bibnamefont
  {Yurkewicz}},\ }\bibfield  {title} {\enquote {\bibinfo {title} {Measurement
  of $e2$ transition strengths in $^{32,34}\mathrm{Mg}$},}\ }\href {\doibase
  10.1103/PhysRevC.72.054320} {\bibfield  {journal} {\bibinfo  {journal} {Phys.
  Rev. C}\ }\textbf {\bibinfo {volume} {72}},\ \bibinfo {pages} {054320}
  (\bibinfo {year} {2005})}\BibitemShut {NoStop}%
\bibitem [{\citenamefont {Elekes}\ \emph {et~al.}(2006)\citenamefont {Elekes},
  \citenamefont {Dombr\'adi}, \citenamefont {Saito}, \citenamefont {Aoi},
  \citenamefont {Baba}, \citenamefont {Demichi}, \citenamefont {F\"ul\"op},
  \citenamefont {Gibelin}, \citenamefont {Gomi}, \citenamefont {Hasegawa},
  \citenamefont {Imai}, \citenamefont {Ishihara}, \citenamefont {Iwasaki},
  \citenamefont {Kanno}, \citenamefont {Kawai}, \citenamefont {Kishida},
  \citenamefont {Kubo}, \citenamefont {Kurita}, \citenamefont {Matsuyama},
  \citenamefont {Michimasa}, \citenamefont {Minemura}, \citenamefont
  {Motobayashi}, \citenamefont {Notani}, \citenamefont {Ohnishi}, \citenamefont
  {Ong}, \citenamefont {Ota}, \citenamefont {Ozawa}, \citenamefont {Sakai},
  \citenamefont {Sakurai}, \citenamefont {Shimoura}, \citenamefont {Takeshita},
  \citenamefont {Takeuchi}, \citenamefont {Tamaki}, \citenamefont {Togano},
  \citenamefont {Yamada}, \citenamefont {Yanagisawa},\ and\ \citenamefont
  {Yoneda}}]{elekes2006}%
  \BibitemOpen
  \bibfield  {author} {\bibinfo {author} {\bibfnamefont {Z.}~\bibnamefont
  {Elekes}}, \bibinfo {author} {\bibfnamefont {Zs.}\ \bibnamefont
  {Dombr\'adi}}, \bibinfo {author} {\bibfnamefont {A.}~\bibnamefont {Saito}},
  \bibinfo {author} {\bibfnamefont {N.}~\bibnamefont {Aoi}}, \bibinfo {author}
  {\bibfnamefont {H.}~\bibnamefont {Baba}}, \bibinfo {author} {\bibfnamefont
  {K.}~\bibnamefont {Demichi}}, \bibinfo {author} {\bibfnamefont {Zs.}\
  \bibnamefont {F\"ul\"op}}, \bibinfo {author} {\bibfnamefont {J.}~\bibnamefont
  {Gibelin}}, \bibinfo {author} {\bibfnamefont {T.}~\bibnamefont {Gomi}},
  \bibinfo {author} {\bibfnamefont {H.}~\bibnamefont {Hasegawa}}, \bibinfo
  {author} {\bibfnamefont {N.}~\bibnamefont {Imai}}, \bibinfo {author}
  {\bibfnamefont {M.}~\bibnamefont {Ishihara}}, \bibinfo {author}
  {\bibfnamefont {H.}~\bibnamefont {Iwasaki}}, \bibinfo {author} {\bibfnamefont
  {S.}~\bibnamefont {Kanno}}, \bibinfo {author} {\bibfnamefont
  {S.}~\bibnamefont {Kawai}}, \bibinfo {author} {\bibfnamefont
  {T.}~\bibnamefont {Kishida}}, \bibinfo {author} {\bibfnamefont
  {T.}~\bibnamefont {Kubo}}, \bibinfo {author} {\bibfnamefont {K.}~\bibnamefont
  {Kurita}}, \bibinfo {author} {\bibfnamefont {Y.}~\bibnamefont {Matsuyama}},
  \bibinfo {author} {\bibfnamefont {S.}~\bibnamefont {Michimasa}}, \bibinfo
  {author} {\bibfnamefont {T.}~\bibnamefont {Minemura}}, \bibinfo {author}
  {\bibfnamefont {T.}~\bibnamefont {Motobayashi}}, \bibinfo {author}
  {\bibfnamefont {M.}~\bibnamefont {Notani}}, \bibinfo {author} {\bibfnamefont
  {T.}~\bibnamefont {Ohnishi}}, \bibinfo {author} {\bibfnamefont {H.~J.}\
  \bibnamefont {Ong}}, \bibinfo {author} {\bibfnamefont {S.}~\bibnamefont
  {Ota}}, \bibinfo {author} {\bibfnamefont {A.}~\bibnamefont {Ozawa}}, \bibinfo
  {author} {\bibfnamefont {H.~K.}\ \bibnamefont {Sakai}}, \bibinfo {author}
  {\bibfnamefont {H.}~\bibnamefont {Sakurai}}, \bibinfo {author} {\bibfnamefont
  {S.}~\bibnamefont {Shimoura}}, \bibinfo {author} {\bibfnamefont
  {E.}~\bibnamefont {Takeshita}}, \bibinfo {author} {\bibfnamefont
  {S.}~\bibnamefont {Takeuchi}}, \bibinfo {author} {\bibfnamefont
  {M.}~\bibnamefont {Tamaki}}, \bibinfo {author} {\bibfnamefont
  {Y.}~\bibnamefont {Togano}}, \bibinfo {author} {\bibfnamefont
  {K.}~\bibnamefont {Yamada}}, \bibinfo {author} {\bibfnamefont
  {Y.}~\bibnamefont {Yanagisawa}}, \ and\ \bibinfo {author} {\bibfnamefont
  {K.}~\bibnamefont {Yoneda}},\ }\bibfield  {title} {\enquote {\bibinfo {title}
  {Proton inelastic scattering studies at the borders of the ``island of
  inversion'': The $^{30,31}\mathrm{Na}$ and $^{33,34}\mathrm{Mg}$ case},}\
  }\href {\doibase 10.1103/PhysRevC.73.044314} {\bibfield  {journal} {\bibinfo
  {journal} {Phys. Rev. C}\ }\textbf {\bibinfo {volume} {73}},\ \bibinfo
  {pages} {044314} (\bibinfo {year} {2006})}\BibitemShut {NoStop}%
\bibitem [{\citenamefont {Michimasa}\ \emph {et~al.}(2014)\citenamefont
  {Michimasa}, \citenamefont {Yanagisawa}, \citenamefont {Inafuku},
  \citenamefont {Aoi}, \citenamefont {Elekes}, \citenamefont {F\"ul\"op},
  \citenamefont {Ichikawa}, \citenamefont {Iwasa}, \citenamefont {Kurita},
  \citenamefont {Kurokawa}, \citenamefont {Machida}, \citenamefont
  {Motobayashi}, \citenamefont {Nakamura}, \citenamefont {Nakabayashi},
  \citenamefont {Notani}, \citenamefont {Ong}, \citenamefont {Onishi},
  \citenamefont {Otsu}, \citenamefont {Sakurai}, \citenamefont {Shinohara},
  \citenamefont {Sumikama}, \citenamefont {Takeuchi}, \citenamefont {Tanaka},
  \citenamefont {Togano}, \citenamefont {Yamada}, \citenamefont {Yamaguchi},\
  and\ \citenamefont {Yoneda}}]{michimasa2014}%
  \BibitemOpen
  \bibfield  {author} {\bibinfo {author} {\bibfnamefont {S.}~\bibnamefont
  {Michimasa}}, \bibinfo {author} {\bibfnamefont {Y.}~\bibnamefont
  {Yanagisawa}}, \bibinfo {author} {\bibfnamefont {K.}~\bibnamefont {Inafuku}},
  \bibinfo {author} {\bibfnamefont {N.}~\bibnamefont {Aoi}}, \bibinfo {author}
  {\bibfnamefont {Z.}~\bibnamefont {Elekes}}, \bibinfo {author} {\bibfnamefont
  {Zs.}\ \bibnamefont {F\"ul\"op}}, \bibinfo {author} {\bibfnamefont
  {Y.}~\bibnamefont {Ichikawa}}, \bibinfo {author} {\bibfnamefont
  {N.}~\bibnamefont {Iwasa}}, \bibinfo {author} {\bibfnamefont
  {K.}~\bibnamefont {Kurita}}, \bibinfo {author} {\bibfnamefont
  {M.}~\bibnamefont {Kurokawa}}, \bibinfo {author} {\bibfnamefont
  {T.}~\bibnamefont {Machida}}, \bibinfo {author} {\bibfnamefont
  {T.}~\bibnamefont {Motobayashi}}, \bibinfo {author} {\bibfnamefont
  {T.}~\bibnamefont {Nakamura}}, \bibinfo {author} {\bibfnamefont
  {T.}~\bibnamefont {Nakabayashi}}, \bibinfo {author} {\bibfnamefont
  {M.}~\bibnamefont {Notani}}, \bibinfo {author} {\bibfnamefont {H.~J.}\
  \bibnamefont {Ong}}, \bibinfo {author} {\bibfnamefont {T.~K.}\ \bibnamefont
  {Onishi}}, \bibinfo {author} {\bibfnamefont {H.}~\bibnamefont {Otsu}},
  \bibinfo {author} {\bibfnamefont {H.}~\bibnamefont {Sakurai}}, \bibinfo
  {author} {\bibfnamefont {M.}~\bibnamefont {Shinohara}}, \bibinfo {author}
  {\bibfnamefont {T.}~\bibnamefont {Sumikama}}, \bibinfo {author}
  {\bibfnamefont {S.}~\bibnamefont {Takeuchi}}, \bibinfo {author}
  {\bibfnamefont {K.}~\bibnamefont {Tanaka}}, \bibinfo {author} {\bibfnamefont
  {Y.}~\bibnamefont {Togano}}, \bibinfo {author} {\bibfnamefont
  {K.}~\bibnamefont {Yamada}}, \bibinfo {author} {\bibfnamefont
  {M.}~\bibnamefont {Yamaguchi}}, \ and\ \bibinfo {author} {\bibfnamefont
  {K.}~\bibnamefont {Yoneda}},\ }\bibfield  {title} {\enquote {\bibinfo {title}
  {Quadrupole collectivity in island-of-inversion nuclei
  ${}^{28,30}\mathrm{Ne}$ and ${}^{34,36}\mathrm{Mg}$},}\ }\href {\doibase
  10.1103/PhysRevC.89.054307} {\bibfield  {journal} {\bibinfo  {journal} {Phys.
  Rev. C}\ }\textbf {\bibinfo {volume} {89}},\ \bibinfo {pages} {054307}
  (\bibinfo {year} {2014})}\BibitemShut {NoStop}%
\bibitem [{\citenamefont {Strayer}\ \emph {et~al.}(1973)\citenamefont
  {Strayer}, \citenamefont {Bassichis},\ and\ \citenamefont
  {Kerman}}]{strayer1973}%
  \BibitemOpen
  \bibfield  {author} {\bibinfo {author} {\bibfnamefont {M.~R.}\ \bibnamefont
  {Strayer}}, \bibinfo {author} {\bibfnamefont {W.~H.}\ \bibnamefont
  {Bassichis}}, \ and\ \bibinfo {author} {\bibfnamefont {A.~K.}\ \bibnamefont
  {Kerman}},\ }\bibfield  {title} {\enquote {\bibinfo {title} {Correlation
  effects in nuclear densities},}\ }\href {\doibase 10.1103/PhysRevC.8.1269}
  {\bibfield  {journal} {\bibinfo  {journal} {Phys. Rev. C}\ }\textbf {\bibinfo
  {volume} {8}},\ \bibinfo {pages} {1269--1274} (\bibinfo {year}
  {1973})}\BibitemShut {NoStop}%
\bibitem [{\citenamefont {Watts}\ and\ \citenamefont
  {Bartlett}(1995)}]{watts1995}%
  \BibitemOpen
  \bibfield  {author} {\bibinfo {author} {\bibfnamefont {J.~D.}\ \bibnamefont
  {Watts}}\ and\ \bibinfo {author} {\bibfnamefont {R.~J.}\ \bibnamefont
  {Bartlett}},\ }\bibfield  {title} {\enquote {\bibinfo {title} {Economical
  triple excitation equation-of-motion coupled-cluster methods for excitation
  energies},}\ }\href {\doibase 10.1016/0009-2614(94)01434-W} {\bibfield
  {journal} {\bibinfo  {journal} {Chem. Phys. Lett.}\ }\textbf {\bibinfo
  {volume} {233}},\ \bibinfo {pages} {81 -- 87} (\bibinfo {year}
  {1995})}\BibitemShut {NoStop}%
\bibitem [{\citenamefont {de~Swiniarski}\ \emph {et~al.}(1969)\citenamefont
  {de~Swiniarski}, \citenamefont {Glashausser}, \citenamefont {Hendrie},
  \citenamefont {Sherman}, \citenamefont {Bacher},\ and\ \citenamefont
  {McClatchie}}]{swiniarski1969}%
  \BibitemOpen
  \bibfield  {author} {\bibinfo {author} {\bibfnamefont {R.}~\bibnamefont
  {de~Swiniarski}}, \bibinfo {author} {\bibfnamefont {C.}~\bibnamefont
  {Glashausser}}, \bibinfo {author} {\bibfnamefont {D.~L.}\ \bibnamefont
  {Hendrie}}, \bibinfo {author} {\bibfnamefont {J.}~\bibnamefont {Sherman}},
  \bibinfo {author} {\bibfnamefont {A.~D.}\ \bibnamefont {Bacher}}, \ and\
  \bibinfo {author} {\bibfnamefont {E.~A.}\ \bibnamefont {McClatchie}},\
  }\bibfield  {title} {\enquote {\bibinfo {title} {Evidence for ${Y}_{4}$
  deformation lin $^{20}\mathrm{Ne}$ and other $s\ensuremath{-}d$ shell
  nuclei},}\ }\href {\doibase 10.1103/PhysRevLett.23.317} {\bibfield  {journal}
  {\bibinfo  {journal} {Phys. Rev. Lett.}\ }\textbf {\bibinfo {volume} {23}},\
  \bibinfo {pages} {317--320} (\bibinfo {year} {1969})}\BibitemShut {NoStop}%
\bibitem [{\citenamefont {Pritychenko}\ \emph {et~al.}(2016)\citenamefont
  {Pritychenko}, \citenamefont {Birch}, \citenamefont {Singh},\ and\
  \citenamefont {Horoi}}]{pritychenko2016}%
  \BibitemOpen
  \bibfield  {author} {\bibinfo {author} {\bibfnamefont {B.}~\bibnamefont
  {Pritychenko}}, \bibinfo {author} {\bibfnamefont {M.}~\bibnamefont {Birch}},
  \bibinfo {author} {\bibfnamefont {B.}~\bibnamefont {Singh}}, \ and\ \bibinfo
  {author} {\bibfnamefont {M.}~\bibnamefont {Horoi}},\ }\bibfield  {title}
  {\enquote {\bibinfo {title} {Tables of $e2$ transition probabilities from the
  first $2^+$ states in even–even nuclei},}\ }\href {\doibase
  10.1016/j.adt.2015.10.001} {\bibfield  {journal} {\bibinfo  {journal} {At.
  Data Nucl. Data Tables}\ }\textbf {\bibinfo {volume} {107}},\ \bibinfo
  {pages} {1--139} (\bibinfo {year} {2016})}\BibitemShut {NoStop}%
\bibitem [{\citenamefont {Poves}\ and\ \citenamefont
  {Zuker}(1981)}]{poves1981}%
  \BibitemOpen
  \bibfield  {author} {\bibinfo {author} {\bibfnamefont {A.}~\bibnamefont
  {Poves}}\ and\ \bibinfo {author} {\bibfnamefont {A.}~\bibnamefont {Zuker}},\
  }\bibfield  {title} {\enquote {\bibinfo {title} {Theoretical spectroscopy and
  the fp shell},}\ }\href {\doibase 10.1016/0370-1573(81)90153-8} {\bibfield
  {journal} {\bibinfo  {journal} {Phys. Rep.}\ }\textbf {\bibinfo {volume}
  {70}},\ \bibinfo {pages} {235 -- 314} (\bibinfo {year} {1981})}\BibitemShut
  {NoStop}%
\bibitem [{\citenamefont {Horoi}\ \emph {et~al.}(2007)\citenamefont {Horoi},
  \citenamefont {Gour}, \citenamefont {W\l{}och}, \citenamefont {Lodriguito},
  \citenamefont {Brown},\ and\ \citenamefont {Piecuch}}]{horoi2007}%
  \BibitemOpen
  \bibfield  {author} {\bibinfo {author} {\bibfnamefont {M.}~\bibnamefont
  {Horoi}}, \bibinfo {author} {\bibfnamefont {J.~R.}\ \bibnamefont {Gour}},
  \bibinfo {author} {\bibfnamefont {M.}~\bibnamefont {W\l{}och}}, \bibinfo
  {author} {\bibfnamefont {M.~D.}\ \bibnamefont {Lodriguito}}, \bibinfo
  {author} {\bibfnamefont {B.~A.}\ \bibnamefont {Brown}}, \ and\ \bibinfo
  {author} {\bibfnamefont {P.}~\bibnamefont {Piecuch}},\ }\bibfield  {title}
  {\enquote {\bibinfo {title} {Coupled-cluster and configuration-interaction
  calculations for heavy nuclei},}\ }\href {\doibase
  10.1103/PhysRevLett.98.112501} {\bibfield  {journal} {\bibinfo  {journal}
  {Phys. Rev. Lett.}\ }\textbf {\bibinfo {volume} {98}},\ \bibinfo {pages}
  {112501} (\bibinfo {year} {2007})}\BibitemShut {NoStop}%
\bibitem [{\citenamefont {Novario}\ \emph {et~al.}(2021)\citenamefont
  {Novario}, \citenamefont {Gysbers}, \citenamefont {Engel}, \citenamefont
  {Hagen}, \citenamefont {Jansen}, \citenamefont {Morris}, \citenamefont
  {Navr\'atil}, \citenamefont {Papenbrock},\ and\ \citenamefont
  {Quaglioni}}]{novario2021}%
  \BibitemOpen
  \bibfield  {author} {\bibinfo {author} {\bibfnamefont {S.}~\bibnamefont
  {Novario}}, \bibinfo {author} {\bibfnamefont {P.}~\bibnamefont {Gysbers}},
  \bibinfo {author} {\bibfnamefont {J.}~\bibnamefont {Engel}}, \bibinfo
  {author} {\bibfnamefont {G.}~\bibnamefont {Hagen}}, \bibinfo {author}
  {\bibfnamefont {G.~R.}\ \bibnamefont {Jansen}}, \bibinfo {author}
  {\bibfnamefont {T.~D.}\ \bibnamefont {Morris}}, \bibinfo {author}
  {\bibfnamefont {P.}~\bibnamefont {Navr\'atil}}, \bibinfo {author}
  {\bibfnamefont {T.}~\bibnamefont {Papenbrock}}, \ and\ \bibinfo {author}
  {\bibfnamefont {S.}~\bibnamefont {Quaglioni}},\ }\bibfield  {title} {\enquote
  {\bibinfo {title} {Coupled-cluster calculations of neutrinoless
  double-$\ensuremath{\beta}$ decay in $^{48}\mathrm{Ca}$},}\ }\href {\doibase
  10.1103/PhysRevLett.126.182502} {\bibfield  {journal} {\bibinfo  {journal}
  {Phys. Rev. Lett.}\ }\textbf {\bibinfo {volume} {126}},\ \bibinfo {pages}
  {182502} (\bibinfo {year} {2021})}\BibitemShut {NoStop}%
\end{thebibliography}
%

\end{document}